\documentclass[10pt]{article}
\usepackage{lineno}
\usepackage{graphicx}
\usepackage{amssymb}
\usepackage{fancyhdr}
\usepackage{xcolor}
\usepackage{url}
\usepackage{mathtools}
\usepackage{amsmath}
\usepackage[utf8]{inputenc}
\usepackage[english]{babel}
\usepackage{breakcites}
\usepackage{tabu}
\usepackage{mathrsfs}
\usepackage{dsfont}
\usepackage{booktabs}
\usepackage{multirow}
\usepackage{amsthm}
\usepackage{hyperref}
\usepackage[capitalize]{cleveref}
\usepackage{enumitem}
\usepackage{rotating}
\usepackage{graphicx}
\usepackage{lscape}
\usepackage{bbm}
\usepackage[]{interval}
\usepackage{setspace}
\usepackage{bm}
\usepackage{geometry}
\usepackage{scalerel,stackengine}
\intervalconfig{
  soft open fences,
  separator symbol =;,
}
\definecolor{cornellred}{rgb}{0.7, 0.11, 0.11}
\hypersetup{colorlinks,linkcolor={red},citecolor={cornellred},urlcolor={red}}

\allowdisplaybreaks

\newtheoremstyle{break}
  {}{}
  {\itshape}
  {}
  {\bfseries}
  {}
  {\newline}
  {}%

\newtheoremstyle{tightassumption}%
  {6pt}      
  {6pt}      
  {\itshape} 
  {}         
  {\bfseries}
  {.}        
  {0.3em}    
  {}         

\theoremstyle{break}
\theoremstyle{break}
\newtheorem{assumption}{Assumption E}
\newtheorem{theorem}{Theorem}

\newtheorem{rem}{Remark}
\newtheorem{lemma}{Lemma}

\newlist{thmitem}{enumerate}{1}
\setlist[thmitem]{
    label=(\roman*),
    ref={\theparentenv(\roman*)},
    before={\xdef\theparentenv{\csname the\@currentlabel\endcsname}}
}

\newlist{assumitem}{enumerate}{1}
\setlist[assumitem]{
    label=(\roman*),
    ref={\theparentenv(\roman*)},
    before={\xdef\theparentenv{\csname the\@currentlabel\endcsname}}
}

\crefformat{thmitemi}{Theorem #2#1#3}
\crefformat{assumitemi}{Assumption #2#1#3}

\DeclareMathOperator{\spargel}{sp}
\DeclareMathOperator{\cspargel}{\overline{\spargel}}

\DeclareMathOperator{\plim}{plim}

\DeclareMathOperator{\diag}{diag}

\DeclareMathOperator{\E}{\mathbb E}
\DeclareMathOperator{\V}{\mathbb V}

\DeclareMathOperator{\proj}{proj}

\DeclareMathOperator{\tr}{tr}

\newcommand\utimes{\mathbin{\ooalign{$\cup$\cr%
   \hfil\raise0.42ex\hbox{$\scriptscriptstyle\times$}\hfil\cr}}}
\newcommand\bigutimes{\mathop{\ooalign{$\bigcup$\cr%
   \hfil\raise0.36ex\hbox{$\scriptscriptstyle\boldsymbol{\times}$}\hfil\cr}}}

\makeatletter
\renewenvironment{proof}[1][\proofname]{%
  \par\pushQED{\qed}\normalfont%
  \topsep6\p@\@plus6\p@\relax
  \trivlist\item[\hskip\labelsep\bfseries#1\@addpunct{.}]%
  \ignorespaces
}{%
  \popQED\endtrivlist\@endpefalse
}
\makeatother

\stackMath
\newcommand\reallywidehat[1]{%
\savestack{\tmpbox}{\stretchto{%
  \scaleto{%
    \scalerel*[\widthof{\ensuremath{#1}}]{\kern-.6pt\bigwedge\kern-.6pt}%
    {\rule[-\textheight/2]{1ex}{\textheight}}
  }{\textheight}%
}{0.5ex}}%
\stackon[1pt]{#1}{\tmpbox}%
}

\DeclarePairedDelimiter\abs{\lvert}{\rvert}%
\DeclarePairedDelimiter\norm{\lVert}{\rVert}%

\makeatletter
\let\oldabs\abs
\def\abs{\@ifstar{\oldabs}{\oldabs*}}
\let\oldnorm\norm
\def\norm{\@ifstar{\oldnorm}{\oldnorm*}}
\makeatother

\usepackage[round]{natbib}

\makeatletter
\newcommand{\bianca}{\renewcommand\NAT@open{[}\renewcommand\NAT@close{]}}
\makeatother

\providecommand{\keywords}[1]{\textbf{\textit{Index terms---}} #1}

\providecommand{\keywords}[1]{\textbf{\textit{Index terms---}} #1}

\usepackage{accents}
\newcommand{\ubar}[1]{\underaccent{\bar}{#1}}

\usepackage{xcolor}
\hypersetup{
    colorlinks,
    linkcolor={red!50!black},
    citecolor={blue!50!black},
    urlcolor={blue!80!black}
}

\usepackage{comment}
\excludecomment{ext}\excludecomment{comment}

\usepackage{tabu}

\newcolumntype{L}[1]{>{\raggedright\let\newline\\\arraybackslash\hspace{0pt}}m{#1}}
\newcolumntype{C}[1]{>{\centering\let\newline\\\arraybackslash\hspace{0pt}}m{#1}}
\newcolumntype{R}[1]{>{\raggedleft\let\newline\\\arraybackslash\hspace{0pt}}m{#1}}

\DeclareMathAlphabet\mathbfcal{OMS}{cmsy}{b}{n}

\newcommand{\beq}{\begin{equation}}
\newcommand{\eeq}{\end{equation}}
\newcommand{\bea}{\begin{eqnarray}}
\newcommand{\eea}{\end{eqnarray}}
\newcommand{\ba}{\begin{array}}
\newcommand{\ea}{\end{array}}
\newcommand{\bit}{\begin{itemize}}
\newcommand{\eit}{\end{itemize}}
\newcommand{\ben}{\begin{enumerate}} 
\newcommand{\een}{\end{enumerate}}
\newcommand{\bpm}{\begin{pmatrix}}
\newcommand{\epm}{\end{pmatrix}}
\newcommand{\bbm}{\begin{bmatrix}}
\newcommand{\ebm}{\end{bmatrix}}

\renewcommand{\l}{\left}
\renewcommand{\r}{\right}

\newtheoremstyle{break}
  {\topsep}{\topsep}%
  {\itshape}{}%
  {\bfseries}{}%
  {\newline}{}%

\newcommand{\bA}{\bm{A}}

\newcommand{\bC}{\bm{C}}
\newcommand{\bD}{\bm{D}}
\newcommand{\bF}{\bm{F}}

\newcommand{\bH}{\bm{H}}
\newcommand{\bI}{\bm{I}}
\newcommand{\bK}{\bm{K}}
\newcommand{\bM}{\bm{M}}
\newcommand{\bP}{\bm{P}}

\newcommand{\bW}{\bm{W}}

\newcommand{\bsa}{\bm{a}}
\newcommand{\bsb}{\bm{b}}

\newcommand{\bse}{\bm{e}}

\newcommand{\bsk}{\bm{k}}
\newcommand{\bss}{\bm{s}}

\newcommand{\bsu}{\bm{u}}
\newcommand{\bsv}{\bm{v}}
\newcommand{\bsp}{\bm{p}}
\newcommand{\bsx}{\bm{x}}
\newcommand{\bsy}{\bm{y}}

\newcommand{\bmcF}{\bm{\mathcal F}}
\newcommand{\bmcH}{\bm{\mathcal H}}

\newcommand{\bmcK}{\bm{\mathcal K}}

\newcommand{\bbeta}{\bm{\beta}}
\newcommand{\bGamma}{\bm{\Gamma}}

\newcommand{\bdelta}{\bm{\delta}}
\newcommand{\bvarepsilon}{\bm{\varepsilon}}

\newcommand{\bTheta}{\bm{\Theta}}

\newcommand{\bLambda}{\bm{\Lambda}}

\newcommand{\bchi}{\bm{\chi}}

\newcommand{\bseta}{\bm{\eta}}

\newcommand{\bXi}{\bm{\Xi}}
\newcommand{\bxi}{\bm{\xi}}

\newcommand{\bOmega}{\bm{\Omega}}

\begin{document}
\onehalfspacing
\title{
A Distributed Lag Approach \\
to the Generalized Dynamic Factor Model 
%
%
}
\author{
\textsc{Philipp Gersing}\footnote{Department of Statistics and Operations Research, University of Vienna, Kolingasse 14-16, 1090 Vienna, Austria, philipp.gersing@univie.ac.at}, 
%
%
%
%
%
%
%
%
}
\maketitle
\begin{abstract}
We propose a simple estimator for the dynamic decomposition of the Generalized Dynamic Factor Model that avoids frequency-domain methods. First, we show that it is a reasonable approximation to assume that the dynamic common component of the Generalized Dynamic Factor Model admits a representation in terms of current and lagged statically pervasive factors. Then, assuming finite lag order, this simplification reduces estimation to a regression of the observed variables on estimated factors \textit{and their lags}, where the factors are extracted via static principal components. The proposed approach naturally accommodates weak, non-pervasive factors within the dynamic common space. We establish consistency and asymptotic normality for both the dynamic and weak common components under a new asymptotic framework that allows for such weak factors. In an application to three high-dimensional time series panels of European macroeconomic data we detect a sizeable weak common component share in several key macroeconomic indicators.
\end{abstract}
%
%
%
%
%
%
%
%
%
\section{Introduction}
%
%
We consider a high-dimensional time series panel as double indexed zero mean stationary process  $(y_{it}: i \in \mathbb N, t \in \mathbb Z) \equiv (y_{it})$. The Generalized Dynamic Factor Model (GDFM), introduced by \cite{forni2000generalized, forni2001generalized, hallin2013factor}, relies on a decomposition of the form
\begin{align}
    y_{it} = \chi_{it} + \xi_{it} = \sum_{j =0}^\infty \bK_i(j) \bvarepsilon_{t-j} + \xi_{it},  \quad \bvarepsilon_t \sim WN(\bI_q), \label{eq: GDFM rep}
\end{align}
where $(\bvarepsilon_t)$ is an orthonormal white noise process of the common innovations, $\ubar \bK_i(L):= \sum_{j = 0}^\infty \bK_i (j)L^j$ are pervasive causal square summable filters while $L$ denotes the lag operator,
$(\chi_{it})$ is the dynamic common component - driven by the same $q$ shocks $(\bvarepsilon_t)$ for all $i\in \mathbb N$, and $(\xi_{it})$ is the dynamic idiosyncratic component, which is weakly correlated over time and cross-section and orthogonal to $(\chi_{it})$ at all leads and lags. Note, that in view of \cite{hallin2013factor}, stationarity is sufficient for obtaining a dynamic decomposition - with possibly an infinite number $q$ of common shocks. 

The GDFM has proven to be a powerful tool in many applications \citep[see][]{forni2005generalized, forni2004generalized, forni2017dynamic, forni2018dynamic, barigozzi2020consistent, barigozzi2021time, barigozzi2024fnets, barigozzi2025general, anderson2008generalized} and allows for the presence of weak non-pervasive factors in the dynamic common space \citep{gersing2026weak}. 

In addition to the dynamic factor structure \eqref{eq: GDFM rep}, we assume that $(y_{it})$ also has a static factor structure as standard in the literature commencing with \cite{chamberlain1983arbitrage, chamberlain1983funds, stock2002forecasting, bai2002determining}:
\begin{align}
       y_{it} = C_{it} + e_{it} = \bLambda_i \bF_t + e_{it}, \label{eq: static factor rep}
\end{align}
where $\bF_t$ is an $r\times 1$ vector of statically pervasive factors loaded with deterministic loadings $\bLambda_i$ into the static common component $C_{it}$, and $(e_{it})$ is the static idiosyncratic component, weakly correlated in the cross-section and contemporaneously orthogonal to $(\bF_t)$. The number of static factors $r$ is uniquely determined by the number of diverging signal eigenvalues in the variance matrix of $(\bsy_t^n) = (y_{1t}, ..., y_{nt})'$ for $n\to \infty$ \citep[see][]{chamberlain1983arbitrage}.

Available tools for estimation of the dynamic factor decomposition in \eqref{eq: GDFM rep} are \citet{forni2001generalized, hallin2007determining, forni2005generalized, forni2017dynamic, barigozzi2024inferential, barigozzi2024fnets, barigozzi2024algebraic} and mostly designed for the case of \textit{infinite dimensional} factor spaces, meaning that $\dim \cspargel(\chi_{it}: i \in \mathbb N) = \infty$ for all $t\in \mathbb Z$, where $\cspargel(\cdot)$ denotes the closed span. These methods include spectral density estimation and several steps to achieve a representation that is one-sided in the observed variables and therefore fit for forecasting. In this paper, we consider a restricted, finite-dimensional, version of the GDFM with the aim to provide simpler tools for estimation and inference. We propose the following approximation to the GDFM decomposition \eqref{eq: GDFM rep}:
\begin{align}
    y_{it} &= \chi_{it} + \xi_{it} \nonumber \\
           &= \bbeta_{i0} \bF_t + \bbeta_{i1} \bF_{t-1} + ... + \bbeta_{ip} \bF_{t-p} + \xi_{it} = \bbeta_i \bsx_t + \xi_{it} \label{eq: GDFM stat rep}
\end{align}
where $\bsx_t := (\bF_t', ..., \bF_{t-p}')'$ is of dimension $r_\chi = r (p+1)$ and $(\xi_{it})$ is dynamically idiosyncratic \citep{forni2001generalized, hallin2013factor}, orthogonal to $(\bF_t)$ at all leads and lags. In section \ref{sec: structure theory}, we provide detailed arguments why it is reasonable to approximate the general GDFM representation in \eqref{eq: GDFM rep} by the simpler finite distributed lag representation in \eqref{eq: GDFM stat rep}. Estimation of \eqref{eq: GDFM stat rep} proceeds by regressing on estimated factors extracted via principal components. Therefore the use of spectral density methods and problems associated to two-sidedness are avoided. 

The central point of this paper is that the dynamic common component $(\chi_{it})$ in model \eqref{eq: GDFM rep} captures in general richer dynamics than the static common component $(C_{it})$ in \eqref{eq: static factor rep}: In particular, $\chi_{it}$ - as opposed to $(C_{it})$ - incorporates additional feedback from lagged values of $\bF_t$. This feedback cannot be retrieved by using $r_\chi $ instead of $r$ principal components as only $r< r_\chi$ eigenvalues of $\bGamma_{\bsy}^n$ diverge with $n\to \infty$. As proved by \cite{onatskiy2012asymptotics} and demonstrated in the simulation section of this paper, using $r_\chi$ principal components results in inconsistent estimates.

Next, the distributed lags representation \eqref{eq: GDFM stat rep} differs also from specifying dynamics for $\bF_t$ while imposing the static decomposition \eqref{eq: static factor rep} as e.g. in \cite{forni2009opening, doz2011two, barigozzi2019quasi}. More broadly in view of \eqref{eq: GDFM rep}, the dynamic common component represents the part that is dynamically common - that is, the projection of observed variables onto the infinite past of the common shocks \citep[see][]{gersing2026existence}. By contrast, the static common component captures the contemporaneous co-movement as $(C_{it})$ emerges from cross-sectional aggregations holding $t$ fixed.

The general theory linking the GDFM decomposition \eqref{eq: GDFM rep} and the static factor decomposition \eqref{eq: static factor rep} is introduced in \cite{gersing2023reconciling, gersing2026weak} and subsequently considered in the time-domain-framework in \cite{barigozzi2026dynamic}. Both models can be nested within the following canonical decomposition:
\begin{align}
    y_{it} &=\lefteqn{\underbrace{\phantom{C_{it} + e_{it}^\chi}}_{\chi_{it}}}  C_{it} + \overbrace{e_{it}^\chi + \xi_{it}}^{e_{it}} = \bLambda_i \bF_t + \bLambda_i^w \bF_t^w + \xi_{it} \label{eq: can decomp univar} \\
    \bsy_t^n &= \bC_t^n + \bse_t^{\chi, n} + \bxi_t^n \nonumber \\
    &= \underbrace{\bLambda^n \bF_t + \bLambda^{w, n} \bF_t^w}_{\bchi_t^n}  + \bxi_t^n 
    =\begin{bmatrix}
        \bLambda^n & \bLambda^{w, n}
    \end{bmatrix}\begin{bmatrix}
        \bF_t \\[0.5em] \bF_t^w
    \end{bmatrix} + \bxi_t^n , \label{eq: canonical rep weak and strong factors}
\end{align}
The static common component $(C_{it})$ is driven by statically pervasive, or briefly, strong factors, i.e. associated with thick loading columns, usually expressed in terms of $\bLambda^{n'}\bLambda^n / n \to \bGamma_{\bLambda} > \bm{0}$ for $n\to \infty$. On the other hand the weak common component $(e_{it}^\chi)$ is driven by weak non-pervasive factors, i.e. associated with non-divergent loadings $\sup_{n\in \mathbb N} \text{max eval}\l(\bLambda^{w, n'} \bLambda^{w, n} \r) < \infty$ as pioneered in \cite{onatskiy2012asymptotics}. All three terms in the canonical decomposition \eqref{eq: can decomp univar} are \textit{contemporaneously orthogonal}. However, while $(\xi_{it})$ is orthogonal to $(\bF_t)$ and $(\bF_t^w)$ at all leads and lags, $(\bF_t)$ and $(e_{it})$ may correlate at leads and lags via $(\bF_t^w)$. Note that, in view of the GDFM \eqref{eq: GDFM rep}, both $(\bF_t)$ and $(\bF_t^w)$ are driven by the same shocks $(\bvarepsilon_t)$. 

The weak factors $\bF_t^w$ are to be distinguished from rate-weak factors associated to loadings, i.e. $\bLambda^{w, n'} \bLambda^{w, n} / n^\alpha \to \bGamma_{\Lambda^w} > \bm{0}$ with $\alpha \in (0, 1)$ \citep[see][]{demol2008forecasting, freyaldenhoven2022factor, bai2023approximate}.\footnote{In \cite{barigozzi2026dynamic}, it is shown that rate-weak factors are not possible under cross-sectional exchangeability.} In this paper we do not consider rate-weak factors, however the theory could probably be extended to include rate-weak factors, which would be absorbed in $C_{it}$ \citep[see][for details]{gersing2026weak}. 

We obtain the canonical decomposition \eqref{eq: canonical rep weak and strong factors} from the distributed lags representation \eqref{eq: GDFM stat rep} as follows: First, we project out $\bF_t$ from its lags. For this let $\bdelta$ be the population regression parameter from a regression of $\bF_t^- :=(\bF_{t-1}', ..., \bF_{t-p}')'$ onto $\bF_t$ and set $\bbeta_i^-:= (\bbeta_{i1}, ...., \bbeta_{ip})$, then from \eqref{eq: GDFM stat rep} we obtain
\begin{align}
    y_{it} &= \bbeta_i \bsx_t + \xi_{it} 
    = \underbrace{\l(\bbeta_{i0} + \bbeta_i^-\bdelta\r)}_{\bLambda_i} \bF_t + \underbrace{\bbeta_i^-}_{\bLambda_i^w}\underbrace{\l(\bF_t^- - \bdelta \bF_t\r)}_{\bF_t^w} + \xi_{it} = \underbrace{\bLambda_i \bF_t}_{C_{it}} + \underbrace{\bLambda_i^w \bF_t^w}_{e_{it}^\chi} + \xi_{it}. \label{eq: FWL GDFM static rep}
\end{align}

Consequently, we solve the problem raised by \citet{onatskiy2012asymptotics}, who showed that the principal component estimator is inconsistent for weak factors associated with non-divergent eigenvalues. To the best of our knowledge, the only alternative approach is that of \citet{lettau2020estimating, lettau2020factors}, who propose a modified PCA estimator based on a covariance matrix with an overweighted mean to amplify the signal of such factors. The framework developed here allows the influence of $r_w := r_\chi - r$ weak factors to be estimated consistently. In the distributed lags representation \eqref{eq: GDFM stat rep}, the weak factors arise simply from feedback to lagged strong factors, which themselves are recovered from static principal components with a strong cross-sectional signal. In this paper, we focus on inference for the common components $\chi_{it}$ and $e_{it}^{\chi} = \chi_{it} - C_{it}$, rather than the weak factors $\bF_t^w$ themselves.

An interesting interpretation of \eqref{eq: FWL GDFM static rep} is that weak factors are associated to factor loadings that ``taper off'' for lags of strong factors. While the first $r$ factors are pervasive as $\bbeta_0^{n'}\bbeta_0^n \to \infty$, the influence of lags becomes weaker going further in the past as $\bbeta^{-, n'}\bbeta^{-, n} = \bLambda^{w, n'}\bLambda^{w, n} < \infty$. It is plausible that fewer series, are influenced by information from the past, e.g. if they are more (or less) persistent relative to the majority of series in the panel. Implications to impulse response functions have been studied as well in \cite{gersing2026weak}.

Notably, the distributed lag representation \eqref{eq: GDFM stat rep} resembles the factor-augmented regression (FAVAR) framework commencing with \citet{stock2002forecasting, bernanke2005measuring, bai2006confidence}, but requires a different asymptotic analysis. While \citet{bai2006confidence} assume independence between $(e_{it})$ and $(\bF_t)$, the GDFM allows correlation between $(e_{it})$ and $(\bF_{t})$ at leads and lags, while being contemporaneously orthogonal. This introduces additional terms in the asymptotic expansions which be accommodated in the proofs. FA-VAR frameworks that relax independence typically avoid this issue by excluding lagged factors from the regression \citep[e.g.,][]{gonccalves2014bootstrapping}.

The contributions of this paper are as follows: Firstly, we provide representation theory that illuminates why the finite distributed lags representation \eqref{eq: GDFM stat rep} is a reasonable approximation to the GDFM, thereby illuminating the role of feedback from lagged static factors as accounting for weak factors and its resemblance to factor augmented regressions.

Secondly, we establish consistency and derive asymptotic confidence intervals for the dynamic common component, $\chi_{it}$, as well as for the weak common component, $e_{it}^\chi$, in \eqref{eq: can decomp univar} in a framework that allows for the presence of weak factors. 

Thirdly, by doing so, we develop the first asymptotic theory for the GDFM that entirely avoids frequency-domain methods and avoids estimation of the number of dynamic factors $q$ in \eqref{eq: GDFM rep}. Our framework additionally also accommodates time-varying heteroskedasticity in the idiosyncratic component, extending the scope of existing GDFM estimation theory.

Fourthly, while still relying heavily on techniques employed in the seminal work of \cite{bai2006confidence}, we provide alternative, simpler proofs for consistency and asymptotic normality which commence from consistent second moment estimation (Assumption E\ref{A: simply the sample covariances}) and employ the idea of static averaging from \cite{chamberlain1983arbitrage, forni2001generalized, gersing2023reconciling} (see the role of $\bmcK_j$ in the respective Lemmata) by using \cite{yu2015useful} \citep[see also][who proves consistency of the sample-eigenvectors]{barigozzi2022principal}. This can be of independent interest.

Fifthly, in the empirical application we show that there is empirical evidence for sizeable weak common components in key macroeconomic variables, like e.g. Sentiment Indicators (ESENTIX EA), Industrial Production (IPMN FR), Total Hours Worked (THOURS EA) and Gross-Investment Share of Non-Financial Corporations (GNFCIR DE) coming from three different macroeconomic time series panels about the Euro Area. 

The paper is organised as follows. Section \ref{sec: structure theory} develops representation or structure theory \citep{hannan2012statistical} motivating the approximation of the dynamic common component by the distributed lags representation. Section \ref{sec: estimation} provides methods for estimation and inference after introducing suitable assumptions. Section \ref{sec: simulation} presents numerical results, showing consistency of the proposed estimator, failure of the stacking approach based on $r_\chi$ principal components, and coverage properties of the asymptotic confidence intervals. Finally, Section \ref{sec: empirical} applies the suggested methods in practice.
\section{Representation Theory: Identification from Static Factors}\label{sec: structure theory}
Let $\mathcal P = (\Omega, \mathcal A, \mathbb P)$ be a probability space and $L_2(\mathcal P, \mathbb R)$ be the Hilbert space of square integrable real-valued, zero-mean, random-variables defined on $\Omega$ equipped with the inner product $\langle u, v\rangle =  \E [u v]$ for $u, v \in L_2(\mathcal P, \mathbb R)$. If $\mathbb M\subset L_2(\mathcal P, \mathbb R)$ is a linear subspace, we denote by $\proj(\bsu \mid \mathbb M)$ the orthogonal projection of $\bsu$ onto $\mathbb M$ \citep[see e.g.][Theorem 1.2]{deistler2022time}. Now suppose that $(y_{it})$ lives in $L_2(\mathcal P, \mathbb R)$ and set $\mathbb H(\bsy):=\cspargel(y_{it}: i \in \mathbb N, t \in \mathbb Z)$ and $\mathbb H_t(\bsy):=\cspargel(y_{is}: i \in \mathbb N, s \leq t)$. 

Recall that in view of \cite{hallin2013factor} the dynamic decomposition in equation \eqref{eq: GDFM rep} can be obtained in the time domain, only assuming stationarity. Let $\mathbb D(\bsy)$ be the Hilbert space spanned by all dynamic aggregates, i.e. averages over time and cross-section with squared weights tending to zero \citep[see e.g.][for rigorous definitions]{forni2001generalized, hallin2013factor, gersing2026weak}: The dynamic common component of the GDFM is the orthogonal projection $\chi_{it} = \proj(y_{it} \mid \mathbb D(\bsy))$ - provided that $\mathbb D(\bsy) = \mathbb H(\bsu)$ is spanned by $(\bsu_t)$ orthonormal white noise shocks of dimension $q < \infty$. Note that projecting on $\mathbb D(\bsy)$ is a time domain projection which renders residuals that are orthonormal to $\mathbb D(\bsy)$ at all leads and lags (thus the dynamic idiosyncratic component $(\xi_{it})$).

Next, in \cite{gersing2026existence} it is shown that in this projection - except for pathological edge cases - we can replace $\mathbb D(\bsy) = \mathbb H(\bsu)$ by $\mathbb H_t(\bvarepsilon)$ while $\mathbb H_t(\bvarepsilon) \subseteq \mathbb H_t(\bsy)$ and $(\bvarepsilon_t)$ is $q$-dimensional orthonormal white noise process that we may call the ``common innovations'' (the innovations of the dynamic common component), so 
\begin{align}
\chi_{it} = \proj(y_{it} \mid \mathbb H_t(\bvarepsilon)). \label{eq: proj rep chi eps}
\end{align}
In the following we argue, why it is reasonable to assume that $\mathbb H_t(\bvarepsilon) = \mathbb H_t(\bF)$, in order to replace $\mathbb H_t(\bvarepsilon)$ by $\mathbb H_t(\bF)$ in the projection \eqref{eq: proj rep chi eps}. 

Complementary to $\mathbb D(\bsy)$ there is a static aggregation space $\mathbb S_t(\bsy)$ spanned by all cross-sectional aggregates, i.e. averages over $\{y_{it}:i\in \mathbb N\}$ holding time $t$ fixed, with squared weights tending to zero \citep[see][for rigorous definitions]{gersing2026existence, barigozzi2026dynamic}. By \cite{chamberlain1983arbitrage, gersing2023reconciling, gersing2026weak, barigozzi2026dynamic}, we obtain $C_{it} = \proj(y_{it} \mid \mathbb S_t(\bsy))$, while analogously to the dynamic case, the classical static literature assumes that there exists an $r\times 1$ process of static factors, such that $\cspargel(\bF_t) = \mathbb S_t(\bsy)$ for all $t\in \mathbb Z$, consequently $C_{it} = \bLambda_i \bF_t$ for all $t\in \mathbb Z$ and the residuals from that projection are \textit{contemporaneously} orthogonal to $\mathbb S_t(y)$ (thus the static idiosyncratic component $(e_{it})$). 

By the proof of Theorem 5 in \cite{gersing2026weak}, we know that the space spanned by static aggregates of $(y_{it})$ at fixed time $t$ is the same as the space spanned by static aggregates of $(\chi_{it})$. In particular we have the following relations: $\cspargel(\bF_t) = \mathbb S_t(\bsy) = \mathbb S_t(\bchi) \subset \mathbb H_t(\bvarepsilon)$. It follows that $\bF_t$ can be represented as a causal transformation of the common innovations $(\bvarepsilon_t)$. Suppose that $(\bF_t)$ is an ARMA process:
\begin{align}
    \bF_t = \ubar \bsa^{-1}(L) \ubar \bsb (L) \bvarepsilon_t = \ubar \bsk_{\bF} (L) \bvarepsilon_t. \label{eq: F_t ARMA}
\end{align}
Write $\ubar \bsb (z)$ as a polynomial function with $z\in \mathbb C$. If $q = r$ and $\det \ubar \bsb(z) \neq 0$ for all $|z| < 1$, then we say that $\ubar \bsk_{\bF}(z)$ is minimum phase \citep[][section 6.1]{deistler2022time} and there exists causal left inverse of $\ubar \bsk_{\bF} (z)$, say $\ubar \bsk_{\bF, \text{left}} (z) :=  \ubar \bsb^{-1} (z) \ubar \bsa (z)$ such that $\bvarepsilon_t = \sum_{j = 0}^\infty \bK_{\bF, \text{left}} \bF_{t-j}$. Consequently $\bvarepsilon_t \in \mathbb H_t(\bF)$ and therefore $\mathbb H_t(\bvarepsilon) = \mathbb H_t(\bF)$.
On the other hand, if $q < r$, then by Theorem 3 in \cite{anderson2016structure}, generically, i.e. on a set that is open and dense in the ARMA parameter space of $(\bF_t)$, there exists an autoregressive representation $\ubar \bsa (L) \bF_t = \bsb \bvarepsilon_t$ with $\bsb \in \mathbb R^{r \times q}$ and $\bsb$ having rank $q$. So, generically $\mathbb H_t(\bvarepsilon) = \mathbb H_t(\bF)$. This argument has also been used in \cite{anderson2008generalized, forni2015dynamic, forni2025common}. 

Thus, replacing $\mathbb H_t(\bF) = \mathbb H_t(\bvarepsilon)$ in \eqref{eq: proj rep chi eps}, we conclude that if $q = r$ and $\det \bsb(z)\neq 0$ for $|z| < 1$ or if $q<r$ generically, there exists for every $i \in \mathbb N$ a causal filter $\bbeta_i(L)$ such that we can write
\begin{align}
       \chi_{it} = \ubar \bbeta_i(L) \bF_t = \sum_{j = 0}^\infty \bbeta_{ij} \bF_{t-j}. \label{eq: chi as infinite lags of F_t}
\end{align}
To provide an intuition, consider the following examples: Let $(\varepsilon_t)$ be scalar white noise with unit variance:\\ 
\noindent
{
\begin{tabular}{@{}m{0.3\textwidth}@{\quad}m{0.3\textwidth}@{\quad}m{0.3\textwidth}@{}}
\vspace{-\baselineskip} 
\begin{equation}
  \bchi_t = \begin{pmatrix}
        \varepsilon_t \\
        \hline
        (1-0.5L)\varepsilon_t \\
        (1-0.5L)\varepsilon_t \\
        \vdots
    \end{pmatrix}
    \label{eq: exmp zeros outside}
  \end{equation}
  &
  \vspace{-\baselineskip} 
      \begin{equation}
\bchi_t = \begin{pmatrix}
         \varepsilon_t \\
         \hline
        (1-3L) \varepsilon_t \\
        (1-3L) \varepsilon_t \\
        \vdots
    \end{pmatrix} 
    \label{eq: exmp non-retrievable}
  \end{equation}
  &
  \vspace{-\baselineskip} 
  \begin{equation}
 \bchi_t = \begin{pmatrix}
        \varepsilon_t \\
        \hline
        (1-3L) \varepsilon_t \\
        (1-2L) \varepsilon_t \\
        \hline
        (1-3L) \varepsilon_t \\
        (1-2L) \varepsilon_t \\
        \hline
        \vdots 
    \end{pmatrix}. 
    \label{eq: exmp retrievable}
  \end{equation}
\end{tabular}
}
\\
\noindent 
\underline{\textit{Example \eqref{eq: exmp zeros outside}}:}
In this case $q = 1, r = 1, r_w = 1$ with a statically pervasive factor given by $F_t = (1-0.5L) \varepsilon_t$, while $\chi_{1t} = \varepsilon_t$ contains a weak factor. Since here, $\ubar k_{F}(z) = (1-0.5 z)$ has only one zero outside the unit circle at $z_0 = 1/0.5 = 2$, it has a causal inverse and $\mathbb H_t(\varepsilon) = \mathbb H_t(F)$. Note that, the canonical decomposition corresponding to $\chi_{1t}$ can be obtained by orthogonalisation: $\chi_{1t} = \proj(\chi_{1t} \mid F_t) + e_{1t}^\chi = 1/1.25 F_t + (\varepsilon_t - 1/1.25(\varepsilon_t - 0.5 \varepsilon_{t-1})) = 0.8 F_t + (0.2 \varepsilon_t + 0.4 \varepsilon_{t-1})$. For more details see \cite{gersing2026weak}, Proof of Theorem 5 and Appendix B.\\
\noindent \underline{\textit{Example \eqref{eq: exmp non-retrievable}}}: Here $q = 1, r = 1, r_w = 1$ and the statically pervasive factor is given by $F_t = (1-3L)\varepsilon_t$, while $\chi_{1t} = \varepsilon_t$ contains a weak factor. Here $\ubar k_F(z)$ has a zero at $z_0 = 1/3$ \textit{inside} the unit circle. There is only one static factor and we cannot retrieve the common shock $\varepsilon_t$ from the static factor $F_t$. Here, our approach fails and we would need to apply a spectral based approach to retrieve $(\varepsilon_t)$ \citep{forni2000generalized, barigozzi2024inferential, gersing2026weak}. Note, however that this case is highly special, as almost all units of $(\chi_{it})$ have exactly the same zero at $z_0 = 1/3$.\\
\noindent \underline{\textit{Example \eqref{eq: exmp retrievable}}}: We have $q = 1, r = 2, r_w = 1$ with $\bF_t = (F_{1t}, F_{2_t})' = ((1-3L)\varepsilon_t, (1-2L)\varepsilon_t)'$. Although there appear zeros inside the unit circle in the individual rows of $\ubar \bsk_{\bF}(z)) = ((1-3z), (1-2z))'$, it has full rank everywhere and therefore $\bF_t$ has an auto-regressive representation. This implies again that $\mathbb H_t(\varepsilon) = \mathbb H_t(\bF)$. For instance $\varepsilon_t = -2F_{1t} + 3F_{2t}$. Consequently whenever there should be a potential non-fundamentalness such as in \eqref{eq: exmp non-retrievable}, if $r > q$ this is most-likely to be offset by another static factor.

Some comments in order. Firstly, note that the minimum-phase condition, i.e. $\det \bsb(z)\neq 0$ for $|z|< 1$ employed for the case that $q = r$, is standard in factor analysis, and has always been implicitly imposed where $(\bvarepsilon_t)$ in combination with $(\bF_t)$ play a role such as \cite{bai2007determining, stock2011theoxford, forni2009opening}. Secondly, note that potential zeros inside the unit circle are generically ``removed'' by other factors as in example \eqref{eq: exmp retrievable}, an argument that has been also employed in \cite{forni2025common}. There is quite some empirical evidence that the number of static factors usually exceeds the number of dynamic shocks $q$ \citep[see][among others]{bai2007determining, forni2025common}. Thirdly, note that mostly in the econometrics literature $(\bF_t)$ is assumed to be autoregressive \citep[see e.g.][]{doz2011two, forni2005generalized, forni2009opening} in which case no such issues arise.

We conclude that it is reasonable to assume that the dynamic common component admits a representation in terms of static factors \eqref{eq: chi as infinite lags of F_t} and, for tractability, approximate it by a finite lag order $p$ as in \eqref{eq: GDFM stat rep}. Importantly with this approach, we limit the number of static and weak factors to be finite $r <\infty, r_w < \infty$ as opposed to the more general approach in \cite{forni2001generalized, forni2017dynamic}. 

\section{Estimation}\label{sec: estimation}

For a real valued matrix $\bA \in \mathbb R^{n \times m}$ we denote by $\bA'$ the transpose, by $\mu_j(\bA)$ the $j$-th largest eigenvalue of a square matrix $\bA$, by $\norm{\bA} = \sqrt{\mu_1(\bA \bA')}$ the spectral norm and by $\norm{\bA}_F = \sqrt{\tr(\bA \bA')}$ the Frobenius norm. For a vector $\bsv \in \mathbb R^n$ we write $\norm{\bsv}$ to denote the Euclidean norm. 

If $\bsu_t, \bsv_t$ are stochastic vector processes of zero-mean random variables with fixed dimensions $n_1, n_2$, we denote the $n_1\times n_2$-dimensional covariance matrix by $\E\l[\bsu_t \bsv_{t-h}'\r]=: \bGamma_{\bsu\bsv}(h)$ and the sample covariance by $\widehat \bGamma_{\bsu \bsv}(h) := (T-h)^{-1}\sum_{t = h+1}^T \bsu_t \bsv_{t-h}'$ and $\widehat \bGamma_{\bsu\bsv}(0)=: \widehat \bGamma_{\bsu\bsv}$, $\widehat \bGamma_{\bsu}(-h)' = \widehat \bGamma_{\bsu}(h)$. If either $\bsu_t$ or $\bsv_t$ is of dimension $n$ with $n\to\infty$ we add a superscript $n$, and write e.g. $\bGamma_{\bsu\bsv}^n$. Also, we abbreviate $\V \bsv_t := \E\l[\bsv_t \bsv_t'\r] = \bGamma_{\bsv\bsv} =: \bGamma_{\bsv}$. If $\E\l[\bsv_t \bsv_t'\r]$ depends on time we write $\bGamma_{\bsv_t}(h):=\E\l[\bsv_t \bsv_{t-h}'\r]$ and $\bGamma_{\bsv_t}:=\E\l[\bsv_t \bsv_t'\r]$. 

For a (sample-)variance matrix $\bGamma$ of dimension $n\times n$, we set $\bM(\bGamma) \equiv \diag(\mu_1(\bGamma), ..., \mu_r(\bGamma))$ for $r \leq n$. By $\bP(\bGamma)$ the $r\times n$ matrix consisting of the first orthonormal row eigenvectors (corresponding to the $r$ largest eigenvalues of $\bGamma$) and by $\bP^i(\bGamma)$ the $1\times r$ vector consisting of the entries of the $i$-th row of $\bP(\bGamma)'$. The normalized principal components of $\bsy_t^n$ are 
\begin{align*}
    \bW_t^{y, n} &:= \bM^{-1/2}(\bGamma_{\bsy}^n) \bP(\bGamma_{\bsy}^n) \bsy_t^n = \bmcK(\bGamma_{\bsy}^n) \bsy_t^n \\
    \widehat{\bW}_t^{y, n} &:= \bM^{-1/2}(\widehat \bGamma_{\bsy}^n) \bP(\widehat \bGamma_{\bsy}^n) \bsy_t^n = \bmcK(\widehat \bGamma_{\bsy}^n) \bsy_t^n = \widehat{\bmcK} \bsy_t^n. \\
    \widehat \bsx_t&:= \l(\widehat \bW_t^{y, n'}, ...,\widehat \bW_{t-p}^{y, n'} \r)'
\end{align*}
Clearly the normalized principal components of $\bsy_t^n$ are determined only up to sign. To resolve this sign indeterminacy, we assume (without loss of generality by changing the cross-sectional ordering) that the first $r$ rows of $\bP'(\bGamma_{\bsy}^n) \bM^{1/2}(\bGamma_{\bsy}^n)$ have full rank (from a certain $n$ onwards) and fix the diagonal elements to be positive. This fixes the sign of the eigenvectors and the normalized principal components. We use $\widehat{\bmcK}_j$ is the $j$-th row of $\widehat{\bmcK} := \bM^{-1/2}(\widehat \bGamma_{\bsy}^n) \bP(\widehat \bGamma_{\bsy}^n)$ and $\bmcK_j$ is the $j$-th row of $\bmcK := \bM^{-1/2}(\bGamma_{\bsy}^n) \bP(\bGamma_{\bsy}^n)$.
\subsection{Consistency of Factor- and Loadings-Space}
We employ the following assumptions for estimation which are in line with the standard Assumptions from the theoretical literature on factor models \cite{chamberlain1983arbitrage, forni2001generalized, hallin2013factor}, but in addition impose rates on the divergence, allow for heteroscedasticity over time in the dynamic idiosyncratic component but do not require the existence of the spectral density.   
\begin{assumption}[Asymptotic Properties of the Loadings]\label{A: divergence rates eval}
    We assume that model \eqref{eq: GDFM stat rep} holds, with $(\bF_t)$ being a weakly stationary, zero mean process and orthogonal to $(\xi_{it})$ at all leads and lags. Without loss of generality, we employ the normalisation $\E\l[\bF_t \bF_t'\r] = \bGamma_{\bF} = \bI_r$ and $\E\l[\bF_t^w \bF_t^{w'}\r] = \bGamma_{\bF^w} = \bI_{r^w}$; Furthermore $(\bbeta_i: i \in \mathbb N)$ are deterministic loadings, while $\l(\bLambda_i: i \in \mathbb N\r), \l(\bLambda_i^w: i \in \mathbb N\r)$ are obtained in \eqref{eq: FWL GDFM static rep} with the following features:
    \begin{itemize}
    \item[(i)] We assume that $\E\l[\bsx_t \bsx_t'\r] = \bGamma_{\bsx} > \bm{0}$;
    \item[(ii)] Convergence of the ``loadings variance'': Set $n^{-1}\bLambda^{n'}\bLambda^n  = n^{-1} \sum_{i = 1}^n \bLambda_i' \bLambda_i := \bGamma_{\bLambda}^n$ and suppose that $\norm{\bGamma_{\bLambda}^n- \bGamma_{\bLambda}} = \mathcal O(n^{-1/2})$ for some $\bGamma_{\bLambda} > \bm{0}$; 
    \item[(iii)] The eigenvalues of $\bGamma_{\bLambda}$ are distinct and contained in the diagonal matrix $\bD_{\bLambda}$ sorted from the largest to the smallest; 
    \item[(iv)] There exists $\mathcal B_\Lambda < \infty$, such that $\norm{\bLambda_i} < \mathcal B_\Lambda$ and $\norm{\bLambda_i^w} < \mathcal B_{\Lambda^w}$; 
    \item[(v)] There exists $\mathcal B_\xi$, such that $\E \xi_{it}^2 < \mathcal B_\xi$ for all $i\in \mathbb N$; 
    \item[(vi)] Global bounds: $\sup_{t \in \mathbb Z} \sup_{n\in \mathbb N}\mu_1\l(\bGamma_{\bxi_t}^n\r) < \mathcal B_\xi$ and $\sup_{n\in \mathbb N}\mu_1\l(\bLambda^{w, n'} \bLambda^{w, n}\r) < \mathcal B_{\Lambda^w}$.  
    \end{itemize}
\end{assumption}
First note that the assumption $\bGamma_{\bF} = \bI_r$ is without loss of generality for the results in this paper (see remark \ref{rem: Gamma F = I} in the appendix for details) and is made for simplicity. 
Assumption E\ref{A: divergence rates eval}(vi) allows that dynamic idiosyncratic variance to be time dependent but globally bounded in the first eigenvalue and imposes furthermore the non-pervasiveness of the weak factors.

In the following we use the eigen-decomposition $\bGamma_{\bLambda}^n := \bP_{\bLambda}^{n'} \bD_{\bLambda}^n \bP_{\bLambda}^n$ with eigenvalues sorted from the largest to the smallest in the diagonal matrix $\bD_{\bLambda}^n$ and $\bP_{\bLambda}^n$ being a matrix of orthonormal row eigen-vectors. Analogously we use $\bGamma_{\bLambda} = \bP_{\bLambda}' \bD_{\bLambda} \bP_{\bLambda}$. Assumption E\ref{A: divergence rates eval} implies also that $\sup_{t\in \mathbb Z} \sup_{n\in \mathbb N} \mu_1 \l( \bGamma_{\bse_t}^n  \r) \leq \mathcal B_\xi +\mathcal B_{\Lambda^w}=: \mathcal B_e < \infty$.

Let $\bseta_t^n := n^{-1/2}\bLambda^{n'} \bxi_t^n$ be the $r\times 1$ dimensional weighted average of the idiosyncratic component obtained from the scaled static factor loadings. Note that by Assumption E\ref{A: divergence rates eval}, it holds that $\norm{\E\l[\bseta_t^n \bseta_t^{n'}\r]} < \mathcal B_\xi \norm{n^{-1/2}\bLambda^n}< \mathcal B_\xi \mathcal B_\Lambda$. 

\begin{assumption}[Sample Covariances]\label{A: simply the sample covariances}
We suppose that the sample covariances can be estimated consistently (up to heteroskedasticity) as 
\begin{itemize}
    \item[(i)] $\E\l[\norm{\sqrt{T}\l(\widehat \bGamma_{\bF}(h) - \bGamma_{\bF}(h)\r)}^2\r] \leq \mathcal B_F$ for all $\abs{h} <\infty$; 
    \item[(ii)] $\E\l[\l(\frac{1}{\sqrt{T}}\sum_{t = 1}^T \l\{\xi_{it}\xi_{js} - \E\l[\xi_{it} \xi_{js}\r]\r\} \r)^2\r] \leq \mathcal B_\xi $  independent of $i, j \in \mathbb N$, $t, s\in \mathbb Z$; 
    \item[(iii)] $\E\l[\norm{\frac{1}{\sqrt{T}} \sum_{t = 1}^T \l\{\bseta_t^n \bseta_{t-h}^{n'} - \E\l[\bseta_t^n \bseta_{t-h}^{n'}\r]\r\}}^2\r] \leq \mathcal B_\xi$ for all $\abs{h} < \infty, n\in \mathbb N$;
    \item[(iv)] $\E\l[\norm{\sqrt{T}\widehat \bGamma_{\bF \bxi}(h)\bss_i}^2\r] \leq \mathcal B_{F\xi}$ independent of $i \in \mathbb N$ and $\abs{h} <\infty$;
    \item[(v)] 
    $\E\l[\norm{\frac{1}{\sqrt{T}}\sum_{t = 1}^T \bseta_t^n \bF_{t-h}}^2\r] \leq \mathcal B_{F\xi}$ for $\abs{h} <\infty, n\in \mathbb N$. 
\end{itemize}
\end{assumption}
These conditions establish mean square convergence of the sample covariances. Note that in the standard setup of \cite{chamberlain1983arbitrage}, the variance of \textit{any} cross-sectional average (scaled by $\sqrt{n}$) is bounded in $n$. In particular consider an aggregate $\bsp^{n'}\bxi_t^n$ with any $\bsp^n \in \mathbb R^n$ for all $n \in \mathbb N$, with $\sup_{n\in \mathbb N}\norm{\bsp^n}<\infty$, of which $\bseta_t^n$ is a special case. Then $\E\l[(\bsp^{n'}\bxi_t^n)^2\r]\leq \sup_{n\in \mathbb N}\norm{\bsp^n}\mathcal B_\xi < \infty$ independent of $n \in \mathbb N$. 
\begin{theorem}[Consistency of Factors' and Loadings' Spaces and Common Components]\label{thm: consistency of spaces and CCs}
    Under Assumptions E\ref{A: divergence rates eval} and E\ref{A: simply the sample covariances}, with $\widehat{\bH}= \frac{1}{T}\sum_{t = 1}^T \widehat \bF_t \bF_t' \l(\frac{1}{T} \sum_{t = 1}^T \bF_t \bF_t'\r)^{-1}$ and 
    $\widehat{\bmcH}= \frac{1}{T}\sum_{t = 1}^T \widehat \bsx_t \bsx_t' \l(\frac{1}{T} \sum_{t = 1}^T \bsx_t \bsx_t'\r)^{-1}$
    it follows that
    \begin{itemize}
        \item[(i)] $\norm{\widehat{\bW}_t^{y, n} - \widehat{\bH}\bF_t} = \mathcal O_{P}(\max(n^{-1/2}, T^{-1/2}))$, $\norm{\widehat{\bW}_t^{y, n} - \bP_{\bLambda} \bF_t} = \mathcal O_{P}(\max(n^{-1/2}, T^{-1/2}))$ \\
        and $\norm{\widehat \bsx_t - \widehat{\bmcH} \bsx_t} = \mathcal O_{P}(\max(n^{-1/2}, T^{-1/2}))$, $\norm{\widehat \bsx_t - \l(\bI_{p+1} \otimes \bP_{\bLambda}\r) \bsx_t} = \mathcal O_{P}(\max(n^{-1/2}, T^{-1/2}))$; 
        \item[(ii)] $\norm{\widehat \bLambda_i -  \bLambda_i \widehat{\bH}^{-1}} = \mathcal O_{P}(\max(n^{-1/2}, T^{-1/2}))$, $\norm{\widehat \bLambda_i -  \bLambda_i \bP_{\bLambda}'} = \mathcal O_{P}(\max(n^{-1/2}, T^{-1/2}))$\\
        and $\norm{\widehat \bbeta_i - \bbeta_i \widehat{\bmcH}^{-1}} = \mathcal O_{P}(\max(n^{-1/2}, T^{-1/2}))$, $\norm{\widehat \bbeta_i - \bbeta_i \l(\bI_{p+1} \otimes \bP_{\bLambda}'\r)} = \mathcal O_{P}(\max(n^{-1/2}, T^{-1/2}))$;
        \item[(iii)] $\norm{\widehat C_{it} - C_{it}} = \mathcal O_{P}(\max(n^{-1/2}, T^{-1/2}))$, $\norm{\widehat \chi_{it} - \chi_{it}} = \mathcal O_{P}(\max(n^{-1/2}, T^{-1/2}))$ and $\norm{\widehat e_{it}^\chi - e_{it}^\chi} = \mathcal O_{P}(\max(n^{-1/2}, T^{-1/2}))$ with $\widehat e_{it}^\chi = \widehat \chi_{it} - \widehat C_{it}$.
    \end{itemize}
\end{theorem}
Theorem \ref{thm: consistency of spaces and CCs} states convergence in probability of the sample normalized principal components and the associated sample loadings to the space spanned by the true factors, or the space spanned by the true loadings respectively. 
\subsection{Asymptotic Normality}
The following Assumption about weak convergence is standard in factor analysis \citep[commencing from][Assumption F]{bai2003inferential}. Alternatively, primitive conditions are provided e.g. in \cite{anatolyev2021limit} or \cite{barigozzi2022principal}.
\begin{assumption}[Central Limit Theorem]\label{A: CLTs}
Define $\bXi_t^n:= (\bxi_t^{n'}, ..., \bxi_{t-p}^{n'})'$
\begin{itemize}
    \item[(i)] $T^{-1/2} \bsx'\bxi^i \Rightarrow \mathcal N\left(0, \bOmega_{\bsx\bxi}(i)\right)$ and $\bGamma_{\bsx} > \bm{0}$ and $T^{-1/2} \bF' \bse^i \Rightarrow \mathcal N\left(\bm{0}, \bOmega_{\bF\bse}(i)\right)$;
    \item[(ii)] $\l(\bI_{p+1} \otimes n^{-1/2}\bLambda^{n'} \r) \bXi_t^n \Rightarrow \mathcal N\l(\bm{0}, \bTheta_{\bLambda\bXi}(t)\r)$ and $n^{-1/2}\bLambda^{n'} \bse_t^n \Rightarrow \mathcal N\l(\bm{0}, \bTheta_{\bLambda \bse}(t)\r)$;
    \item[(iii)] The asymptotic covariances are given by
    \begin{align*}
    &\lim_{T\to \infty} \E\l[\l(T^{-1/2}\bsx'\bxi^i\r) \l(T^{-1/2}\bF'\bse^i\r)'\r] = \bOmega_{\bsx\bxi, \bF\bse}(i)\\
    \mbox{and}\quad &\lim_{n \to \infty}\E\l[\l(\l(\bI_{p+1} \otimes n^{-1/2}\bLambda^{n'} \r) \bXi_t^n\r)\l(n^{-1/2}\bLambda^{n'} \bse_t^n\r)'\r] = \bTheta_{\bLambda\bXi, \bLambda \bse}(t), 
    \end{align*}
    while $\bTheta_{\bLambda\bXi, \bLambda \bse}(t) = \bTheta_{\bLambda\bXi, \bLambda\bxi}(t)$ due to orthogonality relations.
\end{itemize}
\end{assumption}
Furthermore to establish asymptotic normality results, we need slightly more restrictive assumptions which refer mostly to the serial- and cross-correlation in the idiosyncratic component.
\begin{assumption}[Idiosyncratic Auto-Covariance]\label{A: more restrictive sample covarainces}\ \\[-3em]
\begin{itemize}
    \item[(i)] 
    Summability in the cross-section: Let 
    \begin{align*}
    &\sup_{n\in \mathbb N} \max_{1 \leq i \leq n}\sum_{j = 1}^n \abs{\E[\xi_{it}\xi_{j,t-k}]} < \mathcal B_\xi, \quad \mbox{and} \ \sup_{n\in \mathbb N} \max_{1 \leq j \leq n}\sum_{i = 1}^n \abs{\E[\xi_{it}\xi_{j,t-k}]} < \mathcal B_\xi \ \mbox{for all} \ t, k \in \mathbb Z,
    \end{align*}
    which implies that $\bmcK \bGamma_{\bxi}^n(h) = \mathcal O(n^{-1})$;
    \item[(ii)] $\bmcK \bLambda^{w, n} = \mathcal O(n^{-1})$ or $\bLambda^{n'} \bLambda^{w, n} = \mathcal O(1)$ which is implied by $\norm{\sum_{i = 1}^n \bLambda_i^{w, n}} < \mathcal B_{\Lambda^w}$;
    \item[(iii)] For all $1\leq j \leq n$, $1 \leq s \leq T$ and $i, t, n, T \in \mathbb N$
    \begin{align*}
        \E\left[\abs{\frac{1}{\sqrt{nT}} \sum_{i = 1}^n \sum_{t = 1}^T \left\{\xi_{is} \xi_{jt} - \E[\xi_{is}\xi_{jt}]\right\} }^2\right] < \mathcal B_\xi;
    \end{align*}
    \item[(iv)] Either $(\xi_{it})$ and $(\bF_t)$ are independent or alternatively for all $t, n, T \in \mathbb N$ and $\abs{h} < \infty$,
    \begin{align*}
     &\E[\xi_{it} \xi_{j, t-h}] \leq \abs{\rho}^h \mathcal B_\xi \ \mbox{for some} \ 0 < \rho < 1, \\
    &\E\norm{   
        \frac{1}{\sqrt{nT}} \sum_{i = 1}^n \sum_{s = 1}^T \bF_s \l[\xi_{is} \xi_{it} - \E\l[\xi_{is} \xi_{it}\r]\r]}^2 < \mathcal B_\xi, \quad \mbox{and} \ \E\norm{\sum_{t = 1}^T \bF_s \rho^{\abs{s-t}}}^2 < \mathcal B_F.
    \end{align*}
    \end{itemize}
\end{assumption}
Assumption E\ref{A: more restrictive sample covarainces} is similar to the conditions used in \cite{barigozzi2022principal} who provides an excellent overview on how the usual Assumptions of the factor model literature are nested in Assumption E\ref{A: more restrictive sample covarainces}. Whereas for consistency, we only need $L^2$ boundedness of dynamic and static idiosyncratic variance, for asymptotic normality we need $L^1$ boundedness reflected in E\ref{A: more restrictive sample covarainces}(i), (ii). Furthermore, additional bounds are needed that restrict serial dependence of the dynamic idiosyncratic part E\ref{A: more restrictive sample covarainces}(iii), (iv).  
\begin{lemma}\label{lem: asymptotic normality loadings residuals}
    Under Assumptions E\ref{A: divergence rates eval}-E\ref{A: more restrictive sample covarainces}, as $\sqrt{T} / n \to 0$,  with $\widehat{\bmcH} = T^{-1}\sum_{t = 1}^T \widehat \bsx_t \bsx_t' \left(T^{-1}\sum_{t = 1}^T  \bsx_t \bsx_t'\right)^{-1}$, we have
    \begin{align*}
        \sqrt{T} \left(\widehat \bbeta_i - \bbeta_i \widehat{\bmcH}^{-1}\right) \Rightarrow \mathcal N\left(\bm{0}, asy\bGamma_{\widehat \bbeta_i}\right)
    \end{align*}
    The asymptotic variance $asy\bGamma_{\widehat \bbeta_i}$ is given by 
    \begin{align}
        &asy\bGamma_{\widehat \bbeta_i} :=  \l(\bI_{p+1} \otimes \bP_{\bLambda}\r) \bGamma_{\bsx}^{-1}\bOmega_{\bsx\bxi}(i) \bGamma_{\bsx}^{-1} \l(\bI_{p+1} \otimes \bP_{\bLambda}'\r) . \label{eq: asymptotic variance hat beta}
    \end{align}
\end{lemma}
Note that $\hat \bmcH^{-1}$ is asymptotically the OLS regression of the estimated on the true loadings (see Lemma \ref{lem: H}). Lemma \ref{lem: asymptotic normality loadings residuals} states that the residuals of that regression are asymptotically normal. 

Following \cite{barigozzi2022principal} for the asymptotic variance in \eqref{eq: asymptotic variance hat beta}, robust to heteroskedasticity, we may use: 
\begin{align}
    %
     \reallywidehat{\l(\bI_{p+1} \otimes \bP_{\bLambda}\r) \bOmega_{\bsx\bxi}(i)\l(\bI_{p+1} \otimes \bP_{\bLambda}\r)'}  &=\frac{1}{T}\sum_{t = 1}^T \sum_{s = 1}^T \widehat \bsx_t \widehat \xi_{it} \widehat \xi_{is}\widehat{\bsx}_s' \kappa(t, s) \label{eq: HAC2 asy beta} \\
     \reallywidehat{\l(\bI_{p+1}\otimes \bP_{\bLambda}\r) \bGamma_{\bsx} \l(\bI_{p+1}\otimes \bP_{\bLambda}\r)'} &=  \frac{1}{T}\sum_{t = 1}^T \widehat \bsx_t \widehat \bsx_t' \label{eq: sample var hat x}
\end{align}
where $\kappa(t, s)$ is a suitable kernel with bandwidth $M_T$ and \eqref{eq: sample var hat x} is from Lemma \ref{lem: H}(vi). The final estimator is given by
\begin{align*}
     asy\widehat \bGamma_{\widehat \bbeta_i} = \left(\frac{1}{T}\sum_{t = 1}^T \widehat \bsx_t \widehat \bsx_t'\right)^{-1}\reallywidehat{\l(\bI_{p+1} \otimes \bP_{\bLambda}\r) \bOmega_{\bsx\bxi}(i)\l(\bI_{p+1} \otimes \bP_{\bLambda}\r)'} \left(\frac{1}{T}\sum_{t = 1}^T \widehat \bsx_t \widehat \bsx_t'\right)^{-1}. 
\end{align*}
\begin{lemma}\label{lem: asymptotic normality of hat bsxt}
Under Assumptions E\ref{A: divergence rates eval}-E\ref{A: more restrictive sample covarainces}, as $\sqrt{n} / T \to 0$, we have
    \begin{align*}
        \sqrt{n}\left(\widehat \bsx_t - \widehat{\bmcH} \bsx_t\right) \Rightarrow \mathcal N(\bm{0}, asy\bGamma_{\widehat \bsx_t})
    \end{align*}
    The asymptotic variance $asy\bGamma_{\widehat \bsx_t}$ is given by 
    \begin{align*}
      asy\bGamma_{\widehat \bsx_t} :=   \l(\bI_{p+1}\otimes \bD_{\bLambda}^{-1} \bP_{\bLambda}\r) \bTheta_{\bLambda\bXi}(t)\l(\bI_{p+1}\otimes \bP_{\bLambda}' \bD_{\bLambda}^{-1}\r). 
    \end{align*}
\end{lemma}
This Lemma states that the residuals from the regression of the estimated on the true factors are asymptotically normal. 
    %
%

%
For estimation of the asymptotic variance we may use \cite{fresoli2024dealing}, which accounts for cross-sectional correlation in the idiosyncratic term: 
\begin{align}    
    \reallywidehat{\bP_{\bLambda} \bTheta_{\bLambda \bxi, h}\bP_{\bLambda}'} &= \frac{1}{n}\sum_{i = 1}^n \sum_{j = 1}^n \widehat \bLambda_i' \widehat \bLambda_j \frac{1}{T - h} \sum_{t = h+1}^{T} \widehat \xi_{it}\widehat \xi_{j,t-h} I\l(\abs{\widehat \sigma_{ij, h}^\xi} \geq c_{ij, h}\r) \label{eq: est cov Lambda xi, h}
\end{align}
with $\widehat \sigma_{ij, h}^\xi := (T-h)^{-1}\sum_{t = h+1}^T \widehat \xi_{it} \widehat \xi_{j,t-h}$, $\widehat{\V}\l[\widehat \xi_{it} \widehat \xi_{j,t-h}\r] := (T-h)^{-1}\sum_{t = h+1}^T \l[\widehat \xi_{it} \widehat \xi_{jt} - \widehat \sigma_{ij, h}^\xi\r]^2$ and $c_{ij, h} := \delta \l[\widehat{\V}\l[\widehat \xi_{it} \widehat\xi_{j,t-h}\r] \log(n) / (T-h)\r]^{1/2}$, while $\delta = [2(2-\gamma)]^{1/2}$ with $\gamma \in (0, 2)$ is a parameter for controlling sparsity. 
\begin{align*}
    asy \widehat \bGamma_{\widehat \bsx_t} = \l(\bI_{p+1} \otimes \l(\widehat \bD_{\bLambda}^n\r)^{-1}\r)
    \begin{pmatrix}
        \reallywidehat{\bP_{\bLambda} \bTheta_{\bLambda\bxi, 0}\bP_{\bLambda}'} & \cdots & \reallywidehat{\bP_{\bLambda} \bTheta_{\bLambda\bxi, p}\bP_{\bLambda}'} \\
                                      &  \ddots      & \\
       \reallywidehat{\bP_{\bLambda} \bTheta_{\bLambda\bxi, -p}\bP_{\bLambda}'} & \cdots & \reallywidehat{\bP_{\bLambda} \bTheta_{\bLambda\bxi, 0}\bP_{\bLambda}'}
    \end{pmatrix} \l(\bI_{p+1} \otimes \l(\widehat \bD_{\bLambda}^n\r)^{-1}\r).
\end{align*}
Using the previous two Lemmata, we obtain asymptotic normality for the common component estimator for $i = 1, ..., n$ and $t = p+1, ..., T$:
\begin{theorem}[Asymptotic Normality of the Dynamic Common Component]\label{thm: asy norm chi}
Under Assumptions E\ref{A: divergence rates eval}-E\ref{A: more restrictive sample covarainces}, as $\sqrt{n} / T \to 0$ and $\sqrt{T} / n \to 0$, we have 
\begin{align*}
    \frac{\widehat \chi_{it} - \chi_{it}}{\sqrt{\frac{1}{T} U_{it} + \frac{1}{n}V_{it}}} \Rightarrow \mathcal N(0, 1), 
\end{align*}
where  $U_{it} := \bsx_t'\bGamma_{\bsx}^{-1} \bOmega_{\bsx\bxi}(i) \bGamma_{\bsx}^{-1}\bsx_t$ and $V_{it} =\bbeta_i\l(\bI_{p+1} \otimes \bGamma_{\bLambda}^{-1}\r) \bTheta_{\bLambda\bXi}(t) \l(\bI_{p+1} \otimes \bGamma_{\bLambda}^{-1}\r) \bbeta_i'$.
\end{theorem}
For estimating the standard deviations $U_{it}$ and $V_{it}$ in Theorem \ref{thm: asy norm chi}, we use:
\begin{align}
    \widehat U_{it} = \widehat \bsx_t' asy \widehat \bGamma_{\widehat \bbeta_i} \widehat \bsx_t, \qquad \widehat V_{it} = \widehat \bbeta_i asy \widehat \bGamma_{\widehat \bsx_t} \widehat \bbeta_i'. \label{eq: est Uit Vit}
\end{align}
Analogously, we obtain the asymptotic normality of the weak common component estimator:
\begin{theorem}[Asymptotic Normality of the Weak Common Component]\label{thm: asy norm $e_{1t}^chi$}
Under Assumptions E\ref{A: divergence rates eval}-E\ref{A: more restrictive sample covarainces}, as $\sqrt{n} / T \to 0$ and $\sqrt{T} / n \to 0$, we have 
\begin{align*}
    \frac{\widehat e_{it}^\chi - e_{it}^\chi}{\sqrt{\frac{1}{T}U_{it} + \frac{1}{n} V_{it}}} \Rightarrow \mathcal N(0, 1), 
\end{align*}
where 
\begin{align*}
    U_{it} &:= \bsx_t'\bGamma_{\bsx}^{-1} \bOmega_{\bsx\bxi}(i) \bGamma_{\bsx}^{-1}\bsx_t + \bF_t' \bOmega_{\bF\bse}(i) \bF_t -2 \bsx_t'\bGamma_{\bsx}^{-1} \bOmega_{\bsx\bxi, \bF\bse}(i) \bF_t \\
    V_{it} &:= \bbeta_i \l(\bI_{p+1} \otimes \bGamma_{\bLambda}^{-1}\r) \bTheta_{\bLambda\bXi}(t) \l(\bI_{p+1} \otimes \bGamma_{\bLambda}^{-1}\r) \bbeta_i' + \bLambda_i \bGamma_{\bLambda}^{-1}\bTheta_{\bLambda \bse}(t)  \bGamma_{\bLambda}^{-1} \bLambda_i'  \\
    & \qquad \qquad \qquad - 2\bbeta_i \l(\bI_{p+1} \otimes \bGamma_{\bLambda}^{-1}\r) \bTheta_{\bLambda \bXi, \bLambda \bse}(t) \bGamma_{\bLambda}^{-1} \bLambda_i'.
\end{align*}
\end{theorem}
The covariance terms can be estimated using: 
\begin{align*}
     \reallywidehat{\l(\bI_{p+1} \otimes \bP_{\bLambda}\r) \bOmega_{\bsx\bxi, \bF\bse}(i)\bP_{\bLambda}'}  &=\frac{1}{T}\sum_{t = 1}^T \sum_{s = 1}^T \widehat \bsx_t \widehat \xi_{it} \widehat e_{is} \widehat \bF_s' \kappa(t, s), 
     %
\end{align*}
while for $\reallywidehat{\bP_{\bLambda} \bTheta_{\bLambda \bxi, \bLambda \bse, h}\bP_{\bLambda}'}$ we use the same formula as in \eqref{eq: est cov Lambda xi, h}, 
so $(T-h)^{-1}\sum_{t = h+1}^{T}\widehat \xi_{it}\widehat \xi_{i, t-h}$ instead $(T-h)^{-1}\sum_{t = h+1}^{T}\widehat \xi_{it}\widehat e_{i, t-h}$ which is justified by the orthogonality between $\bF_t^w$ and $\xi_{is}$ for all leads and lags and $i \in \mathbb N$. The standard deviations are estimated analogously to \eqref{eq: est Uit Vit} using the formulas of Theorem \ref{thm: asy norm $e_{1t}^chi$}.

\section{Simulation Experiments}\label{sec: simulation}
To assess the finite sample properties of the proposed procedures, we consider the prototypical dynamic factor model \citep[see][Remark 9]{gersing2026weak} with data generated as
\begin{align}
    y_{it} &= \chi_{it} + \xi_{it} = \lambda_{i0} f_t + \lambda_{i1} f_{t-1} + \xi_{it}, \ \mbox{with} \ f_t = a f_{t-1} + \varepsilon_t , \quad \varepsilon_t \sim \mathcal N\left(0, (1-a^2)\right)  \label{eq: innocent factor model} 
    %
\end{align}
with $a = 0.8$, $\E[f_t^2] = 1$, one strong and one weak factor for which we set:
\begin{align}
\lambda_{i0}=\l\{
\begin{array}{ccc}
0 &\text{if}& 1\le i\le 10,\\
1 &\text{if}& 11\le i\le 20,\\
\mathcal N(1,1)&\text{if}& i\ge 21,
\end{array}
\r.
\quad
\lambda_{i1}=\l\{
\begin{array}{ccc}
1 &\text{if}& 1\le i\le 10,\\
0 &\text{if}&  i\ge 11.\\
\end{array}
\r.\label{eq:DG1load}
\end{align}
For the dynamic idiosyncratic component $(\xi_{it})$, we consider different versions, based on the specification $\xi_{it} = \alpha_i \xi_{i, t-1} + \varepsilon^\xi_{it}$ with $\varepsilon_{it}^\xi \sim iidN(0, 1)$, independent of $(\varepsilon_t)$, with $\E(\varepsilon_{it}^\xi \varepsilon_{jt}^\xi) = \tau^{\abs{i-j}}, i, j = 1, ..., n$, $\tau \in \{0, 0.5\}$ if $\abs{i-j}\leq 10$ and $\E(\varepsilon_{it}^\xi \varepsilon_{jt}^\xi) = 0$ otherwise; last $\alpha_i = \{0, \delta_i\}$ with $\delta_i \sim iidU(0, \delta)$ and $\delta \in \{0, 0.5\}$. The parameters $\tau$ and $\delta$ are crucial to control the cross-sectional and serial correlation in the dynamic idiosyncratic component, respectively.

So $q = 1, r=1$ and $r_w=1$, and, therefore, $(\chi_{it})$ has covariance $\Gamma_\chi^n$ of rank $r_\chi=2$, but with only the largest eigenvalue diverging as $n$ grows. In particular, the first unit, $i = 1$, decomposes as
\begin{align}
    \chi_{1t} &= a f_t + (f_{t-1} - a f_t) = f_{t-1}, \quad
    C_{1t} =   a f_t , \quad e_{1t}^\chi= f_{t-1} - a f_t,\label{eq: DGP1 C1}
\end{align}
so it has both a non-zero static and weak common component, which are mutually orthogonal by construction.
\paragraph{Consistency of FDL versus Non-Consistency of SPCA.}
We generate data according to \eqref{eq: innocent factor model}. At each replication $j = 1, ..., B$ and for each considered estimator of $\chi_{1t}^{[j]}$, generically denoted as $\widehat \chi_{1t}^{[j]}$, we define $MSE_1 = B^{-1}\sum_{j = 1}^B T^{-1}\sum_{t = 1}^T (\widehat \chi_{1t}^{[j]} - \chi_{1t}^{[j]})^2$ and look at results for different $(n, T)$ combinations with $B = 500$ replications.

\begin{figure}
    \centering
	\begin{tabular}{cc}
    \includegraphics[width=.4\textwidth]{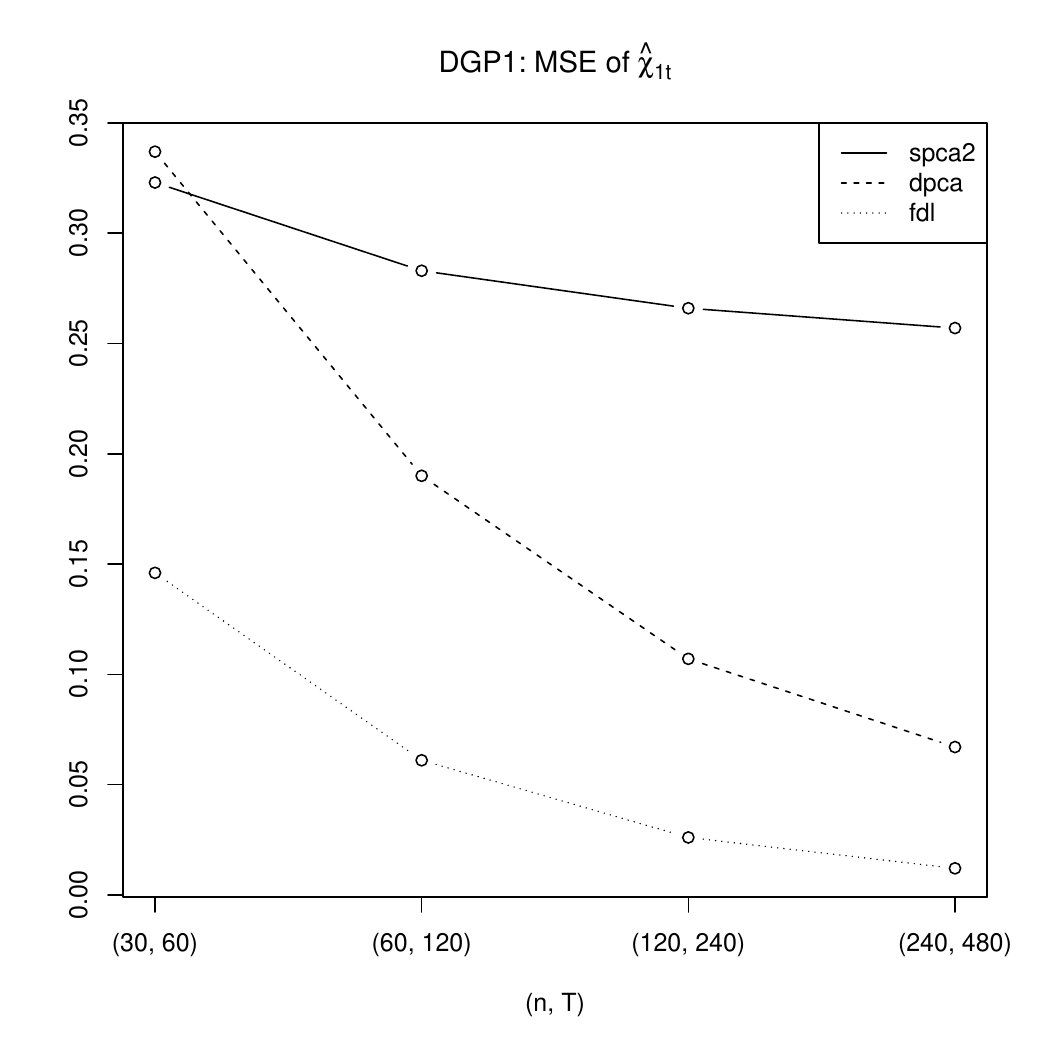}&
    \includegraphics[width=.4\textwidth]{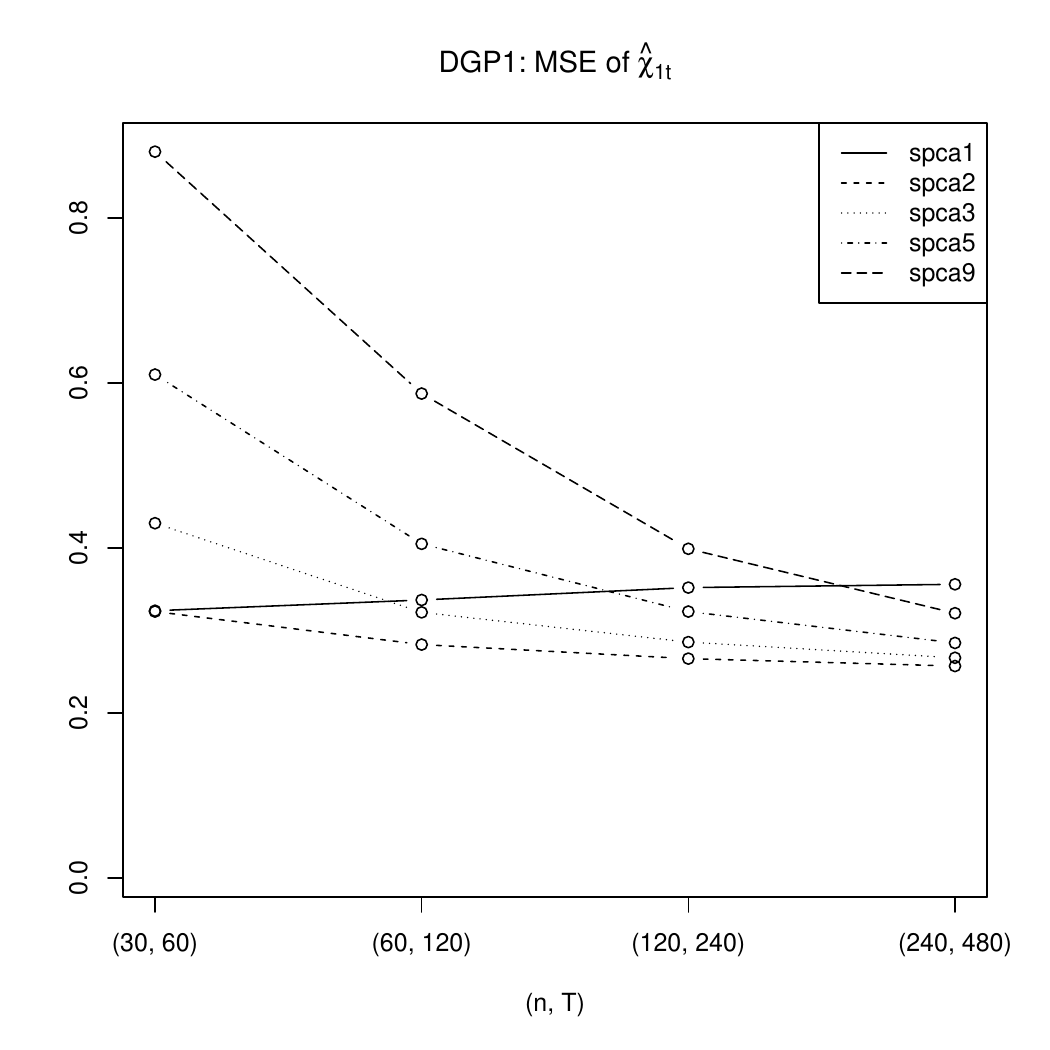}
    \end{tabular}
    \caption{ 
    \footnotesize \textsc{Consistency:}
    Mean Squared Error of $\widehat \chi_{1t}$ over 500 replications under idiosyncratic serial and cross-correlation $(\tau, \delta) = (0.5, 0.5)$. \texttt{spcar}: estimation with static principal components with \texttt{r} $=1,2,3,5,9$, \texttt{dpca}: estimation by dynamic principal components with $q = 1$, \texttt{fdl}: finite distributed lag regression computed by regression on the first principal component and its first lag.}
    \label{fig: inconsistency of weak factors}
\end{figure}

The left plot in Figure \ref{fig: inconsistency of weak factors} shows, that the principal component estimator with $r = 2$ (\texttt{spca2}) is not consistent for $\chi_{1t}$, indeed, $MSE_1$ is not approaching zero as $n$ and $T$ becomes larger \citep[for a theoretical proof see][]{onatskiy2012asymptotics}. The other two dynamic estimators (\texttt{dpca} and \texttt{fdl}) are consistent (Theorem \ref{thm: consistency of spaces and CCs}) as $MSE_1$ is monotonically decreasing with increasing $(n,T)$, while the finite distributed lags approach \texttt{fdl} yields better performance. Furthermore, the right plot of figure \ref{fig: inconsistency of weak factors} shows that we cannot recover $\chi_{1t}$ by static principal components, even if we increase the number of factors $r$, i.e., with \texttt{spcar} for $r = 1, 2, 3, 5, 9$. In other words, the weak factor is \textit{too weak} to be recovered by means of static principal components - though it is important individually, as it explains a large part of the variation for the first ten units. 
Analogous results are obtained for other configurations of the sample size and the idiosyncratic component, see tables \ref{tab: res nonconsist T > n}, \ref{tab: res nonconsist n < T} and \ref{tab: cov rates n = T}. 
\paragraph{Coverage Rates for the Confidence Intervals.}
Next we explore the coverage rates of the asymptotic confidence intervals provided by the CLTs of section \ref{sec: estimation}. For estimating the loadings variance, we use the HAC version from equation \eqref{eq: HAC2 asy beta}, for estimating the factors' variance we use the estimator from \eqref{eq: est cov Lambda xi, h} by \cite{fresoli2024dealing}. Table \ref{tab: cov rates T > n} shows the results for the case where $T>n$. In all cases the sample coverage approaches the nominal rate of 95\% for increasing sample size $(n, T)$, while convergence kicks in about $(n, T) = (120, 240)$ and slightly slower for the case with higher idiosyncratic cross-correlation $(\tau = 0.5)$. Similar results are obtained for different combinations of $(n, T)$ and presented in the appendix.
\begin{table}[!htbp] \centering
\begin{tabular}{@{\extracolsep{5pt}} c|c|c|c|c} 
\\[-1.8ex]\hline 
\hline \\[-1.8ex]
\multicolumn{5}{c}{\textbf{Coverage Rates, $T > n$}} \\[0.5em]
\hline
$(n,T)$ & (30,60) & (60,120) & (120,240) & (240,480) \\
\hline \\[-1.8ex]
\hline \\[-1.8ex]
$e_{1t}^\chi$, $\tau = 0$, $\delta = 0$   & 0.719 & 0.862 & 0.946 & 0.940 \\
$\chi_{1t}$, $\tau = 0$, $\delta = 0$     & 0.864 & 0.890 & 0.950 & 0.942 \\
$C_{1t}$, $\tau = 0$, $\delta = 0$        & 0.826 & 0.910 & 0.930 & 0.936 \\
\hline \\[-1.8ex]
$e_{1t}^\chi$, $\tau = 0.5$, $\delta = 0$ & 0.659 & 0.848 & 0.908 & 0.926 \\
$\chi_{1t}$, $\tau = 0.5$, $\delta = 0$   & 0.798 & 0.884 & 0.900 & 0.930 \\
$C_{1t}$, $\tau = 0.5$, $\delta = 0$      & 0.818 & 0.908 & 0.920 & 0.912 \\
\hline \\[-1.8ex]
$e_{1t}^\chi$, $\tau = 0$, $\delta = 0.5$ & 0.724 & 0.868 & 0.932 & 0.940 \\
$\chi_{1t}$, $\tau = 0$, $\delta = 0.5$   & 0.868 & 0.894 & 0.926 & 0.920 \\
$C_{1t}$, $\tau = 0$, $\delta = 0.5$      & 0.800 & 0.890 & 0.908 & 0.924 \\
\hline \\[-1.8ex]
$e_{1t}^\chi$, $\tau = 0.5$, $\delta = 0.5$ & 0.696 & 0.844 & 0.902 & 0.948 \\
$\chi_{1t}$, $\tau = 0.5$, $\delta = 0.5$   & 0.822 & 0.904 & 0.878 & 0.930 \\
$C_{1t}$, $\tau = 0.5$, $\delta = 0.5$      & 0.784 & 0.908 & 0.892 & 0.916 \\
\hline \\
\end{tabular}
\caption{\footnotesize
\textsc{Coverage rates} for asymptotic $1-\alpha = 95\%$-confidence intervals of $\chi_{1,10}$, $e_{1,10}^{\chi}$ and $C_{1,10}$ over $B = 500$ replications.}
\label{tab: cov rates T > n}
\end{table}

\section{Empirical Application}\label{sec: empirical}
%
To demonstrate our methods empirically, we use different high-dimensional macroeconomic time series panels for the Euro Area (EA) and major EA countries, based on the open source dataset of \cite{barigozzi2024large}. We consider three datasets: a) monthly EA and country-level series ($n = 381$, $T = 309$, $2000:01$ to $2025:09$) b) quarterly EA and country-level series ($n = 595$, $T = 103$, $2000:Q1$ to $2025:Q3$) c) quarterly series for Germany ($n = 63$, $T = 103$, $2000:Q1$ to $2025:Q3$). All series are standardised to zero mean and unit sample variance. Following \cite{barigozzi2024large}, the data are transformed to stationarity and outliers are removed using standard methods. 

The number of factors is estimated using \cite{alessi2010improved}. Following factor selection, model \eqref{eq: GDFM stat rep} is estimated by regressing on the estimated factors, with the lag order determined by the modified BICM of \cite{groen2013model}, specifically designed for regressions involving estimated factors, with a more restrictive penalty term a normal BIC. To retain information from crisis periods, parameter estimation is conducted on outlier removed data, while factor and common component estimation use the standardised raw data combined with eigenvectors and eigenvalues estimated from the cleaned sample covariance matrix. Finally, we obtain the estimates $\widehat \chi_{it}$ from model \eqref{eq: GDFM stat rep} and $\widehat e_{it}^\chi = \widehat \chi_{it} - \widehat C_{it}$.

Let us comment on the results in order: a) For the monthly data, we estimate $\widehat r = 4$ factors, with $130$ series exhibiting lag order $p>0$. Notably, $r = 4$ includes already rate-weak factors \cite{bai2023approximate}. Choosing a larger $r$ would therefore result in inconsistent estimates \citep[see][for a more detailed empirical and theoretical treatment of the idea of over-specifying $r$]{gersing2026weak}. The weak common component (WCC) explains $3.6\%$ of the total variance, while the static common component (SCC) accounts for $19.6\%$. For $55$ variables, the WCC explains more than $10\%$ of the total variance, reaching up to $25.5\%$ for ESENTIX (Economic Sentiment Indicator) of the Euro Area.

\begin{figure}
    \centering
	\begin{tabular}{cc}
    \includegraphics[width=.5\textwidth]{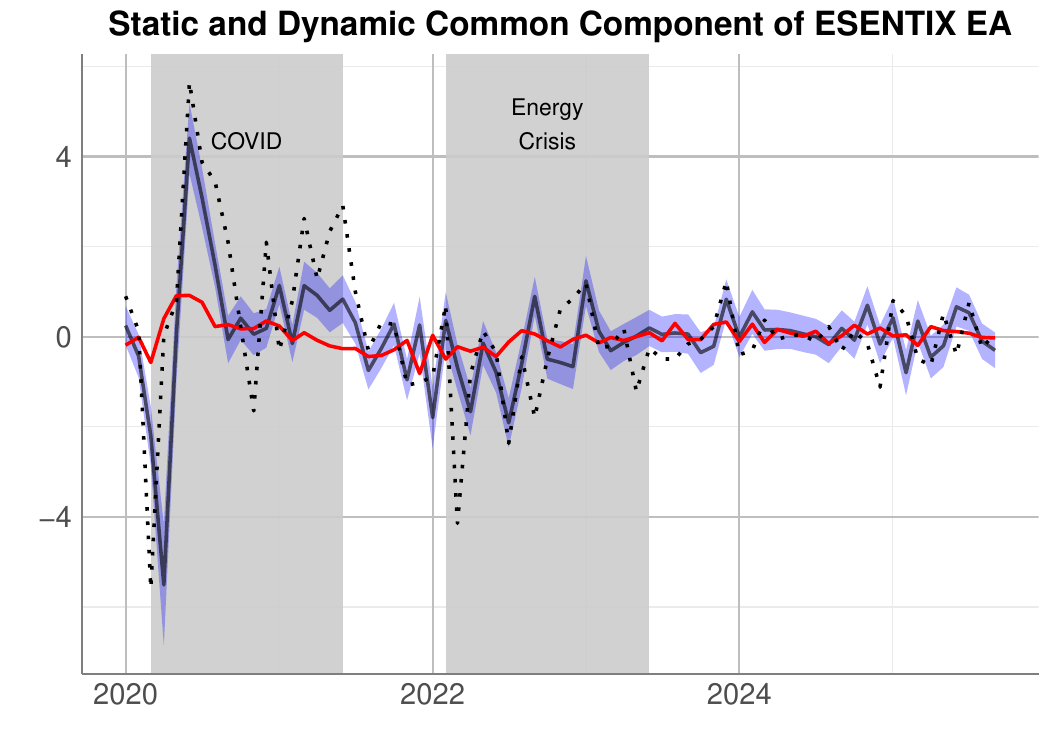}&
    \includegraphics[width=.5\textwidth]{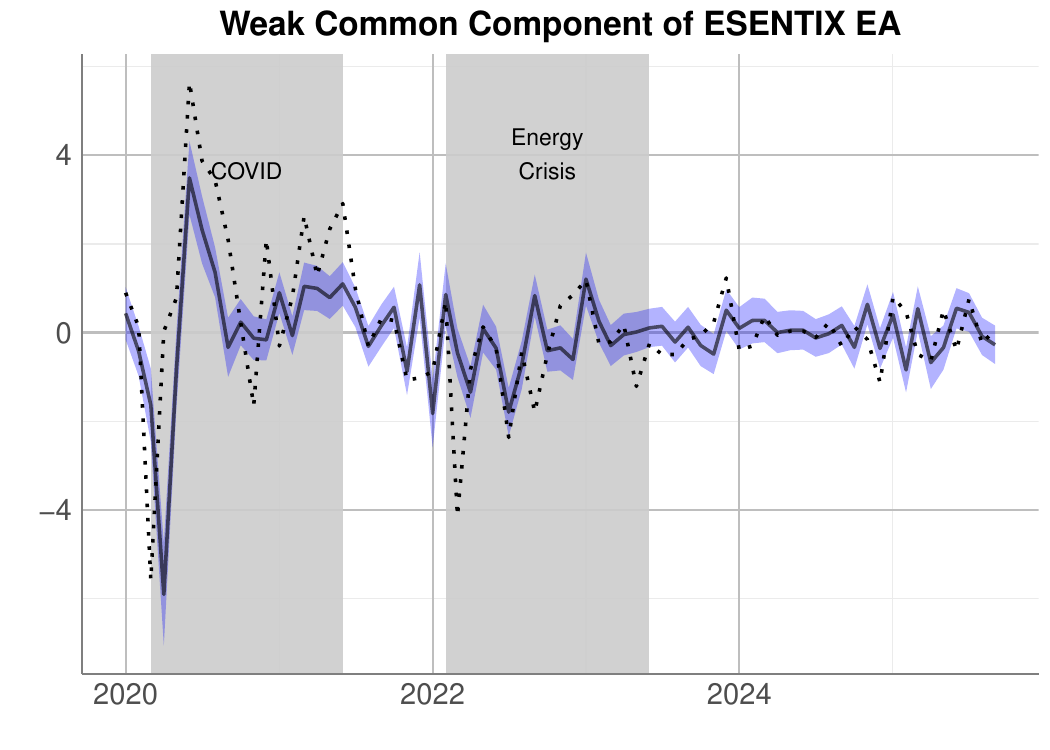}
    \end{tabular}
    \caption{ 
    \footnotesize Monthly Data a): \textsc{Estimates of the Canonical Decomposition} of standardised $\Delta  \text{ESENTIX}_t$ = Economic Sentiment Indicator of the Euro Area. \texttt{Black solid line}: \textit{dynamic common component} (left), or \textit{weak common component} (right), \texttt{red line}: static common component, \texttt{dotted line}: true observed standardised index. Blue area represents 90\% confidence intervals.}
    \label{fig: ESENTIX EA}
\end{figure}
\begin{figure}
    \centering
	\begin{tabular}{cc}
    \includegraphics[width=.5\textwidth]{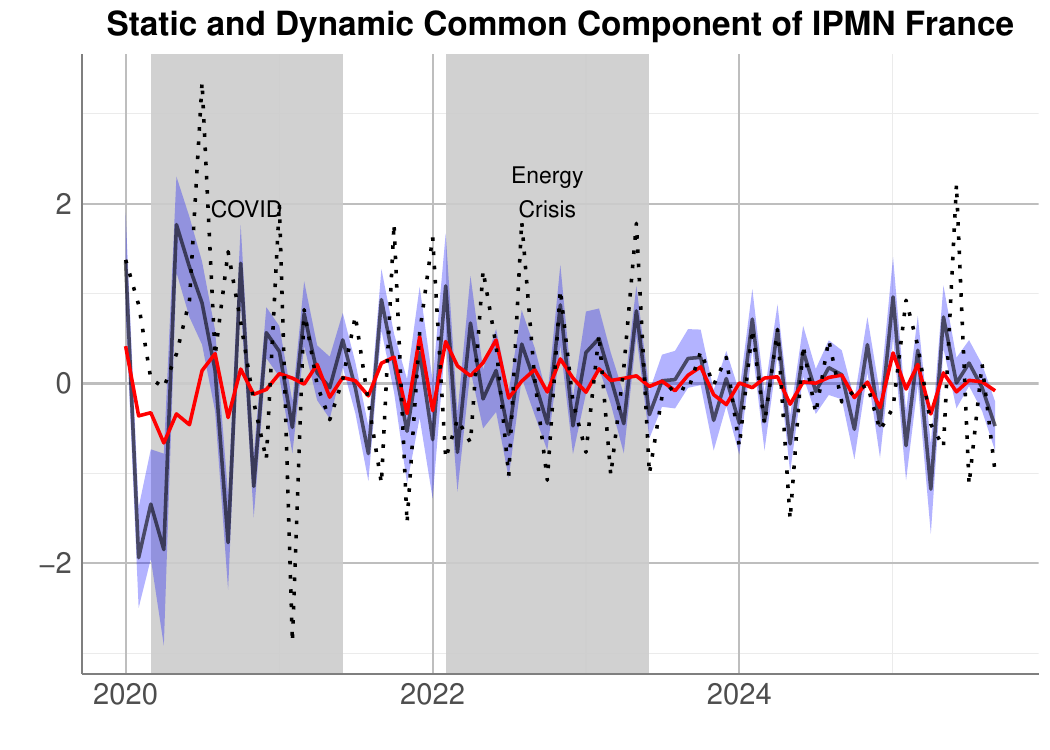}&
    \includegraphics[width=.5\textwidth]{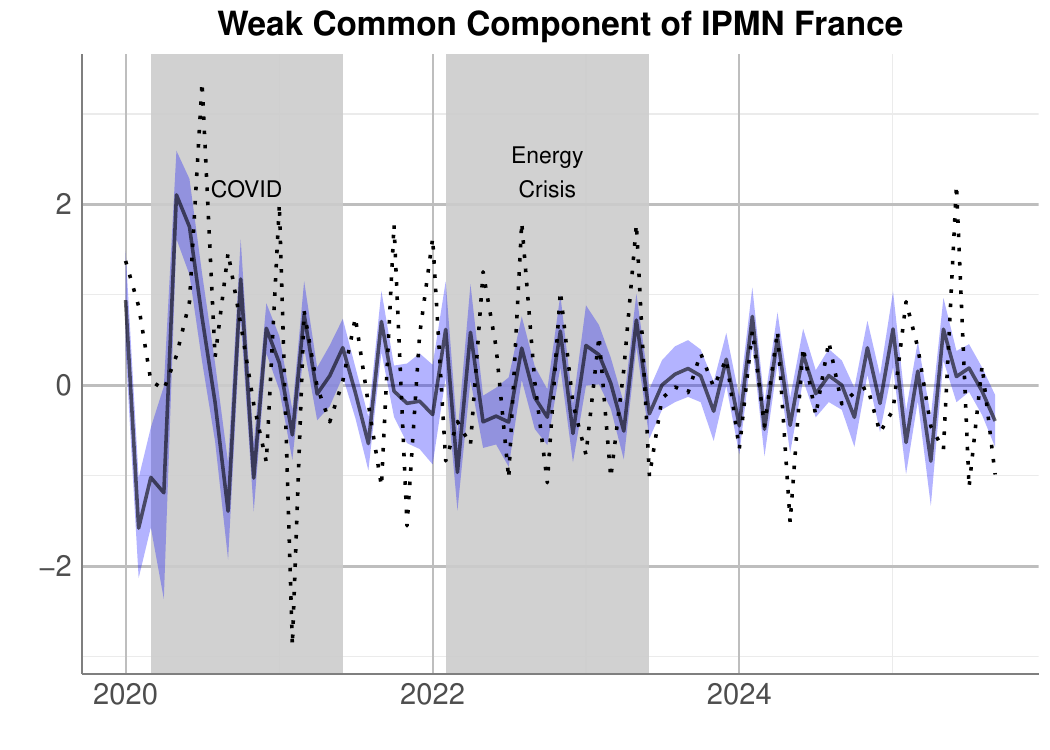}
    \end{tabular}
    \caption{ 
    \footnotesize Monthly Data a): \textsc{Estimates of the Canonical Decomposition} of standardised $\Delta \log \text{IPMN}_t$ = Industrial Production Index: Manufacturing of France. \texttt{Black solid line}: \textit{dynamic common component} (left), or \textit{weak common component} (right), \texttt{red line}: static common component, \texttt{dotted line}: true observed standardised index. Blue area represents 90\% confidence intervals.}
    \label{fig: IPMN France}
\end{figure}
\begin{figure}
    \centering
	\begin{tabular}{cc}
    \includegraphics[width=.5\textwidth]{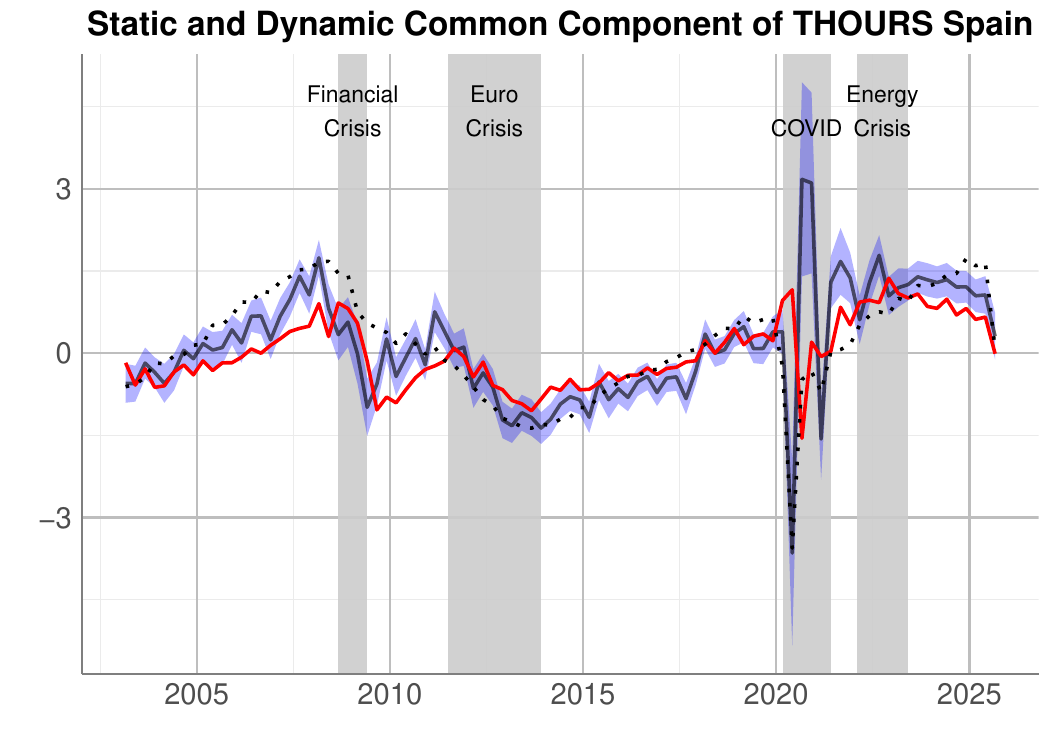}&
    \includegraphics[width=.5\textwidth]{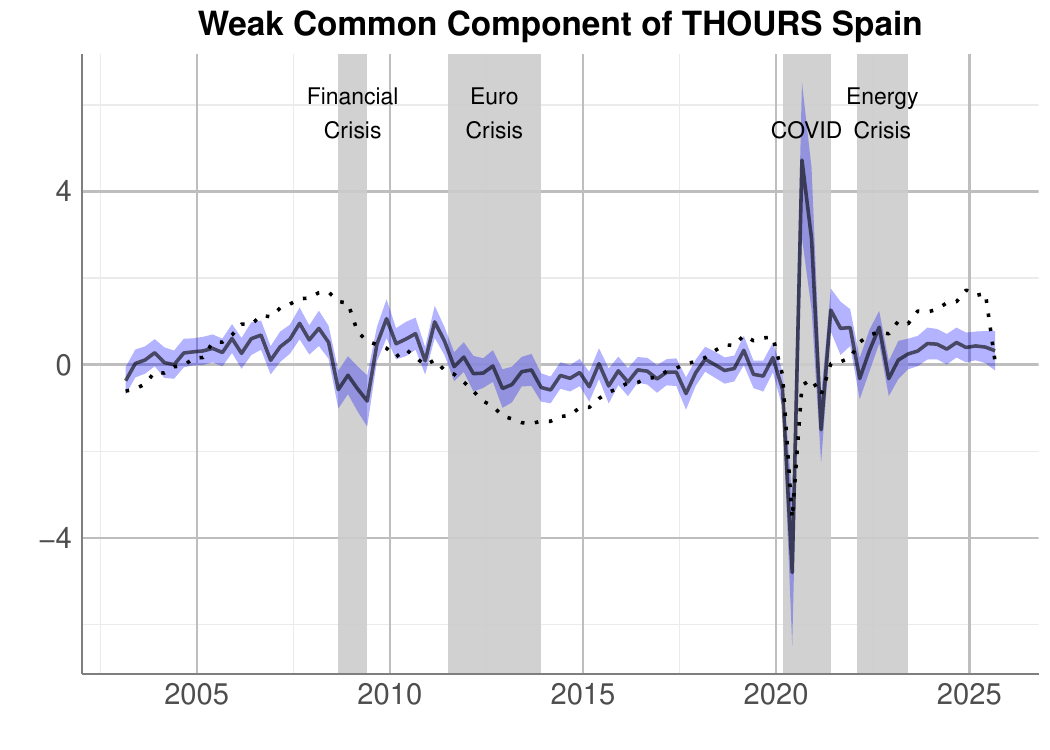}
    \end{tabular}
    \caption{ 
    \footnotesize Quarterly data b): \textsc{Estimates of the Canonical Decomposition} of standardised THOURS = Total Hours Worked for Spain. \texttt{Black solid line}: \textit{dynamic common component} (left), or \textit{weak common component} (right), \texttt{red line}: static common component, \texttt{dotted line}: true observed standardised index. Blue area represents 90\% confidence intervals.}
    \label{fig: TOTHOURS ES}
\end{figure}

For ESENTIX (Figure \ref{fig: ESENTIX EA}), the WCC allows the DCC to track the observed series more closely during the COVID period, while the SCC responds only weakly to the large shocks. Moreover, outside crisis periods, the WCC captures additional dynamics not accounted for by the SCC. Similarly, for the growth rate of French industrial production (Figure \ref{fig: IPMN France}), the DCC exhibits dynamics that differ significantly from those of the SCC. 
Next, consider dataset b) of quarterly series. 
The WCC explains 0.3\% of total variation, whereas the SCC accounts for 38.9\%. Estimates for total hours worked (THOURS) in Spain - alongside unemployment, a key labour market indicator - are shown in Figure \ref{fig: TOTHOURS ES}. Here as well, the WCC appears offset the stronger mean reversion of the SCC relative to the observed series.
\begin{figure}
    \centering
	\begin{tabular}{cc}
    \includegraphics[width=.5\textwidth]{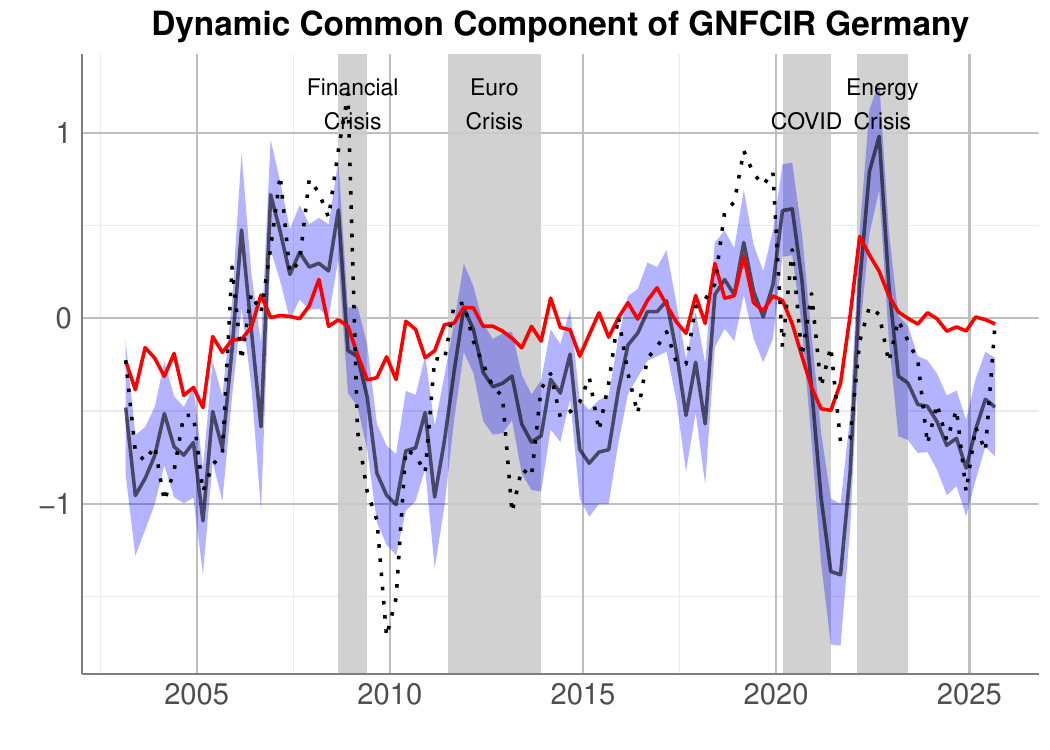}&
    \includegraphics[width=.5\textwidth]{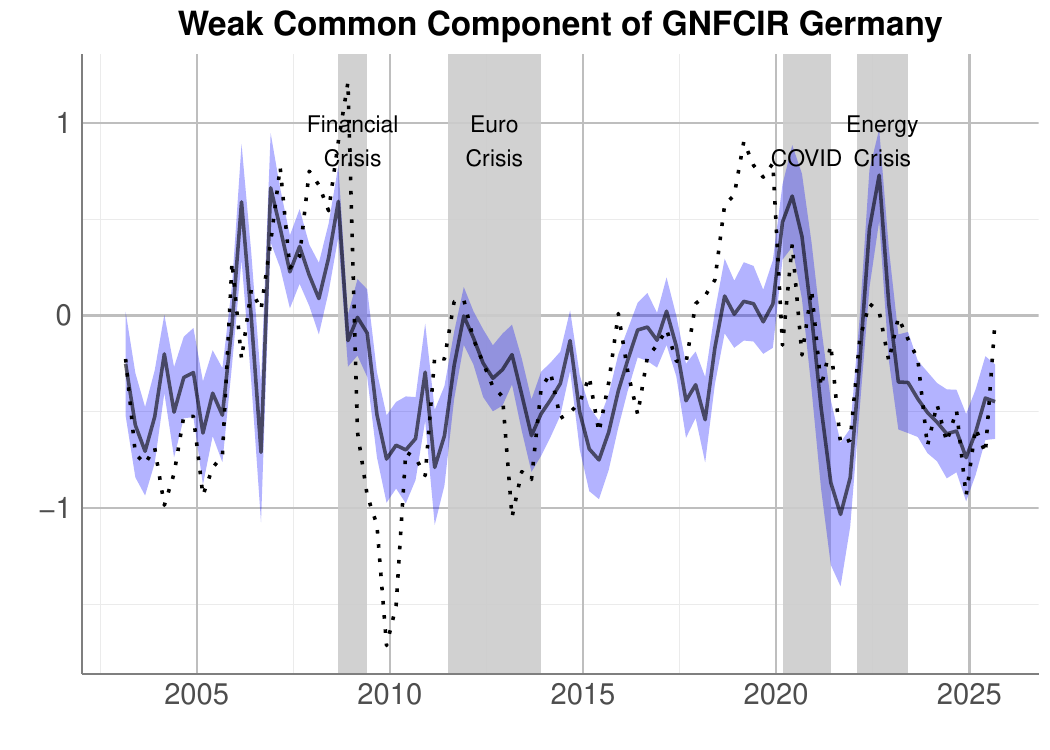}
    \end{tabular}
    \caption{ 
    \footnotesize Quarterly data c): \textsc{Estimates of the Canonical Decomposition} using a time series panel comprising only macroeconomic indicators from Germany of GNFCIR = Gross-Investment Share of Non-Financial Corporations. \texttt{Black solid line}: \textit{dynamic common component} (left), or \textit{weak common component} (right), \texttt{red line}: static common component, \texttt{dotted line}: true observed standardised index. Blue area represents 90\% confidence intervals.}
    \label{fig: GNFCIR Germany}
\end{figure}
Finally, consider dataset c), comprising only German series which is a time series panel of considerably smaller size. Only one variable has $p>0$, namely GNFCIR (Gross Investment Share of Non-Financial Corporations) with $p=2$. Despite, GNFCIR is a key leading indicator of long-term economic growth, reflecting business confidence about future conditions. In this case, most of the dynamics are captured by the WCC (40\%), while the SCC accounts for a smaller share of variation (30\%) relative to the observed series (see Figure \ref{fig: GNFCIR Germany}). 
\section{Conclusion}
We consider the generalized dynamic factor model with finitely many static and weak factors. We show that the dynamic common component admits a finite distributed-lag representation in static factors and develop estimation and inference that accommodate weak non-pervasive factors through lead-lag dependence between factors and the static idiosyncratic component. This extends existing factor-augmented regression results. An application to Euro-Area macroeconomic data reveals a substantial weak common component in several variables.
\section*{Acknowledgments}
{{I am deeply grateful to Manfred Deistler, for his guidance and support. My thanks further go to Nikolaus Hautsch, Luca Gonzato and Yuan Chen.}}

\section*{Funding}
{{This research is supported by the University of Vienna.}}

\section*{Data Availability Statement}
The data that support the findings of this study are openly available on \url{https://doi.org/10.5281/zenodo.17233192} at reference number 17233192.
\bibliographystyle{apalike} 
\bibliography{references.bib}

\clearpage
\begin{center}
{\LARGE \bf Appendix}
\end{center}
\appendix
\renewcommand{\theequation}{\thesection.\arabic{equation}}
Additional Notation: By $\bF:=\l(\bF_1, ..., \bF_T\r)'$ and $\bsx:=\l(\bsx_1, ..., \bsx_T\r)'$ we denote the $T\times r$ and $T \times r_\chi$ matrices of factors or stacked factors. By $\bsy^i:=(y_{i1}, ..., y_{iT})'$ we denote the time observations of the $i$-th variable, analogously for $\bxi^i$ and $\bse^i$. If $\widehat{\bA}$ is a matrix of random variables and $\bA$ a constant matrix we write $\widehat{\bA}_n - \bA = \mathcal O_{ms}(a_n^{-1})$ if $\E\norm{a_n(\widehat{\bA}_n - \bA)}^2 = \mathcal O(1)$ is bounded for $n\to \infty$, analogously for $T\to \infty$ or both. Recall that here $\norm{\cdot}$ is the spectral norm for matrices, while for vectors we use the Euclidean norm.    
\begin{rem}\label{rem: Gamma F = I}
\normalfont Without loss of generality for the results of this paper we may assume that $\bGamma_{\bF} = \bI_r$. Suppose there is a ``true'' underlying factor process $(\tilde \bF_t)$ with $\bGamma_{\tilde F} \neq \bI_r$ and $C_{it} = \tilde \bLambda_i \tilde \bF_t$. First define $\bF_t := \bGamma_{\tilde F}^{-1/2} \tilde \bF_t$, then $C_{it} = \tilde \bLambda_i \bGamma_{\tilde F}^{1/2} \bGamma_{\tilde F}^{-1/2} \tilde \bF_t = \bLambda_i \bF_t$. For consistency and asymptotic normality we show that an estimator, say $\widehat \bF_t$ for $\bF_t$ converges to the true factors in the sense that
\begin{align*}
     \norm{\widehat \bF_t - \widehat{\bH}\bF_t} = o_p(1)  \quad \mbox{or} \quad \widehat \bF_t - \widehat{\bH}\bF_t \Rightarrow \mathcal N(\bm{0}, .), 
     \qquad \mbox{while} \quad \widehat{\bH}:= \widehat{\bF}' \bF \l(\bF' \bF\r)^{-1}.
\end{align*}
Looking at $\tilde{\bH}:= \widehat{\bF} ' \tilde{\bF} \l(\tilde{\bF}' \tilde{\bF}\r)^{-1}$, it is easily verified that $\tilde{\bH} = \widehat{\bH}\bGamma_{\tilde{\bF}}^{-1/2}$ and therefore it immediately follows that also $\norm{\widehat \bF_t - \tilde{\bH} \tilde \bF_t} = o_p(1)$ and $\widehat \bF_t - \tilde{\bH} \tilde \bF_t \Rightarrow \mathcal N(\bm{0}, .)$.
\end{rem}
\section{Proofs of Main Theorems}

\begin{proof}[Proof of Theorem \ref{thm: consistency of spaces and CCs}]
    For part $(i)$, we use the expansion of $\widehat{\bH}$ from \eqref{eq: expansion of hat H} and plug in $\bsy_t^n =\bLambda^n \bF_t + \bse_t^n$: 
    \begin{align*}
        \widehat{\bW}_t^{y, n} - \widehat{\bH}\bF_t &= \widehat{\bmcK}\bLambda^n \bF_t + \widehat{\bmcK} \bse_t^n - \widehat{\bmcK}\bLambda^n \bF_t - \widehat{\bmcK} \widehat \bGamma_{\bse \bF}^n \widehat \bGamma_{\bF}^{-1} \bF_t\\
        &= \l(\widehat{\bmcK} - \bmcK\r) \bse_t^n + \bmcK \bse_t^n - \widehat{\bmcK} \widehat \bGamma_{\bse \bF}^n \widehat \bGamma_{\bF}^{-1} \bF_t.\\
        \norm{\widehat{\bW}_t^{y, n} - \widehat{\bH}\bF_t} &\leq \norm{ \l(\widehat{\bmcK} - \bmcK\r)}n^{1/2}\norm{\frac{\bse_t^n}{\sqrt{n}}}  + \norm{\bmcK \bse_t^n} + \norm{\widehat{\bmcK} \widehat \bGamma_{\bse \bF}^n}\norm{\widehat \bGamma_{\bF}^{-1}} \norm{\bF_t} \\
        &= \mathcal O_{P}(T^{-1/2}n^{-1/2}) \mathcal O(n^{1/2})\mathcal O_{ms}(1) + \mathcal O_{ms}(n^{-1/2}) + \mathcal O(\max(T^{-1}, n^{-1/2}T^{-1/2})),
    \end{align*}
    while the rate for the first term follows from Lemmas \ref{lem: hat evals-vecs} and \ref{lem: Var est}. For the second note that $\E\l[\norm{\bmcK \bse_t^n}^2\r] = \E\l[\sum_{l = 1}^r \l(\bmcK_l \bse_t^n\r)^2\r] = \tr\l(\bmcK \bGamma_{\bse_t}^n \bmcK' \r) \leq \mu_1\l(\bGamma_{\bse_t}^n \r)\norm{\bmcK}^2 = \mathcal O(n^{-1})$. For the third term we use Lemma \ref{lem: killing K}. 
    Furthermore from Lemma \ref{lem: H}(ii), we know that $\norm{\widehat{\bH} - \bP_{\bLambda}} = \mathcal O_{P}(\max(n^{-1/2}, T^{-1/2}))$, so
    \begin{align*}
        \norm{\widehat{\bW}_t^{y, n} - \bP_{\bLambda} \bF_t} \leq  \norm{\widehat{\bW}_t^{y, n} - \widehat{\bH} \bF_t}  + \norm{\widehat{\bH} - \bP_{\bLambda}}\norm{\bF_t}  = 
        \mathcal O_{P}(\max(n^{-1/2}, T^{-1/2})).
    \end{align*}

    For the second part, we proceed analogously, by using the canonical decomposition and rewrite $\widehat \bsx_t$ in terms of 
    \begin{align*}
    \widehat \bsx_t &= \begin{pmatrix}
    \widehat{\bmcK}\begin{bmatrix}\bLambda^n  & \bLambda^{w, n} \end{bmatrix}\\
    \widehat{\bmcK}\begin{bmatrix}\bLambda^n  & \bLambda^{w, n} \end{bmatrix}\\
        \vdots \\
    \widehat{\bmcK}\begin{bmatrix}\bLambda^n  & \bLambda^{w, n} \end{bmatrix}
        \end{pmatrix}
        \begin{pmatrix}
            \begin{matrix}
                \bF_t\\
                \bF_t^w
            \end{matrix}\\
               \begin{matrix}
                \bF_{t-1}\\
                \bF_{t-1}^w
            \end{matrix}\\
            \vdots\\
               \begin{matrix}
                \bF_{t-p}\\
                \bF_{t-p}^w
            \end{matrix}
        \end{pmatrix} + \left(\bI_{p+1} \otimes \widehat{\bmcK}\right)\begin{pmatrix}
            \bxi_t^n \\
            \bxi_{t-1}^n\\
            \vdots \\
            \bxi_{t-p}^n
        \end{pmatrix} \\
        &= \left(\bI_{p+1}\otimes \widehat{\bmcK}\bLambda^n \right) 
        \underbrace{\begin{pmatrix}
            \bF_t \\
            \vdots \\
            \bF_{t-p} 
        \end{pmatrix}}_{\bsx_t} 
        + \left(\bI_{p+1}\otimes \widehat{\bmcK} \bLambda^{w, n}\right)
        \underbrace{\begin{pmatrix}
            \bF_t^w \\
            \vdots \\
            \bF_{t-p}^w 
        \end{pmatrix}}_{\bmcF_t^w} 
        + \left(\bI_{p+1} \otimes \widehat{\bmcK}\right)\bXi_t^n.
        \end{align*}
    using the expansion of $\widehat{\bmcH}$ from \eqref{eq: expansion of mathcal H}, we obtain by Lemmas \ref{lem: Var est} and \ref{lem: killing K} that 
    \begin{align}
        &\widehat \bsx_t - \widehat{\bmcH}\bsx_t  = \left(\bI_{p+1}\otimes \widehat{\bmcK}\bLambda^n \right) \bsx_t
        + \left(\bI_{p+1}\otimes \widehat{\bmcK} \bLambda^{w, n}\right)
        \bmcF_t^w
        + \left(\bI_{p+1} \otimes \widehat{\bmcK}\right)\bXi_t^n  \nonumber\\
        &\qquad \qquad \qquad -  \left(\bI_{p+1}\otimes \widehat{\bmcK}\bLambda^n \right)\bsx_t -  \left(\bI_{p+1}\otimes \widehat{\bmcK} \bLambda^{w, n} \right) \widehat \bGamma_{\bmcF^w \bsx}\widehat \bGamma_{\bsx}^{-1}\bsx_t -  \left(\bI_{p+1}\otimes \widehat{\bmcK} \right) \widehat \bGamma_{\bXi\bsx}^n\widehat \bGamma_{\bsx}^{-1}\bsx_t \nonumber \\
        &= \left(\bI_{p+1}\otimes \widehat{\bmcK} \bLambda^{w, n}\right)
        \bmcF_t^w + \left(\bI_{p+1} \otimes \widehat{\bmcK}\right)\bXi_t^n 
         -  \left(\bI_{p+1}\otimes \widehat{\bmcK} \bLambda^{w, n} \right) \widehat \bGamma_{\bmcF^w \bsx}\widehat \bGamma_{\bsx}^{-1}\bsx_t -  \left(\bI_{p+1}\otimes \widehat{\bmcK} \right) \widehat \bGamma_{\bXi\bsx}^n\widehat \bGamma_{\bsx}^{-1}\bsx_t \label{eq: rep hat xt - hat mathcal H xt} \\
        &=\mathcal O_P(n^{-1/2}) + \mathcal O_P(T^{-1/2}) + \mathcal O_P(n^{-1/2}) + \mathcal O_P(\max(T^{-1}, n^{-1/2}T^{-1/2})).\nonumber
    \end{align}
    Analogously to above, we obtain by Lemma \ref{lem: H}(iv), that $\norm{\widehat \bsx_t - \l(\bI_{p+1} \otimes \bP_{\bLambda}\r) \bsx_t} = \mathcal O_{P}(\max(n^{-1/2}, T^{-1/2}))$.
    
    For the second part of part (ii), we note that $\widehat \bbeta_i$ can be obtained from regressing $y_{it}$ on $\widehat \bsx_t$. Consider the following expansion where we plug in $\bsy^i = \bsx \bbeta_i' + \bxi^i$:
        \begin{align}
        \widehat \bbeta_i' - \widehat{\bmcH}^{-1'}\bbeta_i' &=(\widehat{\bsx}' \widehat{\bsx})^{-1} \widehat{\bsx}' \bsy^i - \widehat{\bmcH}^{-1'} \bbeta_i' = (\widehat{\bsx}' \widehat{\bsx})^{-1} \widehat{\bsx}'\left(\bsx \widehat{\bmcH}' \widehat{\bmcH}^{-1'} \bbeta_i ' + \bxi^i\right) - \widehat{\bmcH}^{-1'} \bbeta_i'\nonumber\\
        &= (\widehat{\bsx}' \widehat{\bsx})^{-1} \widehat{\bsx}' \left( \widehat{\bsx} \widehat{\bmcH}^{-1'} \bbeta_i' + \bxi^i + 
        (\bsx \widehat{\bmcH}' - \widehat{\bsx}) \widehat{\bmcH}^{-1'} \bbeta_i' \right) - \widehat{\bmcH}^{-1'} \bbeta_i'\nonumber \\
        &= \l(T^{-1} \widehat{\bsx}' \widehat{\bsx}\r)^{-1} \l[T^{-1} \widehat{\bsx}' \bxi^i + T^{-1} \widehat{\bsx}' \l(\bsx \widehat{\bmcH}' - \widehat{\bsx} \r)\widehat{\bmcH}^{-1'}\bbeta_i'\r] \nonumber\\
        &= \widehat \bGamma_{\widehat{\bsx}}^{-1} \l[\widehat{\bmcH}  \widehat \bGamma_{\bsx\bxi^i} + T^{-1} (\widehat{\bsx} - \bsx \widehat{\bmcH}')'\bxi^i + T^{-1} \widehat{\bsx}' \l(\bsx \widehat{\bmcH}' - \widehat{\bsx} \r)\widehat{\bmcH}^{-1'}\bbeta_i' \r]  \label{eq: expansion hat beta_i minus H beta_i}\\
        &= \mathcal O_P(\max(T^{-1/2}, n^{-1/2})). \nonumber
\end{align}
by Assumption E\ref{A: simply the sample covariances} and Lemma \ref{lem: x xi and x x}. The first part is proved analogously. 

To show part $(iii)$ we use the previous results: 
\begin{align*}
    \widehat \chi_{it} - \chi_{it} &= \widehat \bbeta_i \widehat \bsx_t - \bbeta_i \widehat{\bmcH}^{-1} \widehat \bsx_t + \bbeta_i \widehat{\bmcH}^{-1} \widehat \bsx_t - \bbeta_i \widehat{\bmcH}^{-1} \widehat{\bmcH} \bsx_t \\
    &= \l(\widehat \bbeta_i - \bbeta _i \widehat{\bmcH}^{-1}\r) \widehat \bsx_t + \bbeta_i \widehat{\bmcH}^{-1} \l(\widehat \bsx_t - \widehat{\bmcH} \bsx_t\r) \\
    &=\l(\widehat \bbeta_i - \bbeta _i \widehat{\bmcH}^{-1}\r)\l(\widehat \bsx_t - \widehat{\bmcH} \bsx_t\r) + \l(\widehat \bbeta_i - \bbeta_i \widehat{\bmcH}^{-1}\r) \widehat{\bmcH} \bsx_t + \bbeta_i \widehat{\bmcH}^{-1} \l(\widehat \bsx_t - \widehat{\bmcH} \bsx_t\r) \\
    &= \mathcal O_P(\max(n^{-1/2}, T^{-1/2})). 
\end{align*}
For the static common component we proceed analogously and finally
\begin{align*}
    \widehat e_{it}^\chi - e_{it}^\chi = \l(\widehat \chi_{it} - \widehat C_{it}\r) - \l(\chi_{it} - C_{it}\r) = \l(\widehat \chi_{it} - \chi_{it}\r) - \l(\widehat C_{it} - C_{it}\r) = \mathcal O_P(\max(n^{-1/2}, T^{-1/2})).
\end{align*}
\end{proof}
\begin{proof}[Proof of Theorem \ref{lem: asymptotic normality loadings residuals}]
    We use the same expansion as in \eqref{eq: expansion hat beta_i minus H beta_i} multiplied by $\sqrt{T}$: 
    \begin{align*}
        &\sqrt{T} \left((\widehat{\bsx}' \widehat{\bsx})^{-1} \widehat{\bsx}' \bsy^i - \widehat{\bmcH}^{-1'} \bbeta_i' \right) \\
        %
        %
        %
        %
        %
        %
        %
        %
        %
        %
        &= \left(T^{-1} \widehat{\bsx}' \widehat{\bsx}\right)^{-1} \left[\widehat{\bmcH} T^{-1/2}  \bsx ' \bxi^i +\underbrace{ T^{-1/2} (\widehat{\bsx} - \bsx \widehat{\bmcH}')'\bxi^i}_{(1)} + \underbrace{T^{-1/2} \widehat{\bsx}' \left(\bsx \widehat{\bmcH}' - \widehat{\bsx} \right)\widehat{\bmcH}^{-1'}\bbeta_i'}_{(2)} \right]
\end{align*}
where $(1) =o_P(1)$ and $(2) = o_P(1)$ by Lemma \ref{lem: x xi and x x}. 

Next by Lemma \ref{lem: H}, we know that 
\begin{align*}
    &\plim_{n, T \to \infty} \widehat{ \bmcH} = \l(\bI_{p+1} \otimes \bP_{\bLambda}\r) \quad \mbox{and} \quad \plim_{n, T\to \infty} \widehat \bGamma_{\widehat{\bsx}} = \l(\bI_{p+1} \otimes \bP_{\bLambda}\r) \bGamma_{\bsx} \l(\bI_{p+1} \otimes \bP_{\bLambda}\r)'\\
    \mbox{so}\quad & \plim_{n, T \to \infty} \widehat \bGamma_{\widehat{\bsx}}^{-1} \widehat{ \bmcH} = \l(\bI_{p+1} \otimes \bP_{\bLambda}\r) \bGamma_{\bsx}^{-1}.
\end{align*}
Since $T^{-1/2} \sum_{t=1}^T x_{it}\xi_{it} \Rightarrow \mathcal N(\bm{0}, \bOmega_{\bsx\bxi}(i))$ it follows that
\begin{align*}
    \sqrt{T}(\widehat \bbeta_i - \bbeta_i \widehat{\bmcH}^{-1})'  \Rightarrow \mathcal N \l(\bm{0}, \l(\bI_{p+1} \otimes \bP_{\bLambda}\r)\bGamma_{\bsx}^{-1} \bOmega_{\bsx\bxi}(i) \bGamma_{\bsx}^{-1}  \l(\bI_{p+1} \otimes \bP_{\bLambda}' \r)\r), 
\end{align*}
which completes the proof.
\end{proof}
\begin{proof}[Proof of Theorem \ref{lem: asymptotic normality of hat bsxt}]
We use the expansion of $(\widehat \bsx_t - \widehat{\bmcH} \bsx_t)$ from equation \eqref{eq: expansion of mtcalHxt - xt}, noting that
    \begin{align}
    \widehat{\bmcK} &= \bM^{-1}(\widehat \bGamma_{\bsy}^n) \bM^{1/2}(\widehat \bGamma_{\bsy}^n) \bP(\widehat \bGamma_{\bsy}^n) = \left(\frac{\bM(\widehat \bGamma_{\bsy}^n)}{n}\right)^{-1} \frac{\widehat \bLambda^{n'}}{n} = \left(\widehat \bD_{\bLambda}^n\right)^{-1} \frac{1}{n}\widehat \bLambda^{n'} \nonumber \\
    &= \left(\widehat \bD_{\bLambda}^n\right)^{-1} \frac{1}{n}\left( \widehat \bLambda^n -\bLambda^n \widehat{\bH}_{\bLambda}\right)' + \left(\widehat \bD_{\bLambda}^n\right)^{-1} \widehat{\bH}_{\bLambda}'\bLambda^{n'} \label{eq: expansion of hat mathcal K}\\
    \widehat{\bH}_{\bLambda} &= \left(\bLambda^{n'}\bLambda^n\right)^{-1}\bLambda^{n'} \widehat \bLambda^n , \nonumber
    \end{align}
    we obtain the following: 
    \begin{align*}
       &\sqrt{n}\left(\widehat \bsx_t - \widehat{\bmcH} \bsx_t\right) \\
       &= \sqrt{n}\l(\left(\bI_{p+1}\otimes \widehat{\bmcK} \bLambda^{w, n} \right)\left(\bmcF_t^w - \widehat \bGamma_{\bmcF^w \bsx}\widehat \bGamma_{\bsx}^{-1} \bsx_t \right) -\left(\bI_{p+1}\otimes \widehat{\bmcK}  \right) \widehat \bGamma_{\bXi\bsx}^n \widehat \bGamma_{\bsx}^{-1} \bsx_t  + \left(\bI_{p+1}\otimes \widehat{\bmcK} \right)  \bXi_t^n \r) \\
       &= \sqrt{n} \bigg[\left(\bI_{p+1}\otimes \widehat{\bmcK} \bLambda^{w, n} \right)\left(\bmcF_t^w - \widehat \bGamma_{\bmcF^w \bsx}\widehat \bGamma_{\bsx}^{-1} \bsx_t \right) - \left(\bI_{p+1}\otimes \widehat{\bmcK}  \right) \widehat \bGamma_{\bXi\bsx}^n \widehat \bGamma_{\bsx}^{-1} \bsx_t  \\
       & \qquad\qquad + \left(\bI_{p+1}\otimes \left(\widehat \bD_{\bLambda}^n\right)^{-1} \frac{1}{n}\left( \widehat \bLambda^n -\bLambda^n \widehat{\bH}_{\bLambda}\right)' \right)  \bXi_t^n  +  \left(\bI_{p+1} \otimes \left(\widehat \bD_{\bLambda}^n\right)^{-1} \widehat{\bH}_{\bLambda}' \frac{1}{n}\bLambda^{n'}\right) \bXi_t^n
            \bigg] \\
        &= \sqrt{n} \bigg[ \mathcal O_P(\max(T^{-1/2}n^{-1/2}, n^{-1})) + \mathcal O_P(\max(T^{-1}, T^{-1/2} n^{-1/2})) +\mathcal O_P(n^{-1/2}T^{-1/2})  \\
        & \qquad\qquad+ \left(\bI_{p+1} \otimes \left(\widehat \bD_{\bLambda}^n\right)^{-1} \widehat{\bH}_{\bLambda}' \frac{1}{n}\bLambda^{n'}\right) \bXi_t^n \bigg] 
    \end{align*}
    where we used Lemma \ref{lem: x xi and x x} and \ref{lem: killing K}. Now since  $\l(\bI_{p+1} \otimes n^{-1/2}\bLambda^{n'} \r) \bXi_t^n \Rightarrow \mathcal N\l(\bm{0}, \bTheta_{\bLambda\bXi}(t)\r)$ and $\widehat{\bH}_{\bLambda} = \bP_{\bLambda}' + o_P(1)$ and $\widehat \bD_{\bLambda}^n = \bD_{\bLambda} + o_P(1)$, we have 
    \begin{align*}
        \sqrt{n}\l(\widehat \bsx_t - \widehat{\bmcH} \bsx_t\r) \Rightarrow \mathcal N\l(\bm{0}, \l(\bI_{p+1}\otimes \bD_{\bLambda}^{-1} \bP_{\bLambda}\r) \bTheta_{\bLambda\bXi}(t)\l(\bI_{p+1}\otimes \bP_{\bLambda}' \bD_{\bLambda}^{-1}\r)\r).
    \end{align*}
\end{proof}
\begin{proof}[Proof of Theorem \ref{thm: asy norm chi}]
By definition, we have 
    \begin{align*}
        \widehat \chi_{it} - \chi_{it} &= \widehat \bbeta_i \widehat \bsx_t - \widehat \bbeta_i \widehat{\bmcH}^{-1} \widehat{\bmcH} \bsx_t \\
        &= \l(\widehat \bbeta_i - \bbeta_i \widehat{\bmcH}^{-1} \r) \widehat \bsx_t + \bbeta_i \widehat{\bmcH}^{-1}\l(\widehat \bsx_t - \widehat{\bmcH} \bsx_t \r).
\end{align*}
Now set $a_{nT} := \frac{\min\l(\sqrt{n}, \sqrt{T}\r)}{\sqrt{T}}$ and $b_{nT} := \frac{\min\l(\sqrt{n}, \sqrt{T}\r)}{\sqrt{n}}$
\begin{align*}
        \min\l(\sqrt{n}, \sqrt{T}\r)\l(\widehat \chi_{it} - \chi_{it}\r) &= a_{nT}\sqrt{T}\l(\widehat \bbeta_i - \bbeta_i \widehat{\bmcH}^{-1} \r) \l(\widehat \bsx_t - \l(\bI_{p+1}\otimes \bP_{\bLambda}\r) \bsx_t\r) \\
        & + a_{nT}\sqrt{T}\l(\widehat \bbeta_i - \bbeta_i \widehat{\bmcH}^{-1}\r)\l(\bI_{p+1}\otimes \bP_{\bLambda}\r) \bsx_t \\
        & + b_{nT}\bbeta_i\l(\widehat{\bmcH}^{-1} - \l(\bI_{p+1}\otimes \bP_{\bLambda}'\r)\r) \widehat{\bmcH}^{-1} \sqrt{n} \l(\widehat \bsx_t - \widehat{\bmcH} \bsx_t \r) \\
        &+ b_{nT}\bbeta_i \l(\bI_{p+1}\otimes \bP_{\bLambda}'\r)\sqrt{n}\l(\widehat \bsx_t - \widehat{\bmcH} \bsx_t \r) \\
        &= a_{nT}\underbrace{\bsx_t'\l(\bI_{p+1}\otimes \bP_{\bLambda}'\r) \sqrt{T}\l(\widehat \bbeta_i' - \widehat{\bmcH}^{-1'}\bbeta_i' \r)}_{\theta_{it, nT}} \\
        & + b_{nT}\underbrace{\bbeta_i \l(\bI_{p+1}\otimes \bP_{\bLambda}'\r)\sqrt{n}\l(\widehat \bsx_t - \widehat{\bmcH} \bsx_t \r)}_{\zeta_{it, nT}} + o_P(1)\\
        &= a_{nT} \theta_{it, nT} + b_{nT} \zeta_{it, nT}+ o_P(1).
    \end{align*}
    Since $\theta_{it, nT}$ is a sum over cross-section and $\zeta_{it, nT}$ is a sum over time and both terms are asymptotically independent. By Theorem \ref{lem: asymptotic normality loadings residuals}, the contribution of the loadings estimation to the asymptotic variance is given by
    \begin{align*}
        \theta_{it, nT} &\Rightarrow \mathcal N\l(0, U_{it}\r), \quad \mbox{with}\ U_{it} = \bsx_t'\bGamma_{\bsx}^{-1} \bOmega_{\bsx\bxi}(i) \bGamma_{\bsx}^{-1}\bsx_t, 
    \end{align*}
   and the by Theorem \ref{lem: asymptotic normality of hat bsxt} contribution of the factor estimation is given by
    \begin{align*}        
        \zeta_{it, nT} &\Rightarrow \mathcal N\l(0, V_{it}\r), \quad \mbox{with} \ V_{it}:=  \bbeta_i\l(\bI_{p+1} \otimes \bGamma_{\bLambda}^{-1}\r) \bTheta_{\bLambda\bXi}(t) \l(\bI_{p+1} \otimes \bGamma_{\bLambda}^{-1}\r) \bbeta_i'.
    \end{align*}

    Finally for $n,T \to \infty$, 
    \begin{align*}
        &\min\l(\sqrt{n}, \sqrt{T}\r) \l(\widehat \chi_{it} - \chi_{it}\r) \Rightarrow \mathcal N\l(0, a_{nT}^2 U_{it} + b_{nT}^2 V_{it}\r)\\
        \mbox{or} \quad & \frac{\min\l(\sqrt{n}, \sqrt{T}\r) \l(\widehat \chi_{it} - \chi_{it}\r)}{\sqrt{a_{nT}^2 U_{it} + b_{nT}^2 V_{it}}} \Rightarrow \mathcal N(0, 1), 
    \end{align*}
    which we can rewrite as 
    \begin{align*}
          \frac{\widehat \chi_{it} - \chi_{it}}{\sqrt{\frac{1}{T} U_{it} + \frac{1}{n}V_{it}}} \Rightarrow \mathcal N(0, 1), 
    \end{align*}
    using that
    \begin{align*}
    \frac{\min\l(\sqrt{n}, \sqrt{T}\r)}{\sqrt{ a_{nT}^2 U_{it} + b_{nT}^2 V_{it}}} =  \frac{\min\l(\sqrt{n}, \sqrt{T}\r)}{\min\l(\sqrt{n}, \sqrt{T}\r) \sqrt{\frac{1}{T}U_{it} + \frac{1}{n} V_{it}}} = \frac{1}{\sqrt{\frac{1}{T}U_{it} + \frac{1}{n} V_{it}}},
    \end{align*}
    which completes the proof.
\end{proof}
\begin{proof}[Proof of Theorem \ref{thm: asy norm $e_{1t}^chi$}]
By definition, we have 
    \begin{align*}
        \widehat e_{it}^\chi - e_{it}^\chi &= \l(\widehat \chi_{it} - \chi_{it}\r) - \l(\widehat C_{it} - C_{it}\r) \\
        &= \l(\widehat \bbeta_i - \bbeta_i \widehat{\bmcH}^{-1} \r) \widehat \bsx_t + \bbeta_i \widehat{\bmcH}^{-1}\l(\widehat \bsx_t - \widehat{\bmcH} \bsx_t \r) \\
        & - \l(\widehat \bLambda_i - \bLambda_i \widehat{\bH}^{-1} \r) \widehat{\bW}_t^{y, n} - \bLambda_i \widehat{\bH}^{-1}\l(\widehat{\bW}_t^{y, n} - \widehat{\bH}\bF_t \r) .
\end{align*}
Following the proof of Theorem \ref{thm: asy norm chi} and setting $a_{nT} := \frac{\min\l(\sqrt{n}, \sqrt{T}\r)}{\sqrt{T}}$ and $b_{nT} := \frac{\min\l(\sqrt{n}, \sqrt{T}\r)}{\sqrt{n}}$, we obtain
\begin{align*}
        \min\l(\sqrt{n}, \sqrt{T}\r)\l(\widehat e_{it}^\chi - e_{it}^\chi\r) &= a_{nT} \underbrace{\l[\bsx_t'\l(\bI_{p+1}\otimes \bP_{\bLambda}'\r) \sqrt{T}\l(\widehat \bbeta_i' - \widehat{\bmcH}^{-1'}\bbeta_i' \r)
        - \bF_t' \bP_{\bLambda}' \sqrt{T}\l(\widehat \bLambda_i' - \widehat{\bH}^{-1}\bLambda_i' \r) \r]}_{\theta_{it, nT}} \\
        &+ b_{nT}\underbrace{\l[\bbeta_i \l(\bI_{p+1}\otimes \bP_{\bLambda}'\r)\sqrt{n}\l(\widehat \bsx_t - \widehat{\bmcH} \bsx_t \r) 
        - \bLambda_i \bP_{\bLambda}' \sqrt{n}\l(\widehat{\bW}_t^{y, n} - \widehat{\bH}\bF_t \r) \r]}_{\zeta_{it, nT}} + o_P(1) \\
        &= a_{nT} \theta_{it, nT} + b_{nT} \zeta_{it, nT} + o_P(1).
    \end{align*}
    Since $\theta_{it, nT}$ is a sum over cross-section and $\zeta_{it, nT}$ is a sum over time and both terms are asymptotically independent. We compute the asymptotic variances by Theorem \ref{lem: asymptotic normality loadings residuals} and \ref{lem: asymptotic normality of hat bsxt} as
    \begin{align*}
        \theta_{it, nT} &\Rightarrow \mathcal N(0, U_{it}), \
        \mbox{while} \quad U_{it} := \bsx_t'\bGamma_{\bsx}^{-1} \bOmega_{\bsx\bxi}(i) \bGamma_{\bsx}^{-1}\bsx_t + \bF_t' \bOmega_{\bF\bse}(i) \bF_t -2 \bsx_t'\bGamma_{\bsx}^{-1} \bOmega_{\bsx\bxi, \bF\bse}(i) \bF_t \\[0.8em]
        \zeta_{it, nT} &\Rightarrow \mathcal N(0, V_{it}), \ \mbox{while} \quad V_{it}:=\bbeta_i \l(\bI_{p+1} \otimes \bGamma_{\bLambda}^{-1}\r) \bOmega_{\bsx\bxi}(i) \l(\bI_{p+1} \otimes \bGamma_{\bLambda}^{-1}\r) \bbeta_i' + \bLambda_i \bGamma_{\bLambda}^{-1} \bOmega_{\bF\bse}(i)  \bGamma_{\bLambda}^{-1} \bLambda_i'\\
        & \hspace{6cm} - 2\bbeta_i \l(\bI_{p+1} \otimes \bGamma_{\bLambda}^{-1}\r) \bOmega_{\bsx\bxi, \bF\bse}(i) \bGamma_{\bLambda}^{-1} \bLambda_i'.
    \end{align*}

    As in the proof of Theorem \ref{thm: asy norm chi} for $n,T \to \infty$, we have
    \begin{align*}
          \frac{\widehat e_{it}^\chi - e_{it}^\chi}{\sqrt{\frac{1}{n} V_{it} + \frac{1}{T}U_{it}}} \Rightarrow \mathcal N(0, 1), 
    \end{align*}
    which completes the proof.
\end{proof}
\section{Asymptotic Normality for Static Factor Decomposition}
\begin{lemma}
     Under Assumptions E\ref{A: divergence rates eval}-E\ref{A: more restrictive sample covarainces}, as $\sqrt{T} / n \to 0$, with $\widehat{\bH}= \frac{1}{T}\sum_{t = 1}^T \widehat{\bW}_t^{y, n} \bF_t' \left( \frac{1}{T}\sum_{t = 1}^T \bF_t \bF_t'\right)^{-1}$
    \begin{align*}
        \sqrt{T}\left(\widehat \bLambda_i - \bLambda_i \widehat{\bH}^{-1}\right) \Rightarrow N\l(0, asy\bGamma_{\bLambda_i}\r).
    \end{align*}
    The asymptotic variance is given by 
    \begin{align*}
        asy\bGamma_{\bLambda_i}:= \bP_{\bLambda} \bOmega_{\bF\bse}(i) \bP_{\bLambda}'.
    \end{align*}
\end{lemma}
\begin{proof}
Set $\widehat{\bW}' = \l(\widehat{\bW}_1^{y, n}, ...,\widehat{\bW}_{T}^{y, n}\r)$ and note that $T^{-1}\widehat{\bW}' \widehat{\bW} = \bI_r$ by construction, then consider the following expansion where $\bsy^i = (y_{i1}, ..., y_{iT})'$ and $\bse^i = (e_{i1}, ..., e_{iT})'$: 
      \begin{align*}
        &\sqrt{T} \left((\widehat{\bW}' \widehat{\bW})^{-1} \widehat{\bW}' \bsy^i - \widehat{\bH}^{-1'} \bLambda_i' \right) \\
        &=\sqrt{T} \left[(\widehat{\bW}' \widehat{\bW})^{-1} \widehat{\bW}'\left(\bF \widehat{\bH}' \widehat{\bH}^{-1'} \bLambda_i ' + \bse^i\right) - \widehat{\bH}^{-1'} \bLambda_i'\right] \\
        &=\sqrt{T} \left[ (\widehat{\bW}' \widehat{\bW})^{-1} \widehat{\bW}' \left( \widehat{\bW} \widehat{\bH}^{-1'} \bLambda_i' + \bse^i + 
        (\bF \widehat{\bH}' - \widehat{\bW}) \widehat{\bH}^{-1'} \bLambda_i' \right) - \widehat{\bH}^{-1'} \bLambda_i' \right] \\
        &= \left[T^{-1/2} \widehat{\bW}' \bse^i + T^{-1/2} \widehat{\bW}' \left(\bF \widehat{\bH}' - \widehat{\bW} \right)\widehat{\bH}^{-1'}\bLambda_i'\right]\\
        &=  \left[\widehat{\bH}T^{-1/2}  \bF ' \bse^i +\underbrace{ T^{-1/2} (\widehat{\bW} - \bF \widehat{\bH}')'\bse^i}_{(1)} + \underbrace{T^{-1/2} \widehat{\bW}' \left(\bF \widehat{\bH}' - \widehat{\bW} \right)\widehat{\bH}^{-1'}\bLambda_i'}_{(2)} \right]
\end{align*}
Now, by Assumption E\ref{A: CLTs}, we know that $T^{-1/2}\bF'\bse^i \Rightarrow \mathcal N\l(\bm{0}, \bOmega_{\bF\bse}(i)\r)$ and $(1) = o_P(1)$ and $(2) = o_p(1)$ by Lemma \ref{lem: x xi and x x}. Consequently as $\widehat{\bH}= \bP_{\bLambda} + o_p(1)$ we obtain that 
\begin{align*}
    \sqrt{T}\l(\widehat \bLambda_i - \bLambda_i \widehat{\bH}^{-1}\r) \Rightarrow \mathcal N\l(\bm{0}, \bP_{\bLambda} \bOmega_{\bF\bse}(i) \bP_{\bLambda}'\r), 
\end{align*}
which completes the proof. 
\end{proof}
\begin{lemma}
    Under Assumptions E\ref{A: divergence rates eval}-E\ref{A: more restrictive sample covarainces}, as $\sqrt{n} / T \to 0$ with $\widehat{\bH}= \frac{1}{T}\sum_{t = 1}^T \widehat{\bW}_t^{y, n} \bF_t' \left( \frac{1}{T}\sum_{t = 1}^T \bF_t \bF_t'\right)^{-1}$
    \begin{align*}
        \sqrt{n}\left(\widehat{\bW}_t^{y, n} - \widehat{\bH}\bF_t\right) \Rightarrow N\l(0, asy\bGamma_{\bF_t}\r).
    \end{align*}
    The asymptotic variance is given by 
    \begin{align*}
        asy\bGamma_{\bF_t} = \bD_{\bLambda}^{-1}\bP_{\bLambda} \bTheta_{\bLambda \bse}(t) \bP_{\bLambda}' \bD_{\bLambda}^{-1}.
    \end{align*}
\end{lemma}
\begin{proof}
    Since $\widehat{\bH}= \widehat{\bmcK}\bLambda^n + \widehat{\bmcK} \widehat \bGamma_{\bse \bF}^n \widehat \bGamma_{\bF}^{-1}$ and using the expansion of $\widehat{\bmcK}$ from \eqref{eq: expansion of hat mathcal K}, we have
    \begin{align*}
        \sqrt{n}\left(\widehat{\bW}_t^{y, n} - \widehat{\bH}\bF_t\right) &= \sqrt{n}\left(\widehat{\bmcK}\bLambda^n \bF_t + \widehat{\bmcK} \bse_t^n - \widehat{\bmcK}\bLambda^n \bF_t - \widehat{\bmcK} \widehat \bGamma_{\bse \bF}^n \widehat \bGamma_{\bF}^{-1} \bF_t\right)\\
        &= \l(\widehat \bD_{\bLambda}^n\r)^{-1} \widehat{\bH}_{\bLambda}' \frac{1}{\sqrt{n}}\bLambda^{n'} \bse_t^n +\underbrace{\l(\widehat \bD_{\bLambda}^n\r)^{-1} \frac{1}{\sqrt{n}}\l(\widehat \bLambda^n -\bLambda^n \widehat{\bH}_{\bLambda}\r)'\bse_t^n}_{(1)}  -\underbrace{\sqrt{n}\widehat{\bmcK} \widehat \bGamma_{\bse \bF}^n \widehat \bGamma_{\bF}^{-1}}_{(2)} \bF_t.
    \end{align*}
For $n,T\to \infty$ by Lemmas \ref{lem: x xi and x x} and \ref{lem: killing K} it follows that $(1) = o_P(1)$ and $(2) = o_P(1)$ and therefore by Assumptions E\ref{A: CLTs}, we obtain with $\widehat{\bH}_{\bLambda}' = \bP_{\bLambda} + o_P(1)$ and $\l(\widehat \bD_{\bLambda}^n\r)^{-1} = \bD_{\bLambda}^{-1} + o_P(1)$ (see Lemmas \ref{lem: H} and \ref{lem: hat evals-vecs}) that
\begin{align*}
     \sqrt{n}\left(\widehat{\bW}_t^{y, n} - \widehat{\bH}\bF_t\right) \Rightarrow \mathcal N\l(\bm{0}, \bD_{\bLambda}^{-1}\bP_{\bLambda} \bTheta_{\bLambda \bse}(t) \bP_{\bLambda}' \bD_{\bLambda}^{-1}\r), 
\end{align*}
which completes the proof. 
\end{proof}
\begin{theorem}[Asymptotic Normality of the Static Common Component]\label{thm: asy norm C}
Under Assumptions E\ref{A: divergence rates eval}-E\ref{A: more restrictive sample covarainces}, as $\sqrt{n} / T \to 0$ and $\sqrt{T} / n \to 0$, we have 
\begin{align*}
    \frac{\widehat C_{it} - C_{it}}{\sqrt{\frac{1}{T}U_{it} + \frac{1}{n} V_{it}}} \Rightarrow \mathcal N(0, 1), 
\end{align*}
where $U_{it} := \bF_t' \bOmega_{Fe}(i) \bF_t$ and $V_{it} := \bLambda_i \bGamma_{\bLambda}^{-1} \bTheta_{\bLambda\bse}(i) \bGamma_{\bLambda}^{-1}\bLambda_i'$.
\end{theorem}
\begin{proof}
    The proof is analogous to the proof of Theorem \ref{thm: asy norm chi}.
\end{proof}
\section{Auxiliary Lemmas}
An indispensable tool for understanding the asymptotic behaviour of population and sample eigenvector is the variant of the Davis-Kahan Theorem proved by \cite{yu2015useful}, which is used in the context of factor models by \cite{barigozzi2022principal}. 
\begin{lemma}[Population Eigenvectors and Eigenvalues]\label{lem: evals-vecs Gammay GammaC}
    Under Assumption E\ref{A: divergence rates eval}, we have
    \begin{itemize}
        \item[(i)] $\norm{\bsp_j\left(\bGamma_{\bsy}^n\right) - \bsp_j(\bGamma_{\bC}^n)} = \mathcal O(n^{-1})$, for $1\leq j \leq r$;
        \item[(ii)] $\abs{\mu_j\left(\bGamma_{\bsy}^n\right) - \mu_j(\bGamma_{\bC}^n)} = \mathcal O(1)$ for $1\leq j \leq r$; 
        \item[(iii)] $\norm{\frac{\bM\left(\bGamma_{\bsy}^n\right)}{n} - \frac{\bM(\bGamma_{\bC}^n)}{n}} = \mathcal O(n^{-1})$;
        \item[(iv)] $\norm{\frac{\bM(\bGamma_{\bC}^n)}{n} - \bD_{\bLambda}} = \mathcal O(n^{-1/2})$ and $\norm{\frac{\bM\left(\bGamma_{\bsy}^n\right)}{n} - \bD_{\bLambda}} = \mathcal O(n^{-1/2})$;
        \item[(v)] $\norm{\left(\frac{\bM(\bGamma_{\bC}^n)}{n}\right)^{-1/2} - \bD_{\bLambda}^{-1/2}} = \mathcal O(n^{-1/2})$ and $\norm{\left(\frac{\bM\left(\bGamma_{\bsy}^n\right)}{n}\right)^{-1/2} - \bD_{\bLambda}^{-1/2}} = \mathcal O(n^{-1/2})$.
    \end{itemize}
\end{lemma}
\begin{proof}
In the following, without loss of generality, we may assume that $\bGamma_{\bsy}^n := \bGamma_{y_t}^n$ is constant over time. 

(i) We apply Theorem 2 from \cite{yu2015useful}, which is a variant of the Davis-Kahan Theorem, then by Assumption E\ref{A: divergence rates eval}:
\begin{align}
  \norm{\bsp_j(\bGamma_{\bsy}^n) - \bsp_j(\bGamma_{\bC}^n)} &\leq \frac{2^{3/2} \mu_1(\bGamma_{\bsy}^n - \bGamma_{\bC}^n)}{\min\left(\mu_{j-1}\left(\bGamma_{\bC}^n\right) - \mu_j\left(\bGamma_{\bC}^n\right), \mu_j\left(\bGamma_{\bC}^n\right) - \mu_{j+1}\left(\bGamma_{\bC}^n\right)\right)} \label{eq: evec Yu pop} \\[0.8em]
  &\leq \frac{\frac{1}{n} 2^{3/2} \mathcal B_e}{\min\left(\frac{\mu_{j-1}\left(\bGamma_{\bC}^n\right)}{n} - \frac{\mu_j\left(\bGamma_{\bC}^n\right)}{n}, \frac{\mu_j\left(\bGamma_{\bC}^n\right)}{n} - \frac{\mu_{j+1}\left(\bGamma_{\bC}^n\right)}{n}\right)}  = \mathcal O(n^{-1}), \nonumber
\end{align}
where $\mu_0(\cdot) = \infty$.

(ii) From Weyl's inequality, we know that 
\begin{align*}
    \mu_j(\bGamma_{\bC}^n + \bGamma_{\bse}^n) &\leq \mu_j(\bGamma_{\bC}^n) + \mu_1(\bGamma_{\bse}^n) \quad 1\leq j \leq r \\
    \mu_j(\bGamma_{\bC}^n + \bGamma_{\bse}^n) - \mu_j(\bGamma_{\bC}^n) & \leq \mu_1(\bGamma_{\bse}^n) \quad 1\leq j \leq r \\
    \abs{\mu_j\left(\bGamma_{\bsy}^n\right) - \mu_j(\bGamma_{\bC}^n)} &\leq \mathcal B_e \quad 1\leq j \leq r, \\[0.8em]
    \abs{\frac{\mu_j\left(\bGamma_{\bsy}^n\right)}{n} - \frac{\mu_j(\bGamma_{\bC}^n)}{n}}&\leq \frac{\mathcal B_e}{n}\quad 1\leq j \leq r
\end{align*}
which also proves part (iii).

For part (iv) note that $\norm{\bGamma_{\bLambda}^n - \bGamma_{\bLambda}} = \mathcal O(n^{-1/2})$ by Assumption E\ref{A: divergence rates eval}, again by Weyl's inequality, we have
\begin{align}
    %
     %
    |\mu_j(\bGamma_{\bLambda}^n) -\mu_j(\bGamma_{\bLambda})| = \mathcal O(n^{-1/2}) \quad 1 \leq j \leq r \label{eq: ineq for Gamma_Lambda}
\end{align}
Note that $\bM(\bGamma_{\bLambda}^n) = M\left(\frac{\bGamma_{\bC}^n}{n}\right)$, which proves the first part. The second part follows with part (iii).

For part (v), note that for $1\leq j \leq r$, we have
\begin{align*}
|\mu_j(\bGamma_{\bLambda}^n)^{-1} - \mu_j(\bGamma_{\bLambda})^{-1}| \leq |\mu_j\left(\bGamma_{\bLambda}^n\right)^{-1}| |\mu_j\left(\bGamma_{\bLambda}^n\right) - \mu_j\left(\bGamma_{\bLambda}\right)||\mu_j(\bGamma_{\bLambda})^{-1}| = \mathcal O(n^{-1/2})
\end{align*}
by equation \eqref{eq: ineq for Gamma_Lambda} and as $|\mu_j(\bGamma_{\bLambda})^{-1}| = \mathcal O(1)$ and $|\mu_j(\bGamma_{\bLambda}^n)^{-1}| = \mathcal O(1)$. This implies by $(a^2-b^2) = (a-b)/ (a+b)$ for $a, b \in \mathbb R, a+b\neq 0$ that for $\bD_{\bLambda}^n = \bM(\bGamma_{\bLambda}^n) = \bM\left(\bGamma_{\bC}^n / n\right)$
\begin{align*}
\norm{(\bD_{\bLambda}^n)^{-1/2} - \bD_{\bLambda}^{-1/2}} = \mathcal O(n^{-1/2}).
\end{align*}
For the second statement, again for $1\leq j \leq r$ we use the inequality
\begin{align*}
    \abs{\left(\frac{\mu_j\left(\bGamma_{\bsy}^n\right)}{n}\right)^{-1} - \left(\frac{\mu_j(\bGamma_{\bC}^n)}{n}\right)^{-1}} 
    &\leq \abs{\left(\frac{\mu_j\left(\bGamma_{\bsy}^n\right)}{n}\right)^{-1}} \abs{\frac{\mu_j\left(\bGamma_{\bsy}^n\right)}{n} - \frac{\mu_j(\bGamma_{\bC}^n)}{n}} \abs{\left(\frac{\mu_j(\bGamma_{\bC}^n)}{n}\right)^{-1}} \\
    &= \mathcal O(1)\mathcal O(n^{-1/2})\mathcal O(1) = \mathcal O(n^{-1/2}), 
\end{align*}
which completes the proof together with the first part of (v). 
\end{proof}
\begin{lemma}[Sample Covariances]\label{lem: Var est}
Under Assumptions E\ref{A: divergence rates eval}-E\ref{A: simply the sample covariances}, we have for $n, T \to \infty$: 
    \begin{itemize}
         \item[(i)] $\norm{\widehat \bGamma_{\bF}(h) - \bGamma_{\bF}(h)} =\mathcal O_{ms}(T^{-1/2})$, $\norm{\widehat \bGamma_{\bsx}(h) - \bGamma_{\bsx}(h)} =\mathcal O_{ms}(T^{-1/2})$ and $\norm{\widehat \bGamma_{\bF}} = \mathcal O_{ms}(1)$, $ \norm{\widehat \bGamma_{\bF}^{-1}} = \mathcal O_{ms}(1)$, $\norm{\widehat \bGamma_{\bsx}} = \mathcal O_{ms}(1)$, $ \norm{\widehat \bGamma_{\bsx}^{-1}} = \mathcal O_{ms}(1)$; 
          \item[(ii)] $\norm{\frac{1}{\sqrt{n}}\widehat \bGamma_{\bF\bxi}^n (h)} = \mathcal O_{ms}(T^{-1/2})$ and $\norm{\frac{1}{\sqrt{n}}\widehat \bGamma_{\bF\bse}^n} = \mathcal O_{ms}(T^{-1/2})$ for all $n \in \mathbb N$, where $\bGamma_{\bF\bse}^n :=  \E \left[\bF_t \bse_t^{n'}\right] = 0$; 
          \item[(iii)] $\norm{\frac{1}{n} \widehat \bGamma_{\bxi}^n (h)} = \mathcal O_{ms}(\max(T^{-1/2}, n^{-1}))$ and  $\norm{\frac{1}{n} \left(\widehat \bGamma_{\bxi}^n - \frac{1}{T}\sum_{t = 1}^T \bGamma_{\bxi_t}^n(h) \right)} = \mathcal O_{ms}(T^{-1/2})$ and $\norm{\frac{1}{n} \left(\widehat \bGamma_{\bse}^n \right)} = \mathcal O_{ms}(\max(T^{-1/2}, n^{-1}))$ and  $\norm{\frac{1}{n} \left(\widehat \bGamma_{\bse}^n - \frac{1}{T}\sum_{t = 1}^T \bGamma_{\bse_t}^n \right)} = \mathcal O_{ms}(T^{-1/2})$; 
          \item[(iv)] $\norm{\frac{1}{\sqrt{n}}\bxi_t^n} = \mathcal O_{ms}(1)$ and $\norm{\frac{1}{\sqrt{n}}\bse_t^n} = \mathcal O_{ms}(1)$.
    \end{itemize}
\end{lemma}
\begin{proof}
   Part $(i)$ is trivial. Note that $\bF_t^w$ is a linear transformation of $\bF_t$ and lags of $\bF_t$ and $\bGamma_{FF^w} = 0$.
   
   For part $(ii)$ without loss of generality we set $h = 0$: 
   \begin{align*}
   \E\l[\norm{\frac{1}{\sqrt{n}} \widehat \bGamma_{\bF \bxi}^n}^2 \r] &\leq \E \l[ \norm{\frac{1}{\sqrt{n}} \widehat \bGamma_{\bF \bxi}^n}_F^2 \r] = \sum_{l = 1}^r  \frac{1}{n} \sum_{i = 1}^n \E \l[ \l( \frac{1}{T} \sum_{t = 1}^T \bF_t \xi_{it} \r)^2 \r] \\
   &\leq r \max_{1\leq i \leq n} \E \l[ \l( \frac{1}{T} \sum_{t = 1}^T \bF_t \xi_{it} \r)^2 \r] \leq \frac{r \mathcal B_{F\xi}}{T}. 
   \end{align*} 
   Since $\widehat \bGamma_{\bF\bse}^n = \widehat \bGamma_{FF^w}\bLambda^{w, n'} + \widehat \bGamma_{\bF\bxi}^n$ and $ \widehat \bGamma_{FF^w}\bLambda^{w, n'} = \mathcal O_{ms}(T^{-1/2})\mathcal O(1)$ the second part follows.
   
    For part $(iii)$, again we may assume without loss of generality that $h = 0$. First consider that
    \begin{align*}
        \norm{\frac{1}{n} \widehat \bGamma_{\bxi}^n} &= \norm{\frac{1}{n} \frac{1}{T}\sum_{t = 1}^T \l\{\bxi_t^n \bxi_t^{n'}  - \bGamma_{\bxi_t}^n + \bGamma_{\bxi_t}^n \r\}} \leq \norm{\frac{1}{n} \frac{1}{T}\sum_{t = 1}^T \l\{\bxi_t^n \bxi_t^{n'}  - \bGamma_{\bxi_t}^n \r\}} + \norm{\frac{1}{n} \frac{1}{T}\sum_{t= 1}^T \bGamma_{\bxi_t}^n}.
        \end{align*}
        The second term is $\mathcal O(n^{-1})$. For the first term we have
        \begin{align*}    
        \E\l[\norm{\frac{1}{n} \frac{1}{T}\sum_{t = 1}^T \l\{\bxi_t^n \bxi_t^{n'}  - \bGamma_{\bxi_t}^n \r\}}^2\r] &\leq \E\l[\norm{\frac{1}{n} \frac{1}{T}\sum_{t = 1}^T \l\{\bxi_t^n \bxi_t^{n'}  - \bGamma_{\bxi_t}^n \r\}}_F^2\r]  \\
        &\leq \frac{1}{n^2}\sum_{i = 1}^n \sum_{j = 1}^n \E\l[\l(\frac{1}{T} \sum_{t = 1}^T \l\{\xi_{it}\xi_{jt} - 
\E\l[\xi_{it} \xi_{jt}\r] \r\}\r)^2 \r]  \\
        &\leq \max_{1\leq i \leq n, 1 \leq j \leq n}\E\l[\l(\frac{1}{T} \sum_{t = 1}^T \l\{\xi_{it}\xi_{jt} - 
\E\l[\xi_{it} \xi_{jt}\r] \r\}\r)^2 \r] \leq \mathcal B_\xi / T, 
    \end{align*}
    where the last inequality follows from Assumption E\ref{A: simply the sample covariances}$(ii)$.
    The second part follows because 
    \begin{align*}        
    \frac{1}{n}\widehat \bGamma_{\bse}^n &= \frac{1}{n}\bLambda^{w, n} \widehat \bGamma_{\bF^w} \bLambda^{w, n'} +\frac{1}{n}\bLambda^{w, n} \widehat \bGamma_{\bF^w \bxi} + \frac{1}{n}\widehat \bGamma_{\bxi \bF^w}\bLambda^{w, n'} + \frac{1}{n}\widehat \bGamma_{\bxi}^n \\
    &= \mathcal O_{ms}(T^{-1/2}n^{-1}) + \mathcal O_{ms}(n^{-1/2}T^{-1/2}) + \mathcal O_{ms}(\max(T^{-1/2}, n^{-1})).
    \end{align*}

    Part $(iv)$ is seen as follows using Assumption E\ref{A: divergence rates eval}: 
    \begin{align*}
        \E\l[\norm{\frac{1}{\sqrt{n}}\bxi_t^n}^2\r] &= \frac{1}{n}\sum_{i = 1}^n \E\l[\xi_{it}^2\r] \leq \max_{1 \leq i \leq n} \E\l[\xi_{it}^2\r] < \mathcal B_\xi \\
        \E\l[\norm{\frac{1}{\sqrt{n}}\bse_t^n}^2\r] &= \frac{1}{n}\sum_{i = 1}^n \E\l[e_{it}^2\r]  \leq 
        \frac{1}{n}\sum_{i = 1}^n \E\l[\xi_{it}^2\r] + \frac{1}{n}\sum_{i = 1}^n \E\l[\l(\bLambda_i^w \bF_t^w\r)^2\r] \\
        &\leq \max_{1 \leq i \leq n} \E\l[\xi_{it}^2\r] + \max_{1 \leq i \leq n} \E\l[\l(\bLambda_i^w \bF_t^w\r)^2\r] < \mathcal B_\xi + (r_\chi - r) \mathcal B_{\Lambda^w}.
    \end{align*}
\end{proof} 
\begin{lemma}[Sample Eigenvectors and Eigenvalues]\label{lem: hat evals-vecs}
    Under Assumptions T\ref{A: divergence rates eval}-E\ref{A: simply the sample covariances}, we have for $1\leq j \leq r$ and independent of $n\in \mathbb N$, and $T\to \infty$ that
    \begin{itemize}
        \item[(i)] $\norm{\bsp_j\left(\widehat\bGamma_{\bsy}^n\right) - \bsp_j\left(\bGamma_{\bC}^n\right)} = \mathcal O_{ms}(\max(T^{-1/2}, n^{-1}))$; \\
        If $\bGamma_{\bse_t}^n = \bGamma_{\bse}^n$ for all $t \in \mathbb Z$, then $\norm{\bsp_j\left(\widehat\bGamma_{\bsy}^n\right) - \bsp_j\left(\bGamma_{\bsy}^n\right)} = \mathcal O_{ms}(T^{-1/2})$; 
        \item[(ii)] $\norm{\frac{\bM\left(\widehat\bGamma_{\bsy}^n\right)}{n}  - \frac{\bM\left(\bGamma_{\bC}^n\right)}{n}} =  \mathcal O_{ms}(\max(T^{-1/2}, n^{-1}))$; \\
        If $\bGamma_{\bse_t}^n = \bGamma_{\bse}^n$ for all $t \in \mathbb Z$, then $\norm{\frac{\bM\left(\widehat\bGamma_{\bsy}^n\right)}{n}  - \frac{\bM\left(\bGamma_{\bsy}^n\right)}{n}} = \mathcal O_{ms}(T^{-1/2})$; 
        \item[(iii)] $\abs{\l(\frac{\mu_j\l(\widehat \bGamma_{\bsy}^n\r)}{n}\r)^{-1/2} - \l(\frac{\mu_j\l(\bGamma_{\bC}^n\r)}{n}\r)^{-1/2}} = \mathcal O_P(\max(T^{-1/2}, n^{-1}))$; \\
        If $\bGamma_{\bse_t}^n = \bGamma_{\bse}^n$ for all $t \in \mathbb Z$, then $\abs{\l(\frac{\mu_j\l(\widehat \bGamma_{\bsy}^n\r)}{n}\r)^{-1/2} - \l(\frac{\mu_j\l(\bGamma_{\bsy}^n\r)}{n}\r)^{-1/2}} = \mathcal O_P(T^{-1/2})$
        %
        %
        %
        \item[(iv)]  $n^{1/2}\norm{\widehat{\bmcK}_j(\widehat \bGamma_{\bsy}^n) - \bmcK_j(\bGamma_{\bC}^n)} = \mathcal O_P(\max(T^{-1/2}, n^{-1}))$; \\
        If $\bGamma_{\bse_t}^n = \bGamma_{\bse}^n$ for all $t \in \mathbb Z$, then $n^{1/2}\norm{\widehat{\bmcK}_j(\widehat \bGamma_{\bsy}^n) - \bmcK_j(\bGamma_{\bsy}^n)} = \mathcal O_P(T^{-1/2})$.
        %
        %
    \end{itemize}
    \end{lemma}
\begin{proof}
To prove the second part of $(i)$, we use again \cite{yu2015useful}, Theorem 2, for $1\leq j \leq r$ with $\mu_0(\cdot) = \infty$
\begin{align*}
    \norm{\bsp_j(\widehat \bGamma_{\bsy}^n) - \bsp_j(\bGamma_{\bsy}^n)} \leq \frac{2^{3/2} \frac{1}{n}\norm{\widehat \bGamma_{\bsy}^n - \bGamma_{\bsy}^n}}{\min\left( \frac{\mu_{j-1}(\bGamma_{\bsy}^n)}{n} -\frac{\mu_j(\bGamma_{\bsy}^n)}{n}, \frac{\mu_{j+1}(\bGamma_{\bsy}^n)}{n} - \frac{\mu_{j}(\bGamma_{\bsy}^n)}{n} \right)}.
\end{align*}
Now using that 
\begin{align*}
    \widehat \bGamma_{\bsy}^n - \bGamma_{\bsy}^n = \bLambda^n \widehat \bGamma_{\bF}\bLambda^{n'} -\bLambda^n\bLambda^{n'} + \bLambda^n \widehat \bGamma_{\bF\bse}^n +  \widehat \bGamma_{\bse \bF}^n\bLambda^{n'} + \widehat \bGamma_{\bse}^n - \bGamma_{\bse}^n,
\end{align*}
we have
\begin{align}
    \frac{1}{n}\norm{\widehat \bGamma_{\bsy}^n  - \bGamma_{\bsy}^n} &\leq \frac{1}{n}\norm{\bLambda^n \left(\widehat \bGamma_{\bF} - \bI_r \right)\bLambda^{n'}} +\frac{1}{n}\norm{\bLambda^n \widehat \bGamma_{\bF\bse}^n} +\frac{1}{n} \norm{ \widehat \bGamma_{\bse \bF}^n\bLambda^{n'}} + \frac{1}{n}\norm{\widehat \bGamma_{\bse}^n - \bGamma_{\bse}^n} \nonumber \\
    &\leq \norm{\frac{\bLambda^n}{{\sqrt{n}}}}^2 \norm{\widehat\bGamma_{\bF}^n - \bI_r} + 2\norm{\frac{\bLambda^n}{\sqrt{n}}}\norm{\frac{1}{\sqrt{n}} \widehat \bGamma_{\bF\bse}^n} + \frac{1}{n}\norm{\widehat \bGamma_{\bse}^n - \bGamma_{\bse}^n} \\
    &= \mathcal O(1) \mathcal O_{ms}(T^{-1/2}) + \mathcal O(1) \mathcal O_{ms}(T^{-1/2}) + \mathcal O_{ms}(T^{-1/2}) \nonumber \\
    &= \mathcal O_{ms}(T^{-1/2}). \label{eq: eval hat Gamma -Gamma_y}
\end{align}

Using the same procedure with $\bGamma_{\bC}^n$ instead of $\bGamma_{\bsy}^n$ yields the first part of $(i)$. 

For the second part of $(ii)$, note that by \eqref{eq: eval hat Gamma -Gamma_y} and Weyl's inequality
\begin{align*}
\abs{\frac{\mu_j(\widehat \bGamma_{\bsy}^n)}{n} - \frac{\mu_j(\bGamma_{\bsy}^n)}{n}} &\leq \frac{1}{n}\norm{\widehat \bGamma_{\bsy}^n - \bGamma_{\bsy}^n} = \mathcal O_{ms}(T^{-1/2}) \\  
\abs{\frac{\mu_j(\widehat \bGamma_{\bsy}^n)}{n} - \frac{\mu_j(\bGamma_{\bC}^n)}{n}} &\leq \frac{1}{n}\norm{\widehat \bGamma_{\bsy}^n - \bGamma_{\bC}^n} = \mathcal O_{ms}(\max(T^{-1/2}, n^{-1})).
\end{align*}

For part (iii), set $\widehat \mu_j := \mu_j\left(\widehat \bGamma_{\bsy}^n\right), \mu_j := \mu_j\left(\bGamma_{\bsy}^n\right)$
\begin{align*}
    \left|\left(\frac{\widehat \mu_j}{n}\right)^{-1} - \left(\frac{\mu_j}{n}\right)^{-1}  \right| \leq \left|\left(\frac{\widehat \mu_j}{n}\right)^{-1} \right| \left|\frac{\widehat \mu_j}{n} - \frac{\mu_j}{n}\right| \left|\left(\frac{\mu_j}{n}\right)^{-1}\right| = \mathcal O_{ms}(1) \mathcal O_{ms}(T^{-1/2}) \mathcal O(1) = \mathcal O_P (T^{-1/2}).
\end{align*}
Finally, since for all $a,b \in \mathbb R$ with $a+b \neq 0$, we have $(a^2 - b^2)/(a+b) = (a-b)$, we conclude
\begin{align*}
\abs{\widehat \mu_j^{-1/2} - \mu_j^{-1/2}} &= n^{-1/2} \abs{\left(\frac{\widehat \mu_j}{n}\right)^{-1/2}- \left(\frac{\mu_j}{n}\right)^{-1/2}} \\
&=n^{-1/2} \abs{\left(\frac{\widehat \mu_j}{n}\right)^{-1} - \left(\frac{\mu_j}{n}\right)^{-1}} \abs{\left(\frac{\widehat \mu_j}{n}\right)^{1/2} +\left(\frac{\mu_j}{n}\right)^{1/2}}^{-1} \\[0.8em]
& = n^{-1/2} \mathcal O_P(T^{-1/2}) \mathcal O_{ms}(1) = \mathcal O_P(n^{-1/2} T^{-1/2}). 
\end{align*}
%
%
%
%

For part (iv), using previous results, for the second statement setting $\widehat \bsp_j := \bsp_j(\widehat \bGamma_{\bsy}^n)$ and $\bsp_j:= \bsp_j(\bGamma_{\bsy}^n)$, we obtain
\begin{align*}
        n^{1/2}\norm{\widehat{\bmcK}_j - \bmcK_j} &= n^{1/2}\norm{\widehat \mu_j^{-1/2} \widehat \bsp_j - \mu_j^{-1/2} \bsp_j} \\
        &\leq \norm{\widehat \bsp_j - \bsp_j} n^{1/2}|\mu_j^{-1/2}| + n^{1/2}|\widehat \mu_j^{-1/2} - \mu_j^{-1/2}| \norm{\bsp_j} + \norm{\widehat \bsp_j - \bsp_j}n^{1/2}|\widehat \mu_j^{-1/2} - \mu_j^{-1/2}| \\
        &= \mathcal O_{ms}(T^{-1/2}) + \mathcal O_P(T^{-1/2}) +
         \mathcal O_{ms}(T^{-1/2})\mathcal O_P(T^{-1/2}) \\
         &= \mathcal O_P(T^{-1/2}).
  \end{align*}  
We proceed analogously for the first statement. 
\end{proof}
In the following, we may either use $\bmcK:= \bmcK(\bGamma_{\bC}^n)$ in the heteroskedastic case and assume $n^{-3/2} \leq n^{-1/2}T^{-1/2}$ to keep the rates simple, or use $\bmcK:= \bmcK(\bGamma_{\bsy}^n)$ in the homoskedastic case. Note that we always use $\widehat{\bmcK} := \bmcK(\widehat \bGamma_{\bsy}^n)$ no matter whether the idiosyncratic component is heteroskedastic or not. 
\begin{lemma}\label{lem: mathcal K}
Recall that $\bP_{\bLambda}$ are the eigenvectors of $\lim_n\bLambda^{n'}\bLambda^n / n = \bGamma_{\bLambda}$: 
    \begin{itemize}
        \item[(i)] Under Assumption E\ref{A: divergence rates eval}, we have  $\norm{\bmcK\l(\bGamma_{\bC}^n\r)\bLambda^n - \bP_{\bLambda}} = \mathcal O(n^{-1/2})$; \\
        If $\bGamma_{\bse_t}^n = \bGamma_{\bse}^n$ for all $t \in \mathbb Z$, then $\norm{\bmcK\l(\bGamma_{\bsy}^n\r)\bLambda^n - \bP_{\bLambda}} = \mathcal O(n^{-1/2})$;
        \item[(ii)] Under Assumptions E\ref{A: divergence rates eval}-E\ref{A: simply the sample covariances}, we have $\norm{\widehat{\bmcK}\bLambda^n - \bP_{\bLambda}} = \mathcal O_P(\max(n^{-1/2}, T^{-1/2}))$;
        \item[(iii)] Under Assumptions E\ref{A: divergence rates eval} and E\ref{A: divergence rates eval}, we have 
        $n^{1/2}\bmcK(\bGamma_{\bC}^n) = \mathcal O(1) \frac{\bLambda^{n'}}{\sqrt{n}}$; \\
        If $\bGamma_{\bse_t}^n = \bGamma_{\bse}^n$ for all $t \in \mathbb Z$, then $n^{1/2}\bmcK(\bGamma_{\bsy}^n) = \mathcal O(n^{-1}) + \mathcal O(1) \frac{\bLambda^{n'}}{\sqrt{n}}$.
    \end{itemize}
\end{lemma}
\begin{proof}
For part $(i)$, note that we can write orthonormal row-eigenvectors of $\bGamma_{\bC}^n$ as follows 
\begin{align*}
    \bP(\bGamma_{\bC}^n) &= \left(\bD_{\bLambda}^n\right)^{-1/2} \bP_{\bLambda}^n \frac{\bLambda^{n'}}{\sqrt{n}}\\
    \mbox{since} \quad &\left(\bD_{\bLambda}^n\right)^{-1/2} \bP_{\bLambda}^n \frac{\bLambda^{n'}}{\sqrt{n}} \frac{\bLambda^n}{\sqrt{n}} \bP_{\bLambda}^{n'} \left(\bD_{\bLambda}^n\right)^{-1/2}  = \bI_r\\
    \mbox{and} \quad & (\bD_{\bLambda}^n)^{-1/2} \bP_{\bLambda}^n \frac{\bLambda^{n'}}{\sqrt{n}}\bGamma_{\bC}^n = (\bD_{\bLambda}^n)^{-1/2} \bP_{\bLambda}^n \frac{\bLambda^{n'}}{\sqrt{n}}\frac{\bLambda^n\bLambda^{n'}}{n} n = (\bD_{\bLambda}^n)^{-1/2} \bP_{\bLambda}^n \bGamma_{\bLambda}^n \frac{\bLambda^{n'}}{\sqrt{n}} n  \\
    &= (\bD_{\bLambda}^n)^{-1/2} \bD_{\bLambda}^n \bP_{\bLambda}^n \frac{\bLambda^{n'}}{\sqrt{n}}n = \bD_{\bLambda}^n n \left(\bD_{\bLambda}^n\right)^{-1/2} \bP_{\bLambda}^n \frac{\bLambda^{n'}}{\sqrt{n}} = \bM(\bGamma_{\bC}^n)  \left(\bD_{\bLambda}^n\right)^{-1/2} \bP_{\bLambda}^n \frac{\bLambda^{n'}}{\sqrt{n}}.
\end{align*}

Consequently, plugging in this expression of $\bP(\bGamma_{\bC}^n)$, results in
\begin{align*}
    \bmcK(\bGamma_{\bC}^n)\bLambda^n = \bM^{-1/2}(\bGamma_{\bC}^n) \bP(\bGamma_{\bC}^n)\bLambda^n &=  \frac{1}{\sqrt{n}}\l(\bD_{\bLambda}^n\r)^{-1/2} \l(\bD_{\bLambda}^n\r)^{-1/2} \bP_{\bLambda}^n \frac{\bLambda^{n'}}{\sqrt{n}}\bLambda^n = \l(\bD_{\bLambda}^n\r)^{-1} \bP_{\bLambda}^n \bGamma_{\bLambda}^n = \bP_{\bLambda}^n. 
\end{align*}
Now using the result from \cite{yu2015useful}, we know that for $1\leq j \leq r$ with $\mu_0(\cdot) = \infty$, we have
\begin{align*}
    \norm{\bsp_j(\bGamma_{\bLambda}^n)-\bsp_j(\bGamma_{\bLambda})} \leq \frac{2^{3/2} \mu_1(\bGamma_{\bLambda}^n - \bGamma_{\bLambda})}{\min\l(\mu_{j-1}\l(\bGamma_{\bLambda}\r) - \mu_j\l(\bGamma_{\bLambda}\r), \mu_j\l(\bGamma_{\bLambda}\r) - \mu_{j+1}\l(\bGamma_{\bLambda}\r)\r)} = \mathcal O(n^{-1/2})
\end{align*}
since by Assumption E\ref{A: divergence rates eval}  the eigenvalues of $\bGamma_{\bLambda}^n$ are distinct and $\mu_1 \l(\bGamma_{\bLambda}^n - \bGamma_{\bLambda} \r) = \mathcal O(n^{-1/2})$ by continuity of the eigenvalues in the matrix entries.

For the second part note first that $\bD_{\bLambda}^{-1/2} \bP_{\bLambda} \bGamma_{\bLambda} = \bD_{\bLambda}^{1/2} \bP_{\bLambda}$. Now
\begin{align*}
    &\bmcK \l(\bGamma_{\bsy}^n\r)\bLambda^n = \bM^{-1/2}(\bGamma_{\bsy}^n) \bP(\bGamma_{\bsy}^n)\bLambda^n = \left(\frac{\bM(\bGamma_{\bsy}^n)}{n}\right)^{-1/2} \bP(\bGamma_{\bsy}^n) \frac{\bLambda^n}{\sqrt{n}} \\
    %
    %
    \mbox{with} \quad &\norm{\left(\frac{\bM(\bGamma_{\bsy}^n)}{n}\right)^{-1/2} - \bD_{\bLambda}^{-1/2}} = \mathcal O(n^{-1/2})\\
    \mbox{and} \quad &\norm{\bP(\bGamma_{\bsy}^n)\frac{\bLambda^n}{\sqrt{n}}- \bD_{\bLambda}^{-1/2} \bP_{\bLambda} \bGamma_{\bLambda}} =  \norm{\bP(\bGamma_{\bsy}^n)\frac{\bLambda^n}{\sqrt{n}} - \bD_{\bLambda}^{1/2}\bP_{\bLambda}} = \mathcal O(n^{-1/2}),
\end{align*}
and we obtain the desired result by standard arguments.

Part $(ii)$ follows from
\begin{align*}
\norm{\widehat{\bmcK}\bLambda^n - \bP_{\bLambda}} &\leq \norm{\left(\widehat{\bmcK} - \bmcK\right)\bLambda^n} + \norm{\bmcK\bLambda^n - \bP_{\bLambda}} \\
& \leq  \norm{\left(\widehat{\bmcK} - \bmcK\right)} \norm{\bLambda^n} + \norm{\bmcK\bLambda^n - \bP_{\bLambda}} \\
&= \mathcal O_P(T^{-1/2} n^{-1/2}) \mathcal O(n^{1/2}) + \mathcal O(n^{-1/2}) = \mathcal O_P(\max(T^{-1/2}, n^{-1/2})).
\end{align*}

Part $(iii)$ is seen as follows: 
\begin{align*}
     \bmcK(\bGamma_{\bC}^n)  &= \bM^{-1/2}(\bGamma_{\bC}^n) \bP(\bGamma_{\bC}^n) = \frac{1}{\sqrt{n}} \l(\bD_{\bLambda}^n\r)^{-1/2} \l(\bD_{\bLambda}^n\r)^{-1/2} \bP_{\bLambda}^n \frac{\bLambda^{n'}}{\sqrt{n}} =\l(\bD_{\bLambda}^n\r)^{-1} \frac{\bLambda^{n'}}{n}  \\
    \bmcK\l(\bGamma_{\bsy}^n\r) &= \bM^{-1/2}\left(\bGamma_{\bsy}^n\right)\left(\bP\left(\bGamma_{\bsy}^n\right) - \bP\left(\bGamma_{\bC}^n\right)\right) +  \bM^{-1/2}\left(\bGamma_{\bsy}^n\right)\bP\left(\bGamma_{\bC}^n\right) \\
    &= \mathcal O(n^{-1/2}) \mathcal O(n^{-1}) + \left(\frac{\bM\left(\bGamma_{\bsy}^n\right)}{n}\right)^{-1/2}\left(\bD_{\bLambda}^n\right)^{-1/2} \bP_{\bLambda}^n \frac{\bLambda^{n'}}{n} \\
    &=\mathcal O(n^{-3/2}) + \mathcal O(1)\frac{\bLambda^{n'}}{n}.
\end{align*}
\end{proof}
\begin{lemma}[Vanishing under $\widehat{\bmcK}$]\label{lem: killing K}
    Under Assumptions E\ref{A: divergence rates eval}-E\ref{A: simply the sample covariances}, we have for $n, T\to \infty$ and $\abs{h} < \infty$: 
    \begin{itemize}
        \item[(i)] $\norm{\widehat{\bmcK} \bLambda^{w, n}} =\mathcal O_P(n^{-1/2})$; \\
        If E\ref{A: more restrictive sample covarainces} holds as well, $\norm{\widehat{\bmcK} \bLambda^{w, n}} =\mathcal O_P(\max(T^{-1/2}n^{-1/2}, n^{-1}))$; 
        \item[(ii)] $\norm{ \widehat{\bmcK} \widehat \bGamma_{\xi x}^n (h)} = \mathcal O_P(\max(T^{-1}, n^{-1/2}T^{-1/2}))$ and $\norm{ \widehat{\bmcK} \widehat \bGamma_{\bse \bF}^n} = \mathcal O_P(\max(T^{-1}, n^{-1/2}T^{-1/2}))$; 
        \item[(iii)] $\norm{\widehat{\bmcK}\widehat \bGamma_{\bxi \bmcF^w}^n(h)} = \mathcal O_P(\max(T^{-1}, n^{-1/2}T^{-1/2}))$; 
        \item[(iv)] $\norm{\widehat{\bmcK} \widehat \bGamma_{\bxi}^n(h) \widehat{\bmcK}'} = \mathcal O_P(\max(T^{-1}, n^{-1}))$ and $\norm{\widehat{\bmcK} \widehat \bGamma_{\bse}^n \widehat{\bmcK}'} = \mathcal O_P(\max(T^{-1}, n^{-1}))$
    \end{itemize}
\end{lemma}
\begin{proof}
    
    For part $(i)$, by Lemma \ref{lem: hat evals-vecs} and Assumption E\ref{A: divergence rates eval} we have 
     \begin{align*}
        \widehat{\bmcK} \bLambda^{w, n} &= (\widehat{\bmcK}- \bmcK) \bLambda^{w, n} + \bmcK \bLambda^{w, n} \\[0.8em]
        &= \begin{cases}
        \mathcal O_P(n^{-1/2} T^{-1/2}) \mathcal O(1) +\mathcal O(n^{-1/2}) \mathcal O(1) \quad \mbox{if E\ref{A: divergence rates eval}(iv) holds but not E\ref{A: more restrictive sample covarainces}(ii),} \\
        \mathcal O_P(n^{-1/2} T^{-1/2}) \mathcal O(1) +\mathcal O(n^{-1}) \quad \mbox{if E\ref{A: more restrictive sample covarainces}(ii) holds.}
        \end{cases}
    \end{align*}

    For part $(ii)$, using Lemmas \ref{lem: Var est}, \ref{lem: hat evals-vecs} and \ref{lem: killing K} and Assumption E\ref{A: simply the sample covariances}(v), we have
    \begin{align*}
        \widehat{\bmcK} \widehat \bGamma_{\xi x}^n(h) &= \left(\widehat{\bmcK} - \bmcK\right) \widehat \bGamma_{\xi x}^n (h) + \bmcK \widehat \bGamma_{\xi x}^n(h) = \mathcal O_P(n^{-1/2}T^{-1/2}) \mathcal O_{ms}(n^{1/2} T^{-1/2}) + \mathcal O_{ms}(n^{-1/2}T^{-1/2}) \\
        &= \mathcal O_P(\max(T^{-1}, n^{-1/2}T^{-1/2})), 
    \end{align*}
    while the second term follows from Assumption E\ref{A: simply the sample covariances}(v) together with Lemma \ref{lem: mathcal K}.

    Part $(iii)$ is proved analogously to part $(ii)$. 

    For part $(iv)$, without loss of generality, we consider the case $h = 0$. By Assumptions E\ref{A: simply the sample covariances}$(iii)$ together with Lemma \ref{lem: mathcal K} and Assumption E\ref{A: divergence rates eval}$(vi)$, we have 
    \begin{align}
        \bmcK \widehat \bGamma_{\bxi}^n \bmcK' &= \bmcK \l(\frac{1}{T} \sum_{t = 1}^T \l\{   \bxi_t^n \bxi_t^{n'} - \bGamma_{\bxi_t}^n + \bGamma_{\bxi_t}^n   \r\} \r) = \mathcal O(1) \frac{1}{nT} \sum_{t = 1}^T \l\{\bseta_t^n \bseta_t^{n'} - \E\l[\bseta_t^n \bseta_t^{n'}\r] \r\} \mathcal O(1) + \bmcK \frac{1}{T}\sum_{t = 1}^T \bGamma_{\bxi_t}^n \bmcK'   \nonumber \\
        &= \mathcal O_{ms}(n^{-1}T^{-1/2}) + \mathcal O(n^{-1}) = \mathcal O_{ms}(n^{-1}). \label{eq: K hat Gamma_xi K'}
    \end{align}
   By Lemma \ref{lem: hat evals-vecs}$(iv)$, \ref{lem: Var est}$(iii)$ together with \eqref{eq: K hat Gamma_xi K'}, it follows that
    \begin{align*}
        \widehat{\bmcK} \widehat \bGamma_{\bxi}^n(h) \bmcK' &= \l(\widehat{\bmcK} - \bmcK\r) \widehat \bGamma_{\bxi}^n(j) \bmcK' + \bmcK \widehat \bGamma_{\bxi}^n(h) \bmcK'  \\
        &= \mathcal O_P(n^{-1/2}T^{-1/2}) \mathcal O_{ms}(nT^{-1/2}) \mathcal O(n^{-1/2}) + \mathcal O_{ms}(n^{-1}) = \mathcal O_P \l(\max(n^{-1}, T^{-1})\r),  \\[0.8em]
        \mbox{thus} \quad \widehat{\bmcK} \widehat \bGamma_{\bxi}^n(h) \widehat{\bmcK}' &= \l(\widehat{\bmcK} - \bmcK\r) \widehat \bGamma_{\bxi}^n(h)  \l(\widehat{\bmcK} - \bmcK\r)' + \widehat{\bmcK} \widehat \bGamma_{\bxi}^n(j) \bmcK' + \bmcK \widehat \bGamma_{\bxi}^n(h) \widehat{\bmcK}' - \bmcK \widehat \bGamma_{\bxi}^n(h) \bmcK' \\
        &= \mathcal O_P(n^{-1/2}T^{-1/2}) \mathcal O_{ms}(nT^{-1/2})\mathcal O_P(n^{-1/2}T^{-1/2})  + \mathcal O_P \l(\max(n^{-1}, T^{-1})\r) + \mathcal O_{ms}(n^{-1}) \\
        &= \mathcal O_P \l(\max(n^{-1}, T^{-1})\r).
    \end{align*}
        %
\end{proof}
Recall that $\widehat{\bH}$ or $\widehat{\bmcH}$ are the regression coefficient matrices from regressing the principal component estimated factors or stacks on the ``true'' factors. As it turns out, if we normalise the true factors to have unit variance, then $\widehat{\bH}$ converges to the eigenvectors of $\bGamma_{\bLambda}$ which is the ``variance matrix'' of the loadings. 
\begin{lemma}\label{lem: H}
Under Assumptions E\ref{A: divergence rates eval}-E\ref{A: simply the sample covariances}, it follows that 
    \begin{itemize}
        \item[(i)] $\norm{\widehat{\bmcK}\bLambda^n - \widehat{\bH}} = \mathcal O_P(\max(T^{-1}, n^{-1/2}T^{-1/2}))$ with $\widehat{\bH}= \frac{1}{T}\sum_{t = 1}^T \widehat{\bW}_t^{y, n} \bF_t' \left( \frac{1}{T}\sum_{t = 1}^T \bF_t \bF_t'\right)^{-1}$; 
        \item[(ii)] $\norm{\widehat{\bH}- \bP_{\bLambda}} =\mathcal O_P(\max(n^{-1/2}, T^{-1/2}))$ and  $\norm{\widehat{\bH}^{-1}- \bP_{\bLambda}'} =\mathcal O_P(\max(n^{-1/2}, T^{-1/2}))$; 
        \item[(iii)] $\norm{\left(\bI_{p+1}\otimes \widehat{\bmcK}\bLambda^n\right) - \widehat{\bmcH}} =  \mathcal O_P(\max(T^{-1/2}n^{-1/2}, n^{-1}, T^{-1}))$; 
        \item[(iv)] $\norm{\left(\bI_{p+1}\otimes \bP_{\bLambda}\right) - \widehat{\bmcH}} =  \mathcal O_P(\max(T^{-1/2}, n^{-1/2}))$ and $\norm{\left(\bI_{p+1}\otimes \bP_{\bLambda}'\right) - \widehat{\bmcH}^{-1}} =  \mathcal O_P(\max(T^{-1/2}, n^{-1/2}))$; 
        \item[(v)] $\norm{\widehat{\bH}_{\bLambda} - \bP_{\bLambda}'} = \mathcal O_P(\max(n^{-1/2}, T^{-1/2}))$;
        \item[(vi)] $\frac{1}{T}\sum_{t = 1}^T \widehat \bsx_t \widehat \bsx_t' = \widehat \bGamma_{\widehat{\bsx}} = \l(\bI_{p+1}\otimes \bP_{\bLambda}\r) \bGamma_{\bsx} \l(\bI_{p+1}\otimes \bP_{\bLambda}\r)' + \mathcal O_P(\max(n^{-1/2}, T^{-1/2}))$;
        %
        %
        %
        %
        \item[(vii)] $\widehat{\bH} = \mathcal O_P(1)$, $\widehat{\bH}^{-1} = \mathcal O_P(1)$ and $\widehat{\bmcH} = \mathcal O_P(1)$, $\widehat{\bmcH}^{-1} = \mathcal O_P(1)$.
    \end{itemize}
\end{lemma}
\begin{proof}
    For part $(i)$ note that 
  \begin{align}
        \widehat{\bH}&= \frac{1}{T}\sum_{t=1}^T \left(\widehat{\bmcK}\bLambda^n \bF_t + \widehat{\bmcK} \bse_t^n\right) \bF_t' \widehat \bGamma_{\bF}^{-1} = \widehat{\bmcK}\bLambda^n + \widehat{\bmcK}\widehat \bGamma_{\bse \bF}^n \widehat \bGamma_{\bF}^{-1} \label{eq: expansion of hat H} \\
        &= \widehat{\bmcK}\bLambda^n + \mathcal O_P(\max(T^{-1}, n^{-1/2}T^{-1/2})) \mathcal O_P(1) \nonumber
    \end{align}

Part $(ii)$ follows immediately from part $(i)$ and Lemma \ref{lem: mathcal K}$(ii)$.

For part $(iii)$ we look at the expansion 
\begin{align}
        \widehat{\bmcH} &= \frac{1}{T} \sum_{t = 1}^T \widehat \bsx_t \bsx_t' \left(\frac{1}{T}\sum_{t = 1}^T \bsx_t \bsx_t'\right)^{-1} \nonumber \\
        &= \frac{1}{T} \sum_{t = 1}^T\left( \left(\bI_{p+1}\otimes \widehat{\bmcK}\bLambda^n \right) \bsx_t +  \left(\bI_{p+1}\otimes \widehat{\bmcK} \bLambda^{w, n} \right)\bmcF_t^w  +  \left(\bI_{p+1}\otimes \widehat{\bmcK} \right) \bXi_t^n\right)\bsx_t' \widehat \bGamma_{\bsx}^{-1} \nonumber \\
        %
        %
        %
        &=\left(\bI_{p+1}\otimes \widehat{\bmcK}\bLambda^n \right) +  \left(\bI_{p+1}\otimes \widehat{\bmcK} \bLambda^{w, n} \right) \widehat \bGamma_{\bmcF^w \bsx}\widehat \bGamma_{\bsx}^{-1} +  \left(\bI_{p+1}\otimes \widehat{\bmcK} \right) \widehat \bGamma_{\bXi\bsx}^n\widehat \bGamma_{\bsx}^{-1} \label{eq: expansion of mathcal H}\\
        &= \left(\bI_{p+1}\otimes \widehat{\bmcK}\bLambda^n \right) + \mathcal O_P(\max(T^{-1/2}n^{-1/2}, n^{-1})) +  \mathcal O_P(\max(T^{-1}, n^{-1/2}T^{-1/2})) \nonumber
    \end{align}
by Lemma \ref{lem: killing K}. The rate follows from part $(i)$. 

Part $(iv)$ follows from parts $(ii)$ and $(iii)$. 

For part $(v)$, we use Lemma \ref{lem: killing K}$(ii)$ and $(iv)$ considering the expansion for $n, T \to \infty$
\begin{align}
     \widehat{\bH}_{\bLambda}' &= \frac{1}{n}\sum_{j = 1}^n \widehat \bLambda_j' \bLambda_j \left(\frac{1}{n}\sum_{j = 1}^n \bLambda_j'\bLambda_j\right)^{-1}=\frac{1}{n}\sum_{j = 1}^n \widehat \bLambda_j' \bLambda_j \left(\bGamma_{\bLambda}^n\right)^{-1} \nonumber \\
        &=\left\{ \widehat{\bmcK}\bLambda^n \widehat \bGamma_{\bF} \bGamma_{\bLambda}^n  + \widehat{\bmcK}\bLambda^n \widehat \bGamma_{\bF\bse}^n \frac{1}{n}\bLambda^n + \widehat{\bmcK}\widehat \bGamma_{\bse \bF}^n \bGamma_{\bLambda}^n + \widehat{\bmcK}\widehat \bGamma_{\bse}^n \frac{1}{n}\bLambda^n  
        \right\} \left(\bGamma_{\bLambda}^n\right)^{-1} \nonumber \\
        &= \widehat{\bmcK}\bLambda^n \widehat \bGamma_{\bF}  + \widehat{\bmcK}\bLambda^n \widehat \bGamma_{\bF\bse}^n \frac{1}{n}\bLambda^n \left(\bGamma_{\bLambda}^n\right)^{-1}  + \widehat{\bmcK}\widehat \bGamma_{\bse \bF}^n + \widehat{\bmcK}\widehat \bGamma_{\bse}^n \frac{1}{n}\bLambda^n \left(\bGamma_{\bLambda}^n\right)^{-1}  \label{eq: expansion hat H_Lambda'}\\
        &= \widehat{\bmcK}\bLambda^n \widehat \bGamma_{\bF} + 
        \mathcal O_P(\max(T^{-1}, n^{-1/2}T^{-1/2})) + \mathcal O_P(\max(T^{-1}, n^{-1/2}T^{-1/2})) + \mathcal O_P(n^{-1/2}T^{-1/2}) \nonumber \\
        &= \bP_{\bLambda} + \mathcal O(\max(n^{-1/2}, T^{-1/2})).  \nonumber
\end{align}

For part $(vi)$, considering the structure of $\widehat \bsx_t$ it is enough to see that 
\begin{align*}
    \frac{1}{T}\sum_{t = 1}^T \widehat{\bW}_t^{y, n} \widehat{\bW}_{t-h}^{y, n'} &= \widehat{\bmcK}\bLambda^n \widehat \bGamma_{\bF}(h) \l(\widehat{\bmcK}\bLambda^n\r)' + \widehat{\bmcK}\bLambda^n \widehat \bGamma_{\bF\bF^w}(h) \l(\widehat{\bmcK}\bLambda^{w, n}\r)'\widehat{\bmcK}\bLambda^n \widehat \bGamma_{\bF \bxi}^n(h)\widehat{\bmcK}' \\
    &+ \widehat{\bmcK}\bLambda^{w, n} \widehat \bGamma_{F^w F}(h) \l(\widehat{\bmcK}\bLambda^n\r)' + \widehat{\bmcK}\bLambda^{w, n} \widehat \bGamma_{\bF^w}(h)\l(\widehat{\bmcK}\bLambda^{w, n}\r)' + \widehat{\bmcK}\bLambda^{w, n} \widehat \bGamma_{\bF^w \bxi}^n(h) \widehat{\bmcK}' \\
    &+ \widehat{\bmcK}\widehat \bGamma_{\bxi \bF}^n(h) \l(\widehat{\bmcK}\bLambda^n\r)' + \widehat{\bmcK}\widehat \bGamma_{\bxi \bF^w}^n(h) \l(\widehat{\bmcK}\bLambda^{w, n}\r)' + \widehat{\bmcK}\widehat \bGamma_{\bxi}^n(h) \widehat{\bmcK}'\\
    &= \widehat{\bmcK}\bLambda^n \widehat \bGamma_{\bF}(h) \l(\widehat{\bmcK}\bLambda^n\r)' +  \mathcal O_P(\max(T^{-1}, n^{-1/2}T^{-1/2})) \mathcal O_P(n^{-1/2}) +  \mathcal O_P(\max(n^{-1}, T^{-1})) \\
    &= \widehat{\bmcK}\bLambda^n \widehat \bGamma_{\bF}(h) \l(\widehat{\bmcK}\bLambda^n\r)' +  \mathcal O_P(\max(n^{-1}, T^{-1})) \\
    &= \bP_{\bLambda}\bGamma_{\bF}(h) \bP_{\bLambda}' +  \mathcal O_P(\max(n^{-1/2}, T^{-1/2})).
\end{align*}
where we used Lemma \ref{lem: killing K}.
Part $(vii)$ is trivial. 
\end{proof}

\subsection{Additional Lemmas for Asymptotic Normality}

\begin{lemma}\label{lem: killing K for asynorm}
    Under Assumptions E\ref{A: divergence rates eval}-E\ref{A: more restrictive sample covarainces}, it holds that 
    \begin{itemize}
        \item[(i)] $\norm{\widehat{\bmcK} \widehat \bGamma_{\xi}(h)\bss_i} = \mathcal O_P(\max(T^{-1/2}n^{-1/2}, T^{-1}, n^{-1}))$, where $\bss_i$ is the vector that selects the $i$-th column from $\bGamma_{\xi}(h)$; 
        \item[(ii)] $\norm{\frac{1}{n}\widehat \bGamma_{\bF\bxi}^n \bxi_{t-h}^n} = \mathcal O_P(\max(n^{-1/2} T^{-1/2}, T^{-1}))$; 
        \item[(iii)] $\norm{\widehat{\bmcK} \widehat \bGamma_{\bxi}^n \frac{1}{\sqrt{n}}} = \mathcal O_P(n^{-1/2}T^{-1/2})$ and $\norm{\widehat{\bmcK} \widehat \bGamma_{\bse}^n \frac{1}{\sqrt{n}}} = \mathcal O_P(n^{-1/2}T^{-1/2})$.
    \end{itemize}
\end{lemma}
\begin{proof}
    For part $(i)$, without loss of generality we consider the case $h = 0$. 
    
    We begin by preparing some terms: First note that by Assumption E\ref{A: simply the sample covariances}$(ii)$ for all $1\leq j \leq n$,
    \begin{align}
        \E\l[ \norm{\frac{1}{\sqrt{n}} \l(\widehat \bGamma_{\bxi}^n - \frac{1}{T} \sum_{t = 1}^T \bGamma_{\bxi_t}^n \r) \bss_j}^2\r] &=  \E\l[\norm{\frac{1}{\sqrt{n}T} \sum_{t = 1}^T \l\{ \bxi_t^n \xi_{jt} - \E\l[\bxi_t^n \xi_{jt}\r] \r\}}^2\r]  \nonumber \\
        &=  \frac{1}{n}\sum_{i = 1}^n \E\l[\l(\frac{1}{T} \sum_{t = 1}^T \l\{ \xi_{it}\xi_{jt} - \E\l[\xi_{it} \xi_{jt}\r]\r\}\r)^2\r] \nonumber \\[0.8em]
        & \leq T^{-1} \mathcal B_\xi = \mathcal O(T^{-1}). \label{eq: rate 1/sqrtn hat Gamma_xi - avg Gamma_xi}
    \end{align}
     While by Assumptions E\ref{A: more restrictive sample covarainces}$(iii)$ and E\ref{A: divergence rates eval}$(iv)$, using the same procedure as in \cite{barigozzi2022principal}, we have
     \begin{align}
        \norm{\frac{\bLambda^{n'}}{n}\left(\widehat \bGamma_{\bxi}^n - \frac{1}{T} \sum_{t = 1}^T \bGamma_{\bxi_t}^n \right)\bss_j} &= \norm{\frac{1}{nT}\sum_{t = 1}^T \sum_{i = 1}^n \bLambda_i' \left\{\xi_{it} \xi_{jt} - \E\left[\xi_{it} \xi_{jt}\right]\right\} 
        } \nonumber \\ 
        &= \left\{ 
        \sum_{l = 1}^r \left(\frac{1}{nT} \sum_{t = 1}^T \sum_{i = 1}^n \left\{\xi_{it} \xi_{jt} - \E\left[\xi_{it} \xi_{jt}\right]\right\} \lambda_{li}\right)^2\right\}^{1/2} \nonumber \\
        &\leq \sqrt{r} \mathcal B_\Lambda \abs{\frac{1}{nT} \sum_{t = 1}^T \sum_{i = 1}^n \left\{\xi_{it} \xi_{jt} - \E\left[\xi_{it} \xi_{jt}\right]\right\}} \nonumber \\
        &= \mathcal O_{ms}(n^{-1/2}T^{-1/2}). \label{eq: bound Lambda hat Gamma xi - avg Gamma xi}
    \end{align}
    Furthermore
    \begin{align}
        \norm{\frac{\bLambda^{n'}}{n} \frac{1}{T} \sum_{t = 1}^T  \bGamma_{\bxi_t}^n \bss_j} &= \norm{\frac{1}{nT}\sum_{t = 1}^T \sum_{i = 1}^n \bLambda_i' \E\left[\xi_{it} \xi_{jt}\right]} = \left\{ 
        \sum_{l = 1}^r \left(\frac{1}{nT} \sum_{t = 1}^T \sum_{i = 1}^n \E\left[\xi_{it} \xi_{jt}\right] \lambda_{li}\right)^2\right\}^{1/2} \nonumber \\
        &\leq \sqrt{r} \mathcal B_\Lambda \abs{\frac{1}{nT} \sum_{t = 1}^T \sum_{i = 1}^n \E\left[\xi_{it} \xi_{jt}\right]} \nonumber \\
        &  \leq \frac{\sqrt{r} \mathcal B_\Lambda}{nT}\sum_{t = 1}^T \sum_{i = 1}^n \abs{\E\left[\xi_{it} \xi_{jt}\right]} \nonumber\\
        & \leq \max_{t = 1, ..., T}\frac{\sqrt{r} \mathcal B_\Lambda}{n}\sum_{i = 1}^n  \abs{\E\left[\xi_{it} \xi_{jt}\right]} \nonumber\\
        & \leq \frac{\sqrt{r} \mathcal B_\Lambda \mathcal B_\xi}{n} \qquad \mbox{by Assumption E\ref{A: more restrictive sample covarainces}(i)} \label{eq: bound Lambda avg Gamma_xi sj}
    \end{align}
    
    Then by Lemma \ref{lem: hat evals-vecs}$(iv)$, Assumption E\ref{A: divergence rates eval}(vi) and \eqref{eq: rate 1/sqrtn hat Gamma_xi - avg Gamma_xi}
    \begin{align*}
        &\widehat{\bmcK} \widehat \bGamma_{\bxi}^n \bss_i = 
        \l(\widehat{\bmcK} - \bmcK\r) \l(\widehat \bGamma_{\bxi}^n - \frac{1}{T} \sum_{t = 1}^T\bGamma_{\bxi_t}^n \r) \bss_i +  \left(\widehat{\bmcK} - \bmcK\right) \frac{1}{T} \sum_{t = 1}^T \bGamma_{\bxi_t}^n \bss_i + \bmcK \widehat \bGamma_{\bxi}^n \bss_i \\
        &= \mathcal O_P(T^{-1/2}n^{-1/2}) \mathcal O_{ms}(n^{1/2}T^{-1/2}) + \mathcal O_P(T^{-1/2}n^{-1/2}) + \mathcal O_{ms}(\max(n^{-1/2}T^{-1/2}, n^{-1})),
    \end{align*}
    while the last term is seen as follows:
    
    We use Lemma \ref{lem: mathcal K}$(iii)$ for the representation of $\bmcK$ in terms of $\bLambda^n$: 
    \begin{align*}
        \bmcK \widehat \bGamma_{\bxi}^n \bss_i &= \mathcal O(n^{-3/2}) \left(\widehat \bGamma_{\bxi}^n - \frac{1}{T} \sum_{t = 1}^T\bGamma_{\bxi_t}^n \right)\bss_i + \mathcal O(1) \frac{\bLambda^{n'}}{n}\left(\widehat \bGamma_{\bxi}^n(h) - \frac{1}{T} \sum_{t = 1}^T \bGamma_{\bxi_t}^n \right)\bss_i \\[0.8em]
        &\qquad \qquad \qquad + \mathcal O(n^{-3/2}) \frac{1}{T} \sum_{t = 1}^T \bGamma_{\bxi_t}^n \bss_i + \mathcal O(1)\frac{\bLambda^{n'}}{n} \frac{1}{T} \sum_{t = 1}^T \bGamma_{\bxi_t}^n \bss_i \\
        \mbox{while} \qquad 
        (1) &= \mathcal O(n^{-3/2}) \mathcal O_{ms}(n^{1/2}T^{-1/2}) = \mathcal O_{ms}(n^{-1}T^{-1/2})\qquad \mbox{by Lemma \ref{lem: mathcal K}(iii) and \eqref{eq: rate 1/sqrtn hat Gamma_xi - avg Gamma_xi}} \\
        (2) &= \mathcal O_{ms}(n^{-1/2}T^{-1/2}) \qquad \mbox{by \eqref{eq: bound Lambda hat Gamma xi - avg Gamma xi}} \\
        (3) &= \mathcal O(n^{-3/2}) \mathcal O(1)  \qquad \mbox{by Assumption E\ref{A: divergence rates eval}$(vi)$} \\
        (4) &= \mathcal O_{ms}(\max(n^{-1})) \qquad \mbox{by \eqref{eq: bound Lambda avg Gamma_xi sj}}.
    \end{align*}
    To prove part $(ii)$, we use Assumption E\ref{A: more restrictive sample covarainces}. Again without loss of generality we suppose $h = 0$, then independence implies \citep[see also][]{barigozzi2022principal}, that
    \begin{align*}
        \E\l[\norm{\frac{\widehat \bGamma_{\bF\bxi}^n \bxi_t^n}{n}}^2\r] &= \E\norm{\frac{1}{nT} \sum_{i = 1}^n \sum_{s = 1}^T \bF_s \xi_{is} \xi_{it}}^2 \\
        &= \frac{1}{n^2 T^2} \sum_{l = 1}^r \sum_{i_1 = 1}^n \sum_{i_2 = 1}^n \sum_{s_1 = 1}^T \sum_{s_2 = 1}^T \E\l[\xi_{i_1, s_1} \xi_{i_2, s_2} \xi_{i_1, t} \xi_{i_2, t} F_{l s_1} F_{l s_2} \r] \\
        &\leq \frac{r}{n^2 T^2} \max_{1\leq l \leq r} \sum_{i_1 = 1}^n \sum_{i_2 = 1}^n \sum_{s_1 = 1}^T \sum_{s_2 = 1}^T \E\l[\xi_{i_1, s_1} \xi_{i_2, s_2} \xi_{i_1, t} \xi_{i_2, t} \r]\E\l[F_{l s_1} F_{l s_2} \r] \\
        &\leq r \l\{ \max_{1\leq l \leq r} \max_{1\leq s_1, s_2 \leq T} \E\l[F_{l s_1} F_{l s_2} \r] \r\} \\
        & \times \bigg\{ \frac{1}{n^2 T^2}
        \sum_{i_1 = 1}^n \sum_{i_2 = 1}^n \sum_{s_1 = 1}^T \sum_{s_2 = 1}^T \l(\E\l[\xi_{i_1, s_1} \xi_{i_2, s_2} \xi_{i_1, t} \xi_{i_2, t}  \r]
        - \E\l[\xi_{i_1, s_1} \xi_{i_2, s_2} \r] \E\l[\xi_{i_1, t} \xi_{i_2, t} \r]\r)\\
        & + 
        \frac{1}{n^2 T^2}
        \sum_{i_1 = 1}^n \sum_{i_2 = 1}^n \sum_{s_1 = 1}^T \sum_{s_2 = 1}^T 
        \E\l[\xi_{i_1, s_1} \xi_{i_2, s_2} \r] \E\l[\xi_{i_1, t} \xi_{i_2, t} \r]
        \bigg\} \\
        &\leq r \mathcal B_F \Bigg\{ \E\l[\l(\frac{1}{nT} \sum_{t = 1}^T \sum_{k = 1}^n \left\{\xi_{kt} \xi_{i, t-j} - \E\left[\xi_{kt} \xi_{i, t}\right]\right\}\r)^2\r] \\
        &+ \max_{1\leq i_1, i_2 \leq n} \abs{\E\l[\xi_{i_1, t} \xi_{i_2, t}\r]}   \frac{1}{n^2 T^2} 
        \sum_{i_1 = 1}^n \sum_{i_2 = 1}^n \sum_{s_1 = 1}^T \sum_{s_2 = 1}^T  \abs{\E\l[\xi_{i_1, s_1} \xi_{i_2, s_2}\r]} \Bigg\} \\
        &\leq \frac{r \mathcal B_F \mathcal B_\xi}{nT} + \frac{r \mathcal B_F \mathcal B_\xi^2}{nT}.
    \end{align*}
    Alternatively, 
    \begin{align*}
        \E\l[\norm{\frac{\widehat \bGamma_{\bF\bxi}^n \bxi_t^n}{n}}^2\r] &\leq \E\norm{\frac{1}{nT} \sum_{i = 1}^n \sum_{s = 1}^T \bF_s\l[ \xi_{is} \xi_{it} - \E\l[ \xi_{is} \xi_{it}\r]\r]}^2 + \E\norm{\frac{1}{T} \sum_{s = 1}^T \bF_s \frac{1}{nT} \sum_{i = 1}^n\E\l[\xi_{is} \xi_{it}\r] }^2 \\
        & \leq \frac{\mathcal B_\xi}{nT} + \mathcal B_\xi^2 \frac{1}{T^2} \E\norm{\sum_{s = 1}^T \bF_s \rho^{\abs{s-t}}}^2 .
    \end{align*}
    For part $(iii)$ we decompose first
    \begin{align*}
    \widehat \bGamma_{\bse}^n &= \frac{1}{T}\sum_{t = 1}^T \l(\bLambda^{w, n} \bF_t^w + \bxi_t^n\r)\l(\bLambda^{w, n} \bF_t^w + \bxi_t^n\r)' = \bLambda^{w, n} \widehat \bGamma_{\bF^w} \bLambda^{w, n'} + \bLambda^{w, n} \widehat \bGamma_{\bF^w \bxi}^n + \widehat \bGamma_{\bxi \bF^w}^n \bLambda^{w, n'} + \widehat \bGamma_{\bxi}^n \\
    \widehat{\bmcK} \frac{\widehat \bGamma_{\bse}^n}{\sqrt{n}} &= \widehat{\bmcK} \bLambda^{w, n} \widehat \bGamma_{\bF^w}\frac{\bLambda^{w, n'}}{\sqrt{n}} + \widehat{\bmcK}\bLambda^{w, n}\frac{ \widehat \bGamma_{\bF^w \bxi}^n }{\sqrt{n}} +\widehat{\bmcK} \widehat \bGamma_{\bxi \bF^w}^n \frac{\bLambda^{w, n'} }{\sqrt{n}} + \widehat{\bmcK}\widehat \bGamma_{\bxi}^n \frac{1}{\sqrt{n}} \\
    &= \mathcal O_P(\max(T^{-1/2}n^{-1/2}, n^{-1})) \mathcal O(n^{-1/2}) + \mathcal O_P(\max(T^{-1/2}n^{-1/2}, n^{-1})) \mathcal O_{ms}(T^{-1/2}) \\
    & + \mathcal O_P(\max(T^{-1/2}n^{-1/2}, T^{-1}))\mathcal O(n^{-1/2}) + \widehat{\bmcK}\widehat \bGamma_{\bxi}^n \frac{1}{\sqrt{n}} \\
    &= \mathcal O_P(\max(T^{-1/2}n^{-1}, n^{-3/2}, T^{-1}n^{-1/2}, n^{-1}T^{-1/2})) +  \widehat{\bmcK}\widehat \bGamma_{\bxi}^n \frac{1}{\sqrt{n}}. 
    \end{align*}

    Furthermore, we note that
    \begin{align*}
        \widehat{\bmcK} \widehat \bGamma_{\bxi}^n \frac{1}{\sqrt{n}} &= \left(\widehat{\bmcK} - \bmcK\right)\left(\widehat \bGamma_{\bxi}^n - \bGamma_{\bxi}^n\right) \frac{1}{\sqrt{n}} +  \left(\widehat{\bmcK} - \bmcK\right) \bGamma_{\bxi}^n \frac{1}{\sqrt{n}} + \bmcK \left(\widehat \bGamma_{\bxi}^n - \bGamma_{\bxi}^n \right)\frac{1}{\sqrt{n}} + \bmcK \bGamma_{\bxi}^n \frac{1}{\sqrt{n}} \\
        &= \mathcal O_P(T^{-1/2}n^{-1/2}) \mathcal O_P(T^{-1/2}n^{1/2}) + \mathcal O_P(T^{-1/2}n^{-1/2}) \mathcal O(n^{-1/2}) + \mathcal O_P(n^{-1/2}T^{-1/2}) + \mathcal O(n^{-1}), 
    \end{align*}
    while we used that
    \begin{align*}
        \E \norm{\bmcK \l( \widehat \bGamma_{\bxi}^n - \bGamma_{\bxi}^n  \r) \frac{1}{\sqrt{n}}}^2 &= \E \norm{\sum_{i = 1}^n  \mathcal O(1) \frac{\bLambda_i'}{n} \frac{1}{T} \sum_{t = 1}^T \l(\xi_{it}\bxi_t^{n'} - \E\l[\xi_{it} \bxi_t^{n'}\r]\r)}^2 \\
        &\leq \mathcal B_\Lambda \frac{1}{(nT)^2}\frac{1}{n} \sum_{j = 1}^n \sum_{i= 1}^n \sum_{t = 1}^T \E \l(\xi_{it} \xi_{jt} - \E \l[\xi_{it} \xi_{jt}\r]\r)^2 \\
        &\leq \mathcal B_\Lambda \frac{1}{n^3 T^2} \sum_{i =1}^n \sum_{j = 1}^n \sum_{t = 1}^T \mathcal B_\xi = \frac{1}{nT}.
    \end{align*}
\end{proof}
\begin{lemma}\label{lem: x xi and x x}
    Under E\ref{A: divergence rates eval}-E\ref{A: more restrictive sample covarainces}, we have 
    \begin{itemize}
        \item[(i)] $\norm{\frac{1}{\sqrt{T}}\left(\widehat{\bsx} - \bsx \widehat{\bmcH}\right)'\bxi^i} = \mathcal O_P(\max(T^{1/2}n^{-1}, n^{-1/2}, T^{-1/2}))$\\ and $\norm{\frac{1}{\sqrt{T}}\left(\widehat{\bW} - \bF \widehat{\bH}\right)'\bse^i} = \mathcal O_P(\max(T^{1/2}n^{-1}, n^{-1/2}, T^{-1/2}))$; 
        \item[(ii)] $\norm{\frac{1}{\sqrt{T}} \widehat{\bsx}' \left(\bsx \widehat{\bmcH }- \widehat{\bsx}\right)} = \mathcal O_P(\max(T^{-1/2}, T^{1/2}n^{-1}))$ \\
        and $\norm{\frac{1}{\sqrt{T}} \widehat{\bW}' \left(\bF \widehat{\bH}- \widehat{\bW}\right)} = \mathcal O_P(\max(T^{-1/2}, T^{1/2}n^{-1}))$; 
        \item[(iii)] $\norm{\frac{1}{\sqrt{n}} \left(\widehat \bLambda^n -\bLambda^n \widehat{\bH}_{\bLambda}\right)'\bxi_{t-j}^n} = \mathcal O_P(T^{-1/2})$ \\
        and $\norm{\frac{1}{\sqrt{n}} \left(\widehat \bLambda^n -\bLambda^n \widehat{\bH}_{\bLambda}\right)'\bse_t^n} = \mathcal O_P(T^{-1/2})$.
    \end{itemize}
\end{lemma}
\begin{proof}
We show part $(i)$. Plugging in the canonical decomposition together with the expansion of $\widehat{\bmcH}$ from \eqref{eq: expansion of mathcal H}, we use the representation for $\widehat \bsx_t - \widehat{\bmcH} \bsx_t$ as in \eqref{eq: rep hat xt - hat mathcal H xt} and obtain
    \begin{align*}
    &\frac{1}{T}\sum_{t = 1}^T \left(\widehat \bsx_t - \widehat{\bmcH} \bsx_t\right) \xi_{it} \\
    &= \frac{1}{T}\sum_{t = 1}^T  \bigg[\left(\bI_{p+1}\otimes \widehat{\bmcK} \bLambda^{w, n} \right)\left(\bmcF_t^w - \widehat \bGamma_{\bmcF^w \bsx}\widehat \bGamma_{\bsx}^{-1} \bsx_t \right) - \left(\bI_{p+1}\otimes \widehat{\bmcK}  \right) \widehat \bGamma_{\bXi\bsx}^n \widehat \bGamma_{\bsx}^{-1} \bsx_t  + \left(\bI_{p+1}\otimes \widehat{\bmcK} \right)  \bXi_t^n 
            \bigg] \xi_{it} \\
    &=\left(\bI_{p+1}\otimes \widehat{\bmcK} \bLambda^{w, n} \right) \widehat \bGamma_{\bmcF^w\bxi} \bss_i - \left(\bI_{p+1}\otimes \widehat{\bmcK} \bLambda^{w, n} \right)  \widehat \bGamma_{\bmcF^w \bsx}\widehat \bGamma_{\bsx}^{-1} \widehat \bGamma_{\bsx\bxi}^n \bss_i -  \left(\bI_{p+1}\otimes \widehat{\bmcK}  \right) \widehat \bGamma_{\bXi\bsx}^n \widehat \bGamma_{\bsx}^{-1} \widehat \bGamma_{\bsx\bxi}^n \bss_i \\
    & + \left(\bI_{p+1}\otimes \widehat{\bmcK} \right) \widehat \bGamma_{\bXi\bxi}^n \bss_i \\
    &= \mathcal O_P(\max(T^{-1/2}n^{-1/2}, n^{-1}, T^{-1})), 
    \end{align*}
    since for the terms in order we have 
    \begin{align*}
    (1) &= \mathcal O_P(n^{-1/2}T^{-1/2}) \quad \mbox{by Lemma \ref{lem: killing K}$(i)$ and Assumption E\ref{A: simply the sample covariances}$(iv)$};\\
    (2) &= \mathcal O_P(n^{-1/2}T^{-1}) \quad \mbox{by Lemma \ref{lem: killing K}$(i)$ and Assumption E\ref{A: simply the sample covariances}$(iv)$}; \\
    (3) &= \mathcal O_P(n^{-1/2}T^{-1}) \mathcal O_P(T^{-1/2}) \quad \mbox{by Lemma \ref{lem: killing K}$(ii)$ and Assumption E\ref{A: simply the sample covariances}$(iv)$};\\
    (4) &= \mathcal O_P(\max(T^{-1/2}n^{-1/2}, T^{-1}, n^{-1})) \quad \mbox{by Lemma \ref{lem: killing K for asynorm}(i)}.
    \end{align*}
For part $(ii)$, consider the expansion 
    \begin{align}
        &\frac{1}{T}\sum_{t = 1}^T \widehat \bsx_t \left(\widehat{\bmcH} \bsx_t - \widehat \bsx_t\right)' \nonumber \\
        &=\frac{1}{T} \sum_{t = 1}^T \left[ \left(\bI_{p+1}\otimes \widehat{\bmcK}\bLambda^n \right) \bsx_t +  \left(\bI_{p+1}\otimes \widehat{\bmcK} \bLambda^{w, n} \right)\bmcF_t^w +  \left(\bI_{p+1}\otimes \widehat{\bmcK} \right)  \bXi_t^n \right]\nonumber\\
        &\times \bigg[
        \left(\bI_{p+1}\otimes \widehat{\bmcK}\bLambda^n \right) \bsx_t + \left(\bI_{p+1}\otimes \widehat{\bmcK} \bLambda^{w, n} \right) \widehat \bGamma_{\bmcF^w \bsx}\widehat \bGamma_{\bsx}^{-1} \bsx_t + \left(\bI_{p+1}\otimes \widehat{\bmcK}  \right) \widehat \bGamma_{\bXi\bsx}^n \widehat \bGamma_{\bsx}^{-1} \bsx_t \nonumber\\
        &  
        \hspace{4cm}- \left(\bI_{p+1}\otimes \widehat{\bmcK}\bLambda^n \right) \bsx_t - \left(\bI_{p+1}\otimes \widehat{\bmcK} \bLambda^{w, n} \right)\bmcF_t^w -  \left(\bI_{p+1}\otimes \widehat{\bmcK} \right)  \bXi_t^n 
        \bigg]'  \nonumber\\
        &=\frac{1}{T}\sum_{t = 1}^T \left[ \left(\bI_{p+1}\otimes \widehat{\bmcK}\bLambda^n \right) \bsx_t +  \left(\bI_{p+1}\otimes \widehat{\bmcK} \bLambda^{w, n} \right)\bmcF_t^w +  \left(\bI_{p+1}\otimes \widehat{\bmcK} \right)  \bXi_t^n \right] \nonumber\\
        &\times \bigg[\left(\bI_{p+1}\otimes \widehat{\bmcK} \bLambda^{w, n} \right)\left( \widehat \bGamma_{\bmcF^w \bsx}\widehat \bGamma_{\bsx}^{-1} \bsx_t - \bmcF_t^w \right) + \left(\bI_{p+1}\otimes \widehat{\bmcK}  \right) \widehat \bGamma_{\bXi\bsx}^n \widehat \bGamma_{\bsx}^{-1} \bsx_t  - \left(\bI_{p+1}\otimes \widehat{\bmcK} \right)  \bXi_t^n 
            \bigg]' \label{eq: expansion of mtcalHxt - xt} \\[0.8em]
        \end{align}
        For the terms of this product in order, we obtain using the Lemmas \ref{lem: Var est} and \ref{lem: killing K}, that
        \begin{align}
        (1, 1) &= \left(\bI_{p+1}\otimes \widehat{\bmcK}\bLambda^n \right)
        \left(\widehat \bGamma_{\bsx} \widehat \bGamma_{\bsx}^{-1} \widehat \bGamma_{\bsx\bmcF^w} -  \widehat \bGamma_{\bsx\bmcF^w} \right) \left(\bI_{p+1}\otimes \widehat{\bmcK} \bLambda^{w, n}\right)' = 0 \nonumber\\
        (1, 2) & 
        = \left(\bI_{p+1}\otimes \widehat{\bmcK}\bLambda^n \right) \widehat \bGamma_{\bsx\bXi}^n \left(\bI_{p+1}\otimes \widehat{\bmcK} \right) ' \nonumber\\
        (1, 3) &= - (1, 2)\nonumber \\
        (2, 1) &= \left(\bI_{p+1}\otimes \widehat{\bmcK} \bLambda^{w, n} \right) \left(\widehat \bGamma_{\bmcF^w \bsx} \widehat \bGamma_{\bsx}^{-1} \widehat \bGamma_{\bsx\bmcF^w} - \widehat \bGamma_{\bmcF^w}\right)\left(\bI_{p+1}\otimes \widehat{\bmcK} \bLambda^{w, n} \right)' =\mathcal O_P(n^{-1})\nonumber \\
        (2, 2) &= \left(\bI_{p+1}\otimes \widehat{\bmcK} \bLambda^{w, n} \right) \widehat \bGamma_{\bmcF^w \bsx}\widehat \bGamma_{\bsx}^{-1} \widehat \bGamma_{\bsx\bXi}^n \left(\bI_{p+1}\otimes \widehat{\bmcK} \right)  = \mathcal O_P(n^{-1/2}) \mathcal O_P(\max(T^{-1}, n^{-1/2}T^{-1/2}))\nonumber \\
        (2, 3) &= \left(\bI_{p+1}\otimes \widehat{\bmcK} \bLambda^{w, n} \right) \widehat \bGamma_{\bmcF^w \bXi}^n \left(\bI_{p+1}\otimes \widehat{\bmcK}\right)' =  \mathcal O_P(n^{-1/2}) \mathcal O_P(\max(T^{-1}, n^{-1/2}T^{-1/2}))  \nonumber\\
        (3, 1) &=\left(\bI_{p+1}\otimes \widehat{\bmcK} \right)\left( \widehat \bGamma_{\bXi\bsx}^n \widehat \bGamma_{\bsx}^{-1} \widehat \bGamma_{\bsx\bmcF^w} -  \widehat \bGamma_{\bXi \bmcF^w}^n \right)\left(\bI_{p+1}\otimes \widehat{\bmcK} \bLambda^{w, n}\right)' = \mathcal O_P(n^{-1/2}) \mathcal O_P(\max(T^{-1}, n^{-1/2}T^{-1/2}))\nonumber\\
        (3,2) &= \left(\bI_{p+1}\otimes \widehat{\bmcK} \right) \widehat \bGamma_{\bXi\bsx}^n\widehat \bGamma_{\bsx}^{-1} \widehat \bGamma_{\bsx\bXi}^n \left(\bI_{p+1}\otimes \widehat{\bmcK} \right)' \mathcal =  O_p(\max(T^{-1}, n^{-1/2}T^{-1/2})^2)\nonumber\\
        (3, 3) &=  \left(\bI_{p+1}\otimes \widehat{\bmcK} \right) \widehat \bGamma_{\bXi}^n  \left(\bI_{p+1}\otimes \widehat{\bmcK} \right)' = \mathcal O_P(\max(T^{-1}, n^{-1})), \nonumber
    \end{align}
    using Lemma \ref{lem: killing K} which implies that 
    \begin{align*}
        \frac{1}{T}\sum_{t = 1}^T \widehat \bsx_t \left(\widehat{\bmcH} \bsx_t - \widehat \bsx_t\right)' = \mathcal O_P(\max(T^{-1}, n^{-1})). 
    \end{align*}

    For part $(iii)$, we have
    \begin{align}
        \widehat \bLambda_j' &= \left(\frac{1}{T}\widehat{\bW}' \widehat{\bW}\right)^{-1} \frac{1}{T}\widehat{\bW}' \bsy^j = \bI_r \frac{1}{T}\sum_{t = 1}^T \widehat{\bW}_t^{y, n}y_{jt} = \frac{1}{T}\sum_{t = 1}^T \left(\widehat{\bmcK}\bLambda^n \bF_t + \widehat{\bmcK} \bse_t^n \right)(\bF_t'\bLambda_j' + e_{jt})  \nonumber \\
        &= \widehat{\bmcK}\bLambda^n \widehat \bGamma_{\bF} \bLambda_j'  + \widehat{\bmcK}\bLambda^n \widehat \bGamma_{\bF\bse}^n \bss_j + \widehat{\bmcK}\widehat \bGamma_{\bse \bF}^n \bLambda_j' + \widehat{\bmcK}\widehat \bGamma_{\bse}^n \bss_j \label{eq: expansion hat Lambda_i}
    \end{align}
    Plugging in the expansions \eqref{eq: expansion hat H_Lambda'} and \eqref{eq: expansion hat Lambda_i}:
    \begin{align*}
        &\frac{1}{n}\sum_{i = 1}^n \left(\widehat \bLambda_i' - \widehat{\bH}_{\bLambda}' \bLambda_i'\right)\xi_{i, t-j} \\
        &= \frac{1}{n}\sum_{i = 1}^n \bigg\{
         \widehat{\bmcK}\bLambda^n \widehat \bGamma_{\bF} \bLambda_i'  + \widehat{\bmcK}\bLambda^n \widehat \bGamma_{\bF\bse}^n \bss_i + \widehat{\bmcK}\widehat \bGamma_{\bse \bF}^n \bLambda_i' + \widehat{\bmcK}\widehat \bGamma_{\bse}^n \bss_i \\
         &\quad -\widehat{\bmcK}\bLambda^n \widehat \bGamma_{\bF}\bLambda_i'  - \widehat{\bmcK}\bLambda^n \widehat \bGamma_{\bF\bse}^n \frac{1}{n}\bLambda^n \left(\bGamma_{\bLambda}^n\right)^{-1} \bLambda_i'  - \widehat{\bmcK}\widehat \bGamma_{\bse \bF}^n \bLambda_i' - \widehat{\bmcK}\widehat \bGamma_{\bse}^n \frac{1}{n}\bLambda^n \left(\bGamma_{\bLambda}^n\right)^{-1} \bLambda_i'\bigg\}
         \xi_{i, t-j} \\
         &= 
        \widehat{\bmcK}\bLambda^n \widehat \bGamma_{\bF\bse}^n \frac{1}{n}\bxi_{t-j}^n  + \widehat{\bmcK}\widehat \bGamma_{\bse}^n \frac{1}{n}\bxi_{t-j}^n  - \widehat{\bmcK}\bLambda^n \widehat \bGamma_{\bF\bse}^n \frac{1}{n}\bLambda^n \left(\bGamma_{\bLambda}^n\right)^{-1} \frac{1}{n}\bLambda^{n'}\bxi_{t-j}^n  - \widehat{\bmcK}\widehat \bGamma_{\bse}^n \frac{1}{n}\bLambda^n \left(\bGamma_{\bLambda}^n\right)^{-1} \frac{1}{n}\bLambda^{n'}\bxi_{t-j}^n \\
        &= \mathcal O_P(n^{-1/2}T^{-1/2}),
    \end{align*}
    while 
    \begin{align*}
        (1) &= \widehat{\bmcK}\bLambda^n \frac{1}{nT}\sum_{s = 1}^T \bF_s \left(\bLambda^{w, n}\bF_s^w  + \bxi_s^n\right)'\bxi_{t-j}^n = \widehat{\bmcK}\bLambda^n\frac{1}{nT} \sum_{s = 1}^T \left(\bF_s \bF_s^w \bLambda^{w, n'} \bxi_{t-j}^n + \bF_s \bxi_s^{n'} \bxi_{t-j}^n\right) \\
            &= \widehat{\bmcK}\bLambda^n \left(\widehat \bGamma_{\bF\bF^w} \frac{1}{n} \bLambda^{w, n'} \bxi_{t-j}^n + \frac{1}{n} \widehat \bGamma_{\bF\bxi}\bxi_{t-j}^n\right)\\
            & = \mathcal O_P(T^{-1/2}) \mathcal O_{ms}(n^{-1}) + \mathcal O_{ms}(n^{-1/2}T^{-1/2}) = \mathcal O_{P}(n^{-1/2}T^{-1/2}) \quad \mbox{by Lemmas \ref{lem: Var est}(i), (iv) and \ref{lem: killing K for asynorm}(i)}\\ 
       (2) &=  \l(\frac{\widehat{\bmcK} \widehat \bGamma_{\bse}^n}{\sqrt{n}} \r) \l(\frac{\bxi_{t-j}^n}{\sqrt{n}}\r) = \mathcal O_P(n^{-1/2} T^{-1/2})  \mathcal O_{ms}(1) \quad \mbox{by Lemmas \ref{lem: killing K for asynorm}(iii) and \ref{lem: Var est}(iv)}
       \\
        (3) &= \l(\frac{\widehat{\bmcK}\bLambda^n}{\sqrt{n}}\r) \left(\frac{\widehat \bGamma_{\bF\bse}^n }{\sqrt{n}}\right) \left(\frac{\bLambda^n }{\sqrt{n}}\right) \left(\bGamma_{\bLambda}^n\right)^{-1} \left(\frac{1}{\sqrt{n}}\bLambda^{n'}\bxi_{t-j}^n\right) \\
        &\hspace{3cm} = \mathcal O_P(n^{-1/2}) \mathcal O_{ms}\left(T^{-1/2}\right) \mathcal O(1) \mathcal O_{ms}(1) \quad \mbox{by Lemmas \ref{lem: mathcal K} and \ref{lem: Var est}(ii),(iv)}. \\
        (4) &= \frac{1}{\sqrt{n}}\left(\widehat{\bmcK}\widehat \bGamma_{\bse}^n \frac{\bLambda^n}{n}\right) \left(\bGamma_{\bLambda}^n\right)^{-1} \left(\frac{1}{\sqrt{n}}\bLambda^{n'}\bxi_{t-j}^n\right) \\
        & \hspace{3cm} = \mathcal O(n^{-1/2})\mathcal O_P(\max(n^{-1}, T^{-1})) \mathcal O(1) \mathcal O_{ms}(1) \quad \mbox{by Lemmas \ref{lem: killing K}(iv) and \ref{lem: Var est}(iv)}.
    \end{align*} 
    The remaining statements are shown analogously.
\end{proof}

\newpage
\section{Additional Simulation Results}

\begin{table}[!htbp] \centering 
\begin{tabular}{@{\extracolsep{5pt}} c|cccc} 
\\[-1.8ex]\hline 
\hline \\[-1.8ex]
\multicolumn{5}{c}{\textbf{MSE Results $\widehat \chi_{1t}$ for DGP1 $T > n$}} \\[0.5em]
\hline
 & (30,60) & (60,120) & (120,240) & (240,480) \\ 
\hline \\[-1.8ex] 
\multicolumn{5}{c}{$\tau = 0, \delta = 0$} \\
\hline
spca2 & $0.151$ & $0.129$ & $0.121$ & $0.114$ \\ 
dpca & $0.316$ & $0.165$ & $0.096$ & $0.057$ \\ 
fdl & $0.078$ & $0.032$ & $0.014$ & $0.007$ \\ 
spca1 & $0.265$ & $0.319$ & $0.346$ & $0.353$ \\ 
spca2 & $0.151$ & $0.129$ & $0.121$ & $0.114$ \\ 
spca3 & $0.228$ & $0.171$ & $0.142$ & $0.124$ \\ 
spca5 & $0.366$ & $0.243$ & $0.187$ & $0.142$ \\ 
spca9 & $0.582$ & $0.370$ & $0.260$ & $0.182$ \\ \hline
\multicolumn{5}{c}{$\tau = 0.5, \delta = 0$} \\
\hline
spca2 & $0.252$ & $0.234$ & $0.232$ & $0.221$ \\ 
dpca & $0.340$ & $0.175$ & $0.102$ & $0.057$ \\ 
fdl & $0.105$ & $0.045$ & $0.018$ & $0.008$ \\ 
spca1 & $0.292$ & $0.325$ & $0.345$ & $0.352$ \\ 
spca2 & $0.252$ & $0.234$ & $0.232$ & $0.221$ \\ 
spca3 & $0.312$ & $0.255$ & $0.242$ & $0.227$ \\ 
spca5 & $0.454$ & $0.310$ & $0.263$ & $0.236$ \\ 
spca9 & $0.695$ & $0.437$ & $0.310$ & $0.259$ \\ 
\hline
\multicolumn{5}{c}{$\tau = 0, \delta = 0.5$} \\
\hline
spca2 & $0.221$ & $0.167$ & $0.155$ & $0.146$ \\ 
dpca & $0.303$ & $0.174$ & $0.108$ & $0.067$ \\ 
fdl & $0.106$ & $0.046$ & $0.021$ & $0.011$ \\ 
spca1 & $0.289$ & $0.324$ & $0.351$ & $0.355$ \\ 
spca2 & $0.221$ & $0.167$ & $0.155$ & $0.146$ \\ 
spca3 & $0.342$ & $0.238$ & $0.189$ & $0.165$ \\ 
spca5 & $0.530$ & $0.349$ & $0.258$ & $0.201$ \\ 
spca9 & $0.772$ & $0.537$ & $0.370$ & $0.267$ \\ 
\hline
\multicolumn{5}{c}{$\tau = 0.5, \delta = 0.5$} \\
\hline
spca2 & $0.322$ & $0.281$ & $0.268$ & $0.256$ \\ 
dpca & $0.321$ & $0.195$ & $0.108$ & $0.063$ \\ 
fdl & $0.139$ & $0.064$ & $0.025$ & $0.012$ \\ 
spca1 & $0.312$ & $0.338$ & $0.352$ & $0.356$ \\ 
spca2 & $0.322$ & $0.281$ & $0.268$ & $0.256$ \\ 
spca3 & $0.429$ & $0.319$ & $0.286$ & $0.265$ \\ 
spca5 & $0.622$ & $0.416$ & $0.322$ & $0.284$ \\ 
spca9 & $0.897$ & $0.597$ & $0.399$ & $0.323$ \\ 
\hline
\end{tabular} 
  \caption{MSE of different estimators for $\chi_{1t}$ for DGP\eqref{eq: innocent factor model} 500 replications: \texttt{spca}-\texttt{r} = static principal components with \texttt{r} factors, all inconsistent, \texttt{dpca} = dynamic principal components with $q = 1$ dynamic factor, \texttt{fdl} = finite distributed lags approach regressing on the first normalized principal component and its first lag.} 
  \label{tab: res nonconsist T > n} 
\end{table} 
\begin{table}[!htbp] \centering 
\begin{tabular}{@{\extracolsep{5pt}} c|cccc} 
\\[-1.8ex]\hline 
\hline \\[-1.8ex]
\multicolumn{5}{c}{\textbf{MSE Results $\widehat \chi_{1t}$ for DGP1 $T < n$}} \\[0.5em]
\hline
 & (60,30) & (120,60) & (240,120) & (480,240) \\ 
\hline \\[-1.8ex] 
\multicolumn{5}{c}{$\tau = 0, \delta = 0$} \\
\hline
spca2 & $0.214$ & $0.185$ & $0.158$ & $0.151$ \\ 
dpca & $0.430$ & $0.272$ & $0.163$ & $0.094$ \\ 
fdl & $0.082$ & $0.039$ & $0.019$ & $0.009$ \\ 
spca1 & $0.341$ & $0.358$ & $0.354$ & $0.359$ \\ 
spca2 & $0.214$ & $0.185$ & $0.158$ & $0.151$ \\ 
spca3 & $0.280$ & $0.216$ & $0.176$ & $0.161$ \\ 
spca5 & $0.405$ & $0.281$ & $0.211$ & $0.178$ \\ 
spca9 & $0.607$ & $0.398$ & $0.281$ & $0.212$ \\ \hline
\multicolumn{5}{c}{$\tau = 0.5, \delta = 0$} \\
\hline
spca2 & $0.318$ & $0.283$ & $0.262$ & $0.253$ \\ 
dpca & $0.499$ & $0.268$ & $0.158$ & $0.094$ \\ 
fdl & $0.099$ & $0.043$ & $0.022$ & $0.010$ \\ 
spca1 & $0.356$ & $0.353$ & $0.354$ & $0.361$ \\ 
spca2 & $0.318$ & $0.283$ & $0.262$ & $0.253$ \\ 
spca3 & $0.376$ & $0.306$ & $0.272$ & $0.257$ \\ 
spca5 & $0.481$ & $0.354$ & $0.297$ & $0.269$ \\ 
spca9 & $0.674$ & $0.458$ & $0.349$ & $0.292$ \\ 
\hline
\multicolumn{5}{c}{$\tau = 0, \delta = 0.5$} \\
\hline
spca2 & $0.391$ & $0.313$ & $0.285$ & $0.279$ \\ 
dpca & $0.481$ & $0.291$ & $0.179$ & $0.102$ \\ 
fdl & $0.142$ & $0.067$ & $0.030$ & $0.016$ \\ 
spca1 & $0.381$ & $0.375$ & $0.362$ & $0.364$ \\ 
spca2 & $0.391$ & $0.313$ & $0.285$ & $0.279$ \\ 
spca3 & $0.476$ & $0.355$ & $0.302$ & $0.278$ \\ 
spca5 & $0.639$ & $0.457$ & $0.347$ & $0.297$ \\ 
spca9 & $0.837$ & $0.620$ & $0.439$ & $0.342$ \\ 
\hline
\multicolumn{5}{c}{$\tau = 0.5, \delta = 0.5$} \\
\hline
spca2 & $0.421$ & $0.361$ & $0.341$ & $0.334$ \\ 
dpca & $0.467$ & $0.285$ & $0.180$ & $0.100$ \\ 
fdl & $0.150$ & $0.067$ & $0.035$ & $0.017$ \\ 
spca1 & $0.379$ & $0.369$ & $0.367$ & $0.366$ \\ 
spca2 & $0.421$ & $0.361$ & $0.341$ & $0.334$ \\ 
spca3 & $0.497$ & $0.395$ & $0.358$ & $0.338$ \\ 
spca5 & $0.625$ & $0.486$ & $0.400$ & $0.357$ \\ 
spca9 & $0.822$ & $0.638$ & $0.486$ & $0.400$ \\ 
\hline \\
\end{tabular}
 \caption{MSE of different estimators for $\chi_{1t}$ for DGP\eqref{eq: innocent factor model} 500 replications: \texttt{spca}-\texttt{r} = static principal components with \texttt{r} factors, all inconsistent, \texttt{dpca} = dynamic principal components with $q = 1$ dynamic factor, \texttt{fdl} = finite distributed lags approach regressing on the first normalized principal component and its first lag.} 
  \label{tab: res nonconsist n < T} 
\end{table}
\begin{table}[!htbp] \centering 
  
\begin{tabular}{@{\extracolsep{5pt}} c|c|c|c|c} 
\\[-1.8ex]\hline 
\hline \\[-1.8ex]
\multicolumn{5}{c}{\textbf{MSE Results $\widehat \chi_{1t}$ for DGP1 $T = n$}} \\[0.5em]
\hline
 & (60,60) & (120,120) & (240,240) & (480,480) \\ 
\hline \\[-1.8ex] 
\multicolumn{5}{c}{$\tau = 0, \delta = 0$} \\
\hline
spca2 & $0.163$ & $0.141$ & $0.132$ & $0.124$ \\ 
dpca & $0.272$ & $0.166$ & $0.097$ & $0.056$ \\ 
fdl & $0.049$ & $0.023$ & $0.010$ & $0.005$ \\ 
spca1 & $0.322$ & $0.345$ & $0.354$ & $0.356$ \\ 
spca2 & $0.163$ & $0.141$ & $0.132$ & $0.124$ \\ 
spca3 & $0.218$ & $0.171$ & $0.145$ & $0.132$ \\ 
spca5 & $0.307$ & $0.224$ & $0.171$ & $0.147$ \\ 
spca9 & $0.474$ & $0.315$ & $0.218$ & $0.174$ \\ \hline
\multicolumn{5}{c}{$\tau = 0.5, \delta = 0$} \\
\hline
spca2 & $0.264$ & $0.242$ & $0.235$ & $0.231$ \\ 
dpca & $0.288$ & $0.167$ & $0.089$ & $0.056$ \\ 
fdl & $0.060$ & $0.026$ & $0.012$ & $0.006$ \\ 
spca1 & $0.329$ & $0.350$ & $0.352$ & $0.358$ \\ 
spca2 & $0.264$ & $0.242$ & $0.235$ & $0.231$ \\ 
spca3 & $0.296$ & $0.255$ & $0.242$ & $0.235$ \\ 
spca5 & $0.372$ & $0.289$ & $0.257$ & $0.243$ \\ 
spca9 & $0.543$ & $0.362$ & $0.290$ & $0.259$ \\ 
\hline
\multicolumn{5}{c}{$\tau = 0, \delta = 0.5$} \\
\hline
spca2 & $0.247$ & $0.204$ & $0.189$ & $0.174$ \\ 
dpca & $0.303$ & $0.165$ & $0.101$ & $0.063$ \\ 
fdl & $0.082$ & $0.036$ & $0.017$ & $0.008$ \\ 
spca1 & $0.351$ & $0.357$ & $0.359$ & $0.360$ \\ 
spca2 & $0.247$ & $0.204$ & $0.189$ & $0.174$ \\ 
spca3 & $0.328$ & $0.244$ & $0.208$ & $0.186$ \\ 
spca5 & $0.471$ & $0.326$ & $0.252$ & $0.210$ \\ 
spca9 & $0.662$ & $0.449$ & $0.332$ & $0.254$ \\ 
\hline
\multicolumn{5}{c}{$\tau = 0.5, \delta = 0.5$} \\
\hline
spca2 & $0.336$ & $0.301$ & $0.288$ & $0.279$ \\ 
dpca & $0.296$ & $0.178$ & $0.102$ & $0.063$ \\ 
fdl & $0.089$ & $0.039$ & $0.020$ & $0.010$ \\ 
spca1 & $0.348$ & $0.362$ & $0.360$ & $0.360$ \\ 
spca2 & $0.336$ & $0.301$ & $0.288$ & $0.279$ \\ 
spca3 & $0.402$ & $0.329$ & $0.300$ & $0.286$ \\ 
spca5 & $0.517$ & $0.386$ & $0.328$ & $0.300$ \\ 
spca9 & $0.727$ & $0.492$ & $0.383$ & $0.329$ \\ 
\hline \\
\end{tabular} 
\caption{MSE of different estimators for $\chi_{1t}$ for DGP\eqref{eq: innocent factor model} 500 replications: \texttt{spca}-\texttt{r} = static principal components with \texttt{r} factors, all inconsistent, \texttt{dpca} = dynamic principal components with $q = 1$ dynamic factor, \texttt{fdl} = finite distributed lags approach regressing on the first normalized principal component and its first lag.} 
  \label{tab: res non consist T = n} 
\end{table}
%


\begin{table}[!htbp] \centering
\begin{tabular}{@{\extracolsep{5pt}} c|c|c|c|c} 
\\[-1.8ex]\hline 
\hline \\[-1.8ex]
\multicolumn{5}{c}{\textbf{Coverage Rates, $n > T$}} \\[0.5em]
\hline
$(n,T)$ & (60,30) & (120,60) & (240,120) & (480,240) \\
\hline \\[-1.8ex]
\hline \\[-1.8ex]
$e_{1t}^\chi$, $\tau = 0$, $\delta = 0$   & 0.846 & 0.898 & 0.914 & 0.956 \\
$\chi_{1t}$, $\tau = 0$, $\delta = 0$     & 0.894 & 0.914 & 0.940 & 0.948 \\
$C_{1t}$, $\tau = 0$, $\delta = 0$        & 0.912 & 0.924 & 0.942 & 0.912 \\
\hline \\[-1.8ex]
$e_{1t}^\chi$, $\tau = 0.5$, $\delta = 0$ & 0.806 & 0.880 & 0.908 & 0.952 \\
$\chi_{1t}$, $\tau = 0.5$, $\delta = 0$   & 0.864 & 0.920 & 0.938 & 0.954 \\
$C_{1t}$, $\tau = 0.5$, $\delta = 0$      & 0.824 & 0.892 & 0.936 & 0.930 \\
\hline \\[-1.8ex]
$e_{1t}^\chi$, $\tau = 0$, $\delta = 0.5$ & 0.878 & 0.920 & 0.930 & 0.958 \\
$\chi_{1t}$, $\tau = 0$, $\delta = 0.5$   & 0.858 & 0.896 & 0.916 & 0.926 \\
$C_{1t}$, $\tau = 0$, $\delta = 0.5$      & 0.840 & 0.872 & 0.910 & 0.916 \\
\hline \\[-1.8ex]
$e_{1t}^\chi$, $\tau = 0.5$, $\delta = 0.5$ & 0.852 & 0.896 & 0.908 & 0.918 \\
$\chi_{1t}$, $\tau = 0.5$, $\delta = 0.5$   & 0.846 & 0.862 & 0.902 & 0.902 \\
$C_{1t}$, $\tau = 0.5$, $\delta = 0.5$      & 0.818 & 0.878 & 0.874 & 0.900 \\
\hline \\
\end{tabular}
\caption{Coverage rates for asymptotic $1-\alpha = 95\%$-confidence intervals of $\chi_{1,10}$, $e_{1,10}^{\chi}$ and $C_{1,10}$ over $B = 500$ replications.}
\label{tab: cov rates n > T}
\end{table}

\begin{table}[!htbp] \centering
\begin{tabular}{@{\extracolsep{5pt}} c|c|c|c|c} 
\\[-1.8ex]\hline 
\hline \\[-1.8ex]
\multicolumn{5}{c}{\textbf{Coverage Rates, $n = T$}} \\[0.5em]
\hline
$(n,T)$ & (60,60) & (120,120) & (240,240) & (480,480) \\
\hline \\[-1.8ex]
\hline \\[-1.8ex]
$e_{1t}^\chi$, $\tau = 0$, $\delta = 0$   & 0.862 & 0.936 & 0.922 & 0.940 \\
$\chi_{1t}$, $\tau = 0$, $\delta = 0$     & 0.924 & 0.934 & 0.938 & 0.944 \\
$C_{1t}$, $\tau = 0$, $\delta = 0$        & 0.896 & 0.920 & 0.950 & 0.952 \\
\hline \\[-1.8ex]
$e_{1t}^\chi$, $\tau = 0.5$, $\delta = 0$ & 0.840 & 0.898 & 0.914 & 0.944 \\
$\chi_{1t}$, $\tau = 0.5$, $\delta = 0$   & 0.892 & 0.924 & 0.932 & 0.936 \\
$C_{1t}$, $\tau = 0.5$, $\delta = 0$      & 0.884 & 0.916 & 0.924 & 0.930 \\
\hline \\[-1.8ex]
$e_{1t}^\chi$, $\tau = 0$, $\delta = 0.5$ & 0.870 & 0.936 & 0.940 & 0.948 \\
$\chi_{1t}$, $\tau = 0$, $\delta = 0.5$   & 0.886 & 0.926 & 0.910 & 0.912 \\
$C_{1t}$, $\tau = 0$, $\delta = 0.5$      & 0.890 & 0.902 & 0.914 & 0.934 \\
\hline \\[-1.8ex]
$e_{1t}^\chi$, $\tau = 0.5$, $\delta = 0.5$ & 0.858 & 0.894 & 0.918 & 0.940 \\
$\chi_{1t}$, $\tau = 0.5$, $\delta = 0.5$   & 0.862 & 0.920 & 0.914 & 0.930 \\
$C_{1t}$, $\tau = 0.5$, $\delta = 0.5$      & 0.860 & 0.898 & 0.936 & 0.920 \\
\hline \\
\end{tabular}
\caption{Coverage rates for asymptotic $1-\alpha = 95\%$-confidence intervals of $\chi_{1,10}$, $e_{1,10}^{\chi}$ and $C_{1,10}$ over $B = 500$ replications.}
\label{tab: cov rates n = T}
\end{table}

\end{document}